\documentclass[aps,floats]{revtex4-2}
\usepackage{amsmath,amssymb}
\usepackage{graphicx,epsfig}
\usepackage[greek,english]{babel}
\usepackage{bbold}

\begin{document}
\bibliographystyle {plain}

\pdfoutput=1
\def\oppropto{\mathop{\propto}} 
\def\opsimeq{\mathop{\simeq}}
\def\opoverderline{\mathop{\overline}}
\def\operarrow{\mathop{\longrightarrow}}
\def\opsim{\mathop{\sim}}

\def\opmin{\mathop{\min}} 
\def\opmax{\mathop{\max}} 
\def\oplim{\mathop{\lim}}

\title{ Using the slowest observable in one-dimensional Markov processes  \\
 to construct quasi-exactly-solvable generators with two explicit levels
 } 


\author{C\'ecile Monthus}
\affiliation{Universit\'e Paris-Saclay, CNRS, CEA, Institut de Physique Th\'eorique, 91191 Gif-sur-Yvette, France}


\begin{abstract}

The construction of Quasi-Exactly-Solvable quantum Hamiltonians where only the first two eigenstates $\Phi_0(x)$ and $\Phi_1(x)$ of energies $E_0$ and $E_1$ are explicit is revisited from the point of view of one-dimensional Markov processes satisfying detailed-balance, whose generators are related to quantum Hamiltonians via similarity transformations. Here, the lowest energy vanishes $E_0=0$ and is associated with the conservation of probability and with the steady state $P_*(x)$, while $E_1>0$ is the rate that governs the exponential relaxation towards the steady-state, and is associated with the slowest observable $L_1(x)$ that corresponds to the ratio  $  \frac{\Phi_1(x) }{\Phi_0(x)}$  of the two quantum eigenstates. Our main conclusion is that the Markov perspective leads to interesting re-interpretations and that the construction of quasi-exactly-solvable models with $N=2$ explicit levels is more intuitive and technically simpler when one takes the slowest observable $L_1(x)$ as the central object from which all the other properties can be reconstructed. This general approach is then applied to Fokker-Planck generators in continuous space and to Markov jump generators on the lattice.

\end{abstract}

\maketitle

\section{ Introduction }

In one-dimensional quantum mechanics, it is important to distinguish between :
 (i) exactly-solvable Hamiltonians when all the eigenstates can be explicitly computed (see the reviews \cite{review_susyquantum,Mielnik_review,sasaki_reviewSusyQM,Review_factorization,UnificationSolvable2022}
with various scopes on supersymmetric quantum mechanics and the factorization method); 
  (ii) quasi-exactly-solvable Hamiltonians where only a finite number $N$ of eigenstates can be written explicitly
  (see the book \cite{QESBook}, the review \cite{TurbinerReview}, the pedagogical introduction \cite{Shifman} and references therein for the various possible constructions for arbitrary $N$).
 Since the case $N=1$ is trivial (indeed one can choose the ground state wave function $\Phi_0(x)$, that should be normalizable with no nodes, and then construct the corresponding scalar potential from the eigenvalue equation), 
various works have focused on the simplest non-trivial case $N=2$ where one wishes to be 
able to write explicitly both the ground state $\Phi_0(x)$ of energy $E_0$ and the first excited state $\Phi_1(x)$ of energy $E_1$\cite{Tkachuk98,Tkachuk98bis,Tkachuk99,Tkachuk01,N2Examples,Quesne}
(see also the related work \cite{3Eigen} for the generalization to $N=3$ eigenstates $\Phi_{n=0,1,2}(x)$).

The goal of the present paper is to revisit these constructions for $N=2$ levels
from the point of view of one-dimensional Markov processes, either for Fokker-Planck generators in continuous space
or for Markov jump generators on the lattice.
Here the lowest energy vanishes $E_0=0$ and is associated with the steady state $P_*(x)$ and with the conservation of probability,
while $E_1>0$ is the rate that governs the exponential relaxation towards the steady-state,
and is associated with the slowest observable $L_1(x)$ that will play a major role in the present paper.

Since Markov generators satisfying detailed-balance are related to hermitian quantum Hamiltonians
via similarity transformations (see the textbooks \cite{gardiner,vankampen,risken} 
and various applications  \cite{glauber,Felderhof,siggia,referee,kimball,peschel,jpb_antoine,pierre,texier,us_eigenvaluemethod,Castelnovo,c_lyapunov,us_gyrator,us_kemeny,c_pearson,c_boundarydriven,c_largedevDBsusy,c_susyFP}), one might consider at first that one just needs to write the dictionary between the two frameworks, but we will see that the Markov perspective leads to both technical simplifications
and interesting re-interpretations with valuable insights:

(i) the factorization of 
supersymmetric quantum Hamiltonians ${\bold H}={\bold Q}^{\dagger} {\bold Q}$ has for counterpart the natural factorization of 
Markov generator ${\bold G}=-\nabla {\bold J} $ into the divergence operator $\nabla $ and the current operator ${\bold J}$
as a consequence of the continuity equation
(note that this factorization of the Markov generator based on the continuity equation is also useful in $d>1$ for non-equilibrium processes breaking detailed-balance
as discussed in the recent work \cite{c_susySVD});

(ii) the supersymmetric quantum partner ${\breve {\bold H}}={\bold Q} {\bold Q}^{\dagger} $ has for counterpart
the Markov partner $\breve{{\bold G}}=- {\bold J} \nabla$ that governs the dynamics of the currents and that has thus a very direct physical meaning in the Markov perspective.

(iii) the rewriting of the quantum partner as ${\breve {\bold H}} = {\bold Q} {\bold Q}^{\dagger} = E_1+ {\bold Q}^{\dagger}_1 {\bold Q}_1$ 
has for counterpart the Doob transformation that produces a genuine Markov generator ${\bold G}^{[1]} $ out of the partner $\breve{{\bold G}} $ using its left eigenvector ${\breve L}_1(x) =    \nabla L_1(x) $.

Our main conclusion is that the left eigenvector $L_1(x)$ of the initial Markov generator ${\bold G}=- \nabla {\bold J} $
and the left eigenvector ${\breve L}_1(x) =    \nabla L_1(x) $ of the Markov partner $\breve{{\bold G}}=- {\bold J} \nabla$
are very useful building blocks to construct quasi-exactly-solvable Markov models with $N=2$ explicit levels.

The paper is organized in three main parts as follows :

$ \bullet$ The general analysis summarized above is described in detail in section \ref{sec_general}
with unifying notations independent of the continuous/discrete nature of the space.

$ \bullet$ All the other sections of the main text concern the specific properties of
Fokker-Planck generators as differential operators involving the positive diffusion coefficient $D(x)$ and the Fokker-Planck force $F(x)$
: section \ref{sec_FPx} describes the analytical counterparts of the algebraic properties described in section \ref{sec_general},
section \ref{sec_2Eigen} is devoted to the construction of Fokker-Planck models with explicit eigenstates for $E_0=0$ and $E_1>0$;
finally, we describe the role of the slowest observable in arbitrary changes of variables $x \to {\mathring x}$ in section \ref{sec_ChangeVariables},
before considering two simplifying new variables, namely the variable $y$ where the slowest observable is linear in section \ref{sec_y},
and the variable $z$ with the diffusion coefficient is unity in section \ref{sec_z}.

$ \bullet$ In Appendix \ref{app_Jump}, the general analysis of section \ref{sec_general}
is applied to Markov jump processes on the one-dimensional lattice, 
where generators correspond to finite-difference operators, instead of the differential operators for the Fokker-Planck dynamics
considered in the main text, in order to stress the similarities and the differences.

Our conclusions are summarized in section \ref{sec_conclusion}.



\section{ Properties of one-dimensional Markov processes with detailed-balance}

\label{sec_general}

In this section, we discuss the general properties of one-dimensional reversible Markov processes
that are independent of the continuous/discrete nature of the space.
Let us first introduce the notations based on the divergence operator $\nabla$ 
and on two potentials $U(x)$ and $U_I(x)$
that simplify the analysis and make obvious many standard properties


\subsection{ Unified notations for continuous/discrete space using the operator $\nabla$ and two potentials $U(x)$ and $U_I(x)$}

\subsubsection{ Continuity Equation for the probability density $P_t(x) $ involving the divergence of the current $J_t(x) $}

The dynamics of the probability density $P_t(x) $ is governed by 
the continuity equation that involves the divergence of the current $J_t(x) $
\begin{eqnarray}
 \partial_t P_t(x)   && =   - \nabla J_t(x)
 \label{Acontinuitynabla}
\end{eqnarray}
where the explicit form of the operator $\nabla $ depends on the continuous or discrete nature of the space
\begin{eqnarray}
\nabla &&\equiv \frac{\partial}{\partial x}  \ \ \ \ \ \ \ \ \ \ \ \ \text{ derivative operator in continuous space }
\nonumber \\
 \nabla &&\equiv e^{ \frac{1}{2} \frac{\partial}{\partial x} } - e^{ - \frac{1}{2} \frac{\partial}{\partial x }} \ \ \ \text{ finite-difference operator in discrete space}  
 \label{Adefnabla}
\end{eqnarray}
The consequences of these two explicit forms will be discussed separately in sections of the main text concerning Fokker-Planck dynamics
in continuous space and in Appendix \ref{app_Jump} concerning Markov jump dynamics in discrete space. 
However in the present section it is useful to analyze first all the common 
consequences
based on the two main properties of the operator $\nabla$, 
namely that $\nabla$ is anti-hermitian and that $\nabla$ annihilates any constant 
\begin{eqnarray}
\nabla^{\dagger} && = -  \nabla
\nonumber \\
 \nabla (constant) && =0
 \label{AnablaProp}
\end{eqnarray}


\subsubsection{ Parametrization of the current operator ${\bold J} $ producing the current $J_t(x) = {\bold J} P_t(x) $}

The current $ J_t(x)$ can be computed from the probability $P_t(x)$ via some current operator ${\bold J}$
\begin{eqnarray}
 J_t(x) = {\bold J} P_t(x)
 \label{AdefJop}
\end{eqnarray}

We will focus on models where the steady state $P_*(x)$ that will be parametrized 
by the potential $U(x)$ and some normalization $Z$
\begin{eqnarray}
  P^*(x)  = \frac{ e^{ -U(x)} }{Z} 
 \label{APstarx}
\end{eqnarray}
is associated with the vanishing steady current $J_*(x)=0$ (Detailed-Balance)
\begin{eqnarray}
0= J_*(x) = {\bold J} P_*(x) =  {\bold J} \frac{ e^{ -U(x)} }{Z} 
 \label{AJstarvanish}
\end{eqnarray}
As a consequence in the present paper, it will be convenient to parametrize the current operator as 
\begin{eqnarray}
{\bold J} && =   - e^{- U_I(x)} \nabla e^{U(x)}
 \label{AJUIU}
\end{eqnarray}
where $U_I(x) $ is another function needed to define the model besides the potential $U(x)$ 
parametrizing the steady state in Eq. \ref{APstarx},
while its adjoint reads
\begin{eqnarray}
{\bold J}^{\dagger}  =  - e^{U(x)}   \nabla^{\dagger} e^{- U_I(x)}=  e^{U(x)}   \nabla e^{- U_I(x)}
 \label{AJUIUdagger}
\end{eqnarray}

\subsubsection{ Discussion }

In summary, for the general analysis of the present section, 
it is convenient to parametrize the one-dimensional reversible Markov models by the two potentials $U(x)$ and $U_I(x)$,
whose exponentials appear around the operator $\nabla$ in the current operator ${\bold J}$ of Eq. \ref{AJUIU}
and its adjoint ${\bold J}^{\dagger}  $ of Eq. \ref{AJUIUdagger}.
The link with more standard parametrizations in terms of two other functions
will be described later for Fokker-Planck dynamics in continuous space in section \ref{sec_FPx}
and for Markov-jump processes in  Appendix \ref{app_Jump}.


\subsection{ Spectral properties of the Markov generator ${\bold G} =  - \nabla  {\bold J} $ governing the dynamics of the probability
$P_t(  x ) $}

The Markov dynamics for the probability density $P_t(  x ) $ alone
obtained from Eqs \ref{Acontinuitynabla} and \ref{AdefJop}
\begin{eqnarray}
 \partial_t P_t( x) =  - \nabla {\bold J} P_t(x) \equiv {\bold G} P_t(x)   
\label{AMarkov}
\end{eqnarray}
is governed by the Markov forward generator ${\bold G} $ 
\begin{eqnarray}
 {\bold G}  \equiv - \nabla  {\bold J}  
 =   \nabla   e^{- U_I(x) } \nabla  e^{ U(x) }  
\label{AgeneratorUIU}
\end{eqnarray}

The averaged-value of the observable $O(x)$ at time $t$ computed from the probability $P_t(x)$
\begin{eqnarray}
{\mathbb E}_t [ O( x ) ]  \equiv \langle O \vert P_t \rangle 
\label{Oxav}
\end{eqnarray}
then follows the dynamics
\begin{eqnarray}
\partial_t {\mathbb E}_t [ O( x ) ] && = \langle O \vert \bigg( \partial_t \vert P_t \rangle \bigg)
=\langle O \vert  {\bold G} \vert P_t \rangle
= \langle {\bold G}^{\dagger} O \vert  P_t \rangle
=  {\mathbb E}_t [ ({\bold G}^{\dagger} O ) ( x) ]  
  \label{ADynObs}
\end{eqnarray}
governed by the adjoint operator
\begin{eqnarray}
{\bold G}^{\dagger}={\bold J}^{\dagger}  \nabla
= e^{U(x)}   \nabla e^{- U_I(x)} \nabla
 \label{AGadjoint}
\end{eqnarray}


\subsubsection{ Similarity transformation ${\bold G}   
=  -  e^{- \frac{U(x) }{2} } {\bold H} e^{ \frac{U(x) }{2} }  $ towards a hermitian supersymmetric quantum Hamiltonian
${\bold H} = {\bold Q}^{\dagger} {\bold Q}$ }

The factorized form of Eq. \ref{AgeneratorUIU}
for the Markov generator ${\bold G} $ suggests the rewriting
\begin{eqnarray}
 {\bold G}   
&& = - e^{- \frac{U(x) }{2} }\left( - e^{ \frac{U(x) }{2} }  \nabla  e^{- \frac{U_I(x) }{2} }  \right) 
 \left(  e^{- \frac{U_I(x) }{2} }  \nabla e^{ \frac{U(x) }{2} }  
 \right) e^{ \frac{U(x) }{2} }  
 \nonumber \\
 && \equiv -  e^{- \frac{U(x) }{2} } {\bold Q}^{\dagger} {\bold Q} e^{ \frac{U(x) }{2} }
 \equiv  -  e^{- \frac{U(x) }{2} } {\bold H} e^{ \frac{U(x) }{2} } 
\label{AgeneratorSimilarity}
\end{eqnarray}
that makes obvious the well-known similarity transformation towards a hermitian supersymmetric quantum Hamiltonian
\begin{eqnarray}
{\bold H} = {\bold H}^{\dagger}= - e^{ \frac{U(x) }{2} }  \nabla  e^{- U_I(x) }  \nabla e^{ \frac{U(x) }{2} } 
= {\bold Q}^{\dagger} {\bold Q}  
\label{AquantumHsusy}
\end{eqnarray}
involving the operator 
\begin{eqnarray}
{\bold Q}  \equiv   e^{- \frac{U_I(x) }{2} }  \nabla  e^{ \frac{U(x) }{2} }
\label{Aqsusy}
\end{eqnarray}
and its adjoint
\begin{eqnarray}
{\bold Q}^{\dagger}  &&\equiv  -e^{ \frac{U(x) }{2} } \nabla e^{- \frac{U_I(x) }{2} } 
\label{Aqdaggersusy}
\end{eqnarray}
whenever detailed-balance holds, as already mentioned in the Introduction
 with the textbooks \cite{gardiner,vankampen,risken} 
and many specific applications  \cite{glauber,Felderhof,siggia,referee,kimball,peschel,jpb_antoine,pierre,texier,us_eigenvaluemethod,Castelnovo,c_lyapunov,us_gyrator,us_kemeny,c_pearson,c_boundarydriven,c_largedevDBsusy,c_susyFP}.


\subsubsection{ Links between the spectral properties of the Markov generator ${\bold G} $ and of
the quantum Hamiltonian ${\bold H} = {\bold Q}^{\dagger} {\bold Q}$ }

The spectral decomposition of the evolution operator $e^{- {\bold H} t} $ associated with the quantum Hamiltonian $ {\bold H} $
\begin{eqnarray}
 e^{- {\bold H} t} 
= \sum_{n=0}^{+\infty} e^{- t E_n} \vert  \Phi_n \rangle \langle  \Phi_n \vert 
\label{Apsispectral}
\end{eqnarray}
involves its real eigenvalues $E_n$, that will be assumed to be all discrete for the present general discussion to simplify the notations, while the corresponding 
real eigenstates $\Phi_n(x)$ satisfy the eigenvalue equations
\begin{eqnarray}
E_n \Phi_n(x) = {\bold H}   \Phi_n(x) =  {\bold Q}^{\dagger} {\bold Q}   \Phi_n(x)
\label{Ahsusyeigen}
\end{eqnarray}
and the orthonormalization
\begin{eqnarray}
\delta_{n n' } = \langle \Phi_n \vert \Phi_{n'} \rangle 
\label{Aorthophin}
\end{eqnarray}
The quantum ground state $\Phi_0(x)$ associated with the vanishing energy $E_0=0$
is simply the square-root of the steady state $P^*(x)$ of Eq. \ref{APstarx}
\begin{eqnarray}
  \Phi_0(x) =   \sqrt{ P^*(x) } = \frac{ e^{ - \frac{ U(x)}{2} } }{\sqrt { Z} }
 \label{Aphi0}
\end{eqnarray}

The similarity transformation of Eq. \ref{AgeneratorSimilarity} yields that the evolution
operator $e^{ {\bold G} t} $ associated with the Markov generator $ {\bold G} $
can be computed from the quantum spectral decomposition of Eq. \ref{Apsispectral}
\begin{eqnarray}
P_t(x\vert x_0) && = \langle x \vert e^{ {\bold G} t}  \vert x_0 \rangle
 =e^{-  \frac{ U(x)}{2}} \langle x \vert e^{- {\bold H} t} \vert x_0 \rangle e^{  \frac{ U(x_0)}{2}} 
 \nonumber \\
 && = \Phi_0(x) \left[  \sum_{n=0}^{+\infty} e^{- t E_n}  \Phi_n (x)  \Phi_n (x_0) \right] \frac{1}{ \Phi_0(x_0)}
  \nonumber \\
 && \equiv   \sum_{n=0}^{+\infty} e^{- t E_n}  R_n (x)  L_n (x_0) 
\label{ASpectralFP}
\end{eqnarray}
and thus involves the same real eigenvalues $E_n$,
while the left eigenvectors $L_n(x)$ and the right eigenvectors $R_n(x)$ given by
\begin{eqnarray}
 L_n ( x )  && \equiv   \frac{\Phi_n(x) }{\Phi_0(x)}
\nonumber \\
R_n(x) && \equiv \Phi_0(x) \Phi_n(x)  = \Phi_0^2(x)L_n(x) = P^*(x) L_n(x) 
\label{Arnlnphin}
\end{eqnarray}
  satisfy the eigenvalue equations
\begin{eqnarray}
 - E_n L_n(x) && =  {\bold G}^{\dagger} L_n(x) = e^{U(x)}   \nabla e^{- U_I(x)} \nabla L_n(x)
 \nonumber \\
 - E_n R_n(x) && = {\bold G} R_n(x) = \nabla   e^{- U_I(x) } \nabla  e^{ U(x) }  R_n(x) 
 \label{AEigenRL}
\end{eqnarray}
and the bi-orthonormalization inherited from Eq. \ref{Aorthophin}
\begin{eqnarray}
\delta_{n n' } = \langle L_n \vert R_{n'} \rangle =  \langle L_n \vert L_{n'} P_* \rangle
 \label{Aortholr}
\end{eqnarray}
where the last expression obtained by replacing $R_{n'}(x) = L_{n'}(x) P^*(x) $ means that the left eigenvectors $L_n(x)$ 
are an orthogonal family with respect to the steady state $P^*(x) $.


\subsubsection{ Discussion : why it is simpler to focus on the left eigenvectors $L_n(x)$ only}

To analyze the spectral properties of the Markov generator ${\bold G}$, 
it is actually simpler to focus on the left eigenvectors $L_n(x)$ only, since they satisfy the simpler eigenvalue Eq. \ref{AEigenRL}, starting with the trivial eigenvector for $n=0$ associated with the conservation of probability
\begin{eqnarray}
 L_{n=0}(x)  = 1
\label{AEigenL0}
\end{eqnarray}
The other left eigenvectors $L_n(x)$ with $n>0$ correspond to simple observables relaxing towards zero
with the single exponentials $e^{-t E_n}$ (instead of all these exponentials for the general observable $O(x)$)
that form an orthogonal family with respect to the steady state $P^*(x) $.

Note that even within the quantum perspective,
these left eigenvectors  $L_n ( x ) =  \frac{\Phi_n(x) }{\Phi_0(x)}$ are actually also simpler 
since they represent the ratios between the excited states $ \Phi_n(x)$ and the ground state $\Phi_0(x)$,
and already in the simplest example of the harmonic oscillator, it is well-known that
 the ratios $ L_n ( x ) =  \frac{\Phi_n(x) }{\Phi_0(x)}$ reduce to polynomials and are thus simpler than $\Phi_n(x) $ 
 that contain the additional exponential ground state $\Phi_0(x)$.


\subsection{ Reminder on supersymmetric pairs of quantum Hamiltonians ${\bold H} = {\bold Q}^{\dagger} {\bold Q} $ and ${\breve{\bold H}} = {\bold Q} {\bold Q}^{\dagger}  $  }

In the field of supersymmetric quantum mechanics (see the reviews \cite{review_susyquantum,Mielnik_review,sasaki_reviewSusyQM,Review_factorization,UnificationSolvable2022}),
it is standard to introduce the supersymmetric partner ${\breve{\bold H}} = {\bold Q} {\bold Q}^{\dagger} $ of the quantum Hamiltonian ${\bold H} = {\bold Q}^{\dagger} {\bold Q} $ of Eq. \ref{AquantumHsusy} that read in our present notations
\begin{eqnarray}
{\breve{\bold H}} = {\bold Q} {\bold Q}^{\dagger} 
 = -  e^{- \frac{U_I(x) }{2} }  \nabla  e^{ U(x) } \nabla e^{- \frac{U_I(x) }{2} } 
\label{AquantumHpartner}
\end{eqnarray}


\subsubsection{ Links between the spectral properties of the quantum Hamiltonian $ {\bold H} = {\bold Q}^{\dagger}  {\bold Q}$ and its partner
${\breve{\bold H}} = {\bold Q} {\bold Q}^{\dagger}$ }

The main output of supersymmetric quantum mechanics (see the reviews \cite{review_susyquantum,Mielnik_review,sasaki_reviewSusyQM,Review_factorization,UnificationSolvable2022})
is that the spectral decomposition of the evolution operator $e^{- {\breve{\bold H}} t} $ associated with the supersymmetric partner $ {\breve{\bold H}} = {\bold Q} {\bold Q}^{\dagger} $
\begin{eqnarray}
 e^{- {\breve{\bold H}} t} 
= \sum_{n=1}^{+\infty} e^{- t E_n} \vert {\breve \Phi}_n \rangle \langle  {\breve \Phi}_n \vert 
\label{Ahatpsispectral}
\end{eqnarray}
involves the non-vanishing eigenvalues $E_{n \ne 0}$ of the Hamiltonian $ {\bold H} = {\bold Q}^{\dagger}  {\bold Q}$,
while the corresponding orthonormalized eigenvectors ${\breve \Phi}_n (x)$ satisfying the eigenvalue equations
\begin{eqnarray}
E_n {\breve \Phi}_n (x)= {\breve{\bold H}}   {\breve \Phi}_n (x)= {\bold Q}  {\bold Q}^{\dagger}  {\breve \Phi}_n(x)
\label{Ahhatsusyeigen}
\end{eqnarray}
 can be obtained from the excited states $\Phi_{n \ne 0} $ of the Hamiltonian ${\bold H} $ 
via the application of the operator ${\bold Q}$ that annihilates the ground state ${\bold Q} \Phi_0(x)=0 $,
while the reciprocal relations involve the adjoint ${\bold Q}^{\dagger} $ 
\begin{eqnarray}
 {\breve \Phi}_n(x) &&= \frac{ {\bold Q}   \Phi_n (x) }{\sqrt{E_n} } 
  = \frac{ 1 }{\sqrt{E_n} } e^{- \frac{U_I(x) }{2} }  \nabla  e^{ \frac{U(x) }{2} }  \Phi_n (x)
 \nonumber \\
  \Phi_n(x) && = \frac{ {\bold Q}^{\dagger}  {\breve \Phi}_n(x) }{\sqrt{E_n} }
  = - \frac{ 1 }{\sqrt{E_n} } e^{ \frac{U(x) }{2} } \nabla e^{- \frac{U_I(x) }{2} }   {\breve \Phi}_n(x)
\label{Ahatphifromphi}
\end{eqnarray}

At the operator level, these relations mean that the excited states $\Phi_{n \ne 0} $ of the Hamiltonian ${\bold H} ={\bold Q}^{\dagger}  {\bold Q}$
and the eigenstates ${\breve \Phi}_n (x) $ of the partner ${\breve{\bold H}}={\bold Q}  {\bold Q}^{\dagger}   $
appears in the Singular Value Decompositions (SVD) of the operators ${\bold Q} $ and ${\bold Q}^{\dagger} $
\begin{eqnarray}
 {\bold Q} && = \sum_{n=1}^{+\infty} \sqrt{E_n} \vert {\breve \Phi}_n \rangle \langle  \Phi_n \vert
 \nonumber \\
 {\bold Q}^{\dagger} && = \sum_{n=1}^{+\infty} \sqrt{E_n} \vert  \rangle \Phi_n \langle   {\breve \Phi}_n\vert
\label{ASVDQQdagger}
\end{eqnarray}


\subsubsection{ Rewriting the quantum partner ${\breve{\bold H}} = {\bold Q} {\bold Q}^{\dagger} $ as 
${\breve{\bold H}} = E_1+ \left( {\bold Q}^{[1]}\right)^{\dagger}  {\bold Q}^{[1]} $ where $ {\bold Q}^{[1]}$ annihilates the ground state ${\breve \Phi}_1(x) $  }

As explained around Eq. \ref{Ahatpsispectral}, the quantum partner ${\breve{\bold H}} = {\bold Q} {\bold Q}^{\dagger} $
has the positive ground state energy $E_1>0$, so that it is useful to introduce  
the new potential  $U^{[1]}(x)$ that parametrizes the corresponding positive ground state $ {\breve \Phi}_1(x) $
\begin{eqnarray}
 {\breve \Phi}_1(x) && =   \frac{  e^{- \frac{ U^{[1]}(x)}{2} } }  {  \sqrt{ Z_1 } } 
\label{Abrevephi1x}
\end{eqnarray}
to mimic the parametrization $ \Phi_0(x) =\frac{ e^{ - \frac{ U(x)}{2} } }{\sqrt { Z} }  $ of Eq. \ref{Aphi0}
for the ground state of ${\bold H}$.

Then one wishes to introduce an operator ${\bold Q}^{[1]} $ 
analogous  to ${\bold Q} $ of Eq. \ref{Aqsusy}  
 that annihilates the ground state ${\breve \Phi}_1(x) $ of Eq. \ref{Abrevephi1x}
\begin{eqnarray}
{\bold Q}^{[1]}  && \equiv e^{- \frac{U_I^{[1]}(x) }{2} }  \nabla  e^{ \frac{U^{[1]}(x) }{2} }
\nonumber \\
\left( {\bold Q}^{[1]} \right)^{\dagger}  &&\equiv  -e^{ \frac{U^{[1]}(x) }{2} } \nabla e^{- \frac{U_I^{[1]}(x) }{2} }
\label{Aq1q1daggersusy}
\end{eqnarray}
where the remaining potential $U_I^{[1]}(x) $ has to be chosen so that
the associated quantum Hamiltonian
\begin{eqnarray}
{\bold H}^{[1]} && \equiv \left( {\bold Q}^{[1]} \right)^{\dagger} {\bold Q}^{[1]} 
= -e^{ \frac{U^{[1]}(x) }{2} } \nabla e^{- U_I^{[1]}(x) }  \nabla  e^{ \frac{U^{[1]}(x) }{2} }
\label{Av1fromu1}
\end{eqnarray}
that has $ {\breve \Phi}_1(x) $ as ground state of vanishing energy
\begin{eqnarray}
\Phi^{[1]}_0(x) && = {\breve \Phi}_1(x) = \frac{  e^{- \frac{ U^{[1]}(x)}{2} } }  {  \sqrt{ Z_1 } } 
\ \ \text{ with } \ \ {\bold H}^{[1]}\Phi^{[1]}_0(x) =0
\label{PHi0GSH1}
\end{eqnarray}
corresponds to the difference between the quantum partner ${\breve{\bold H}} = {\bold Q} {\bold Q}^{\dagger} $
and its ground state energy $E_1$ 
\begin{eqnarray}
{\breve{\bold H}} - E_1 \equiv {\bold Q} {\bold Q}^{\dagger} - E_1 &&= \left( {\bold Q}^{[1]} \right)^{\dagger} {\bold Q}^{[1]}  \equiv  {\bold H}^{[1]}
\nonumber \\
\text{ i.e. } 
 \ \ \ \ \ \ \ \ \ \ \ 
  -  e^{- \frac{U_I(x) }{2} }  \nabla  e^{ U(x) } \nabla e^{- \frac{U_I(x) }{2} } 
 -E_1
&& =-e^{ \frac{U^{[1]}(x) }{2} } \nabla e^{- U_I^{[1]}(x) }  \nabla  e^{ \frac{U^{[1]}(x) }{2} }
\label{ApartnerHsusyrewriting}
\end{eqnarray}
The goal is then to compute the two new potentials $U^{[1]}(x) $ and $U_I^{[1]}(x) $
in terms of the two initial potentials $U(x) $ and $U_I(x) $.

In the next subsection, it is interesting to re-interpret this construction concerning supersymmetric quantum Hamiltonians
by
considering the dynamics of the current of the Markov process.


\subsection{ Dynamics of the current $J_t(x)={\bold J} P_t(x)$ governed by the supersymmetric partner ${\breve {\bold G}} 
=      -  {\bold J}\nabla $ of $ {\bold G}  = -\nabla  {\bold J} $ }

The dynamics of the current $J_t( x )= {\bold J} P_t(x)$ of Eq. \ref{AdefJop}
as obtained from the continuity equation of Eq. \ref{Acontinuitynabla}
\begin{eqnarray}
 \partial_t J_t( x ) =  {\bold J}  \big[ \partial_t P_t(x) \big] =      {\bold J} \big[ - \nabla J_t(x)  \big]
 \equiv {\breve {\bold G}} J_t(x)
\label{Acurrentdyn}
\end{eqnarray}
is governed by the supersymmetric partner
\begin{eqnarray}
{\breve {\bold G}} =      -  {\bold J}\nabla =
 e^{- U_I(x) } \nabla e^{ U(x) }   \nabla
\label{AFPpartner}
\end{eqnarray}
where the two operators $\nabla $ and ${\bold J} $ appear in opposite order with respect to the  generator $ {\bold G}  = -\nabla {\bold J} $ of Eq. \ref{AgeneratorUIU},
while its adjoint reads
\begin{eqnarray}
{\breve {\bold G}}^{\dagger} =  \nabla  {\bold J}^{\dagger} 
= \nabla    e^{ U(x) }  \nabla  e^{ -U_I(x) }  
\label{AFPpartneradjoint}
\end{eqnarray}


\subsubsection{ Similarity transformation between the partners ${\breve {\bold G}} 
=      -  {\bold J}\nabla  $ and  
${\breve{\bold H}} = {\bold Q} {\bold Q}^{\dagger}$ with consequences for their spectral properties}

The partner ${\breve {\bold G}} 
=      -  {\bold J}\nabla  $ introduced in Eq. \ref{AFPpartner}
is related to the quantum partner ${\breve{\bold H}} = {\bold Q} {\bold Q}^{\dagger} $ of Eq. \ref{AquantumHpartner}
via the similarity transformation
\begin{eqnarray}
{\breve {\bold G}} && =    e^{- U_I(x) }  \nabla  e^{ U(x) }   \nabla  
\nonumber \\
&& = - e^{- \frac{U_I(x) }{2} }\left( -  e^{- \frac{U_I(x) }{2} }  \nabla  e^{ U(x) }
 \nabla e^{- \frac{U_I(x) }{2} } \right) e^{ \frac{U_I(x) }{2} }  
 \nonumber \\
&& =  -  e^{- \frac{U_I(x) }{2} } {\breve{\bold H}} e^{ \frac{U_I(x) }{2} }  
\label{AFPpartnersimilarity}
\end{eqnarray}

This similarity transformation yields that the spectral decomposition of 
the evolution operator $e^{ {\breve {\bold G}} t} $ associated with the partner ${\breve {\bold G}} $
can be computed from the spectral decomposition of $e^{- {\breve{\bold H}} t} $ written in Eq. \ref{Ahatpsispectral}
\begin{eqnarray}
\langle x \vert e^{ {\breve {\bold G}} t}  \vert x_0 \rangle
 && = e^{- \frac{U_I(x) }{2} } \langle x \vert e^{- {\breve{\bold H}} t} \vert x_0 \rangle   e^{ \frac{U_I(x_0) }{2} }  
 =  e^{- \frac{U_I(x) }{2} } \left[  \sum_{n=1}^{+\infty} e^{- t E_n}  {\breve \Phi}_n (x)  {\breve \Phi}_n (x_0) \right]   e^{ \frac{U_I(x_0) }{2} } 
 \nonumber \\
 && =     \sum_{n=1}^{+\infty} e^{- t E_n}  {\breve R}_n (x)  {\breve L}_n (x_0) 
\label{ASpectralFPhat}
\end{eqnarray}
and thus involves the same eigenvalues $E_{n \ne 0}$,
while the corresponding left eigenvectors ${\breve L}_n ( x ) $ and right eigenvectors ${\breve R}_n(x)  $
are related to the quantum eigenstates ${\breve \Phi}_n $ via
\begin{eqnarray}
{\breve L}_n ( x )  && \equiv  c_n {\breve \Phi}_n(x) e^{ \frac{U_I(x) }{2} }
\nonumber \\
{\breve R}_n(x) && \equiv \frac{1}{c_n} e^{- \frac{U_I(x) }{2} } {\breve \Phi}_n(x) =\frac{1}{c_n^2}  e^{-U_I(x)} {\breve L}_n ( x ) 
\label{Arnlnphinhat}
\end{eqnarray}
where we have included some constant $c_n$ depending on the level $n$
that will be chosen in the next subsection.


\subsubsection{ Links between the spectral properties of the generator 
$ {\bold G} = - \nabla  {\bold J}$
and of its partner ${\breve {\bold G}} =      -  {\bold J}\nabla  $  }

Let us translate the two relations of Eq. \ref{Ahatphifromphi} between the quantum eigenstates $\Phi_n (x) $ 
and $ {\breve \Phi}_n(x) $ by plugging their expressions in terms of the left eigenvectors $L_n(x) $ and ${\breve L}_n ( x ) $ 
using Eqs \ref{Arnlnphin}
and \ref{Arnlnphinhat}
\begin{eqnarray}
\frac{1}{c_n} e^{- \frac{U_I(x) }{2} } {\breve L}_n(x) && = \frac{ \left[  e^{- \frac{U_I(x) }{2} }  \nabla  e^{ \frac{U(x) }{2} }\right]   \frac{e^{- \frac{U(x) }{2} }}{\sqrt Z} L_n(x)  }{\sqrt{  E_n} }
 \nonumber \\
\frac{e^{- \frac{U(x) }{2} }}{\sqrt Z} L_n(x) && = \frac{ \left[ -e^{ \frac{U(x) }{2} } \nabla e^{- \frac{U_I(x) }{2} }  \right]
\frac{1}{c_n} e^{- \frac{U_I(x) }{2} } {\breve L}_n(x) }{\sqrt{E_n} }
\label{AhatphifromphiTRANSLcalcul}
\end{eqnarray}
that reduce to
\begin{eqnarray}
 {\breve L}_n(x) &&= \frac{    c_n }{\sqrt{ Z E_n} } \nabla L_n(x)
 \nonumber \\
 L_n(x) && 
= - \frac{ \sqrt Z }{ c_n\sqrt{E_n} }
 \left[ e^{ U(x)  } \nabla e^{- U_I(x) }  \right] {\breve L}_n(x) 
 = - \frac{ \sqrt Z }{c_n \sqrt{E_n} } {\bold J}^{\dagger } {\breve L}_n(x)
\label{AhatphifromphiTRANSL}
\end{eqnarray}
where the second line involves the operator $J^{\dagger}$ 
as it should in order to reproduce the two left eigenvalue equations for $L_n(x)$ and for ${\breve L}_n(x) $ respectively
\begin{eqnarray}
- E_n L_n(x) && = {\bold J}^{\dagger } \nabla L_n(x) 
= {\bold G}^{\dagger}  L_n(x)
 \nonumber \\
- E_n  {\breve L}_n(x) && = \nabla {\bold J}^{\dagger } {\breve L}_n(x) 
= {\breve {\bold G}}^{\dagger} {\breve L}_n(x)
\label{ATwoEigenLeft}
\end{eqnarray}

In the present paper, it is convenient to choose the normalization
\begin{eqnarray}
    c_n = \sqrt{ Z E_n} 
\label{choicecn}
\end{eqnarray}
that simplify Eq. \ref{AhatphifromphiTRANSL} into
\begin{eqnarray}
 {\breve L}_n(x) &&=  \nabla L_n(x)
 \nonumber \\
 L_n(x) && = - \frac{ \left[ e^{ U(x)  } \nabla e^{- U_I(x) }  \right] {\breve L}_n(x) }{ E_n }  
 = - \frac{ {\bold J}^{\dagger } {\breve L}_n(x) }{ E_n } 
\label{ChoiceTwoleftviaNabla}
\end{eqnarray}
while the right eigenvectors ${\breve R}_n(x) $ of Eq. \ref{Arnlnphinhat}
read 
\begin{eqnarray}
{\breve R}_n(x) && =\frac{1}{Z E_n}  e^{-U_I(x)} {\breve L}_n ( x )  
\label{ARightEigenPartner}
\end{eqnarray}


\subsubsection{ Doob-transformation from the partner ${\breve {\bold G} } $ towards
the Markov generator ${\bold G}^{[1]}$ with steady state $P_*^{[1]}(x)=  {\breve \Phi}_1^2(x) 
= {\breve L}_1(x) {\breve R}_1(x)$ }

In the field of Markov processes, when an operator like the partner ${\breve {\bold G} } $ appears 
with a non-vanishing ground state energy $E_1>0$ associated with the left eigenvector ${\breve L}_1(x) $ 
and with the right eigenvector ${\breve R}_1(x) $, it is often useful to construct 
the associated genuine Markov generator
${\bold G}^{[1]} $ via
the Doob transformation (introduced in 1957 by Doob \cite{Doob57}, while the more recent  
conditioning of Markov processes on their large deviations is described in detail in
 \cite{chetrite_conditioned,chetrite_optimal} and references therein) :
it is based on the following similarity transformation involving the left eigenvector ${\breve L}_1(x) $
\begin{eqnarray}
 {\bold G}^{[1]} && \equiv {\breve L}_1(x) {\breve {\bold G} } \frac{1}{ {\breve L}_1(x) } + E_1
  \label{ADoobRing}
\end{eqnarray}
with the corresponding transformation for the adjoint operators
\begin{eqnarray}
 \left( {\bold G}^{[1]}  \right)^{\dagger} && = \frac{1}{ {\breve L}_1(x) } {\breve {\bold G} }^{\dagger} {\breve L}_1(x) + E_1
  \label{ADoobRingDagger}
\end{eqnarray}

By construction, the operator ${\bold G}^{[1]} $ has a vanishing eigenvalue $E_0^{[1]} =0 $ associated with
 the trivial left eigenvector unity $L_0^{[1]}(x) =1 $ corresponding to the conservation of probability
\begin{eqnarray}
\left( {\bold G}^{[1]}  \right)^{\dagger} (1 ) && = \frac{1}{ {\breve L}_1(x) } \bigg( - E_1  {\breve L}_1(x) \bigg)+ E_1 =0
  \label{ADoobRingDaggerleft}
\end{eqnarray}
while the corresponding right eigenvector $R_0^{[1]}(x) $ corresponding to the normalized steady state $P_*^{[1]}(x) $
is given by
\begin{eqnarray}
R_0^{[1]}(x) = P_*^{[1]}(x) \equiv   {\breve R}_1(x) {\breve L}_1(x)  = {\breve \Phi}_1^2
  \label{AP1star}
\end{eqnarray}
that is indeed annihilated by $ {\bold G}^{[1]} $
\begin{eqnarray}
 {\bold G}^{[1]} P_*^{[1]}(x) && = {\breve L}_1(x) \bigg( {\breve {\bold G} }   {\breve R}_1(x)  \bigg)+ E_1  {\breve R}_1(x) {\breve L}_1(x)
 =  {\breve L}_1(x) \bigg( -E_1  {\breve R}_1(x)  \bigg)+ E_1  {\breve R}_1(x) {\breve L}_1(x) =0
  \label{ADoobRingp1star}
\end{eqnarray}
 This new Markov generator $ {\bold G}^{[1]} $ is related to the new Hamiltonian $ {\bold H}^{[1]} $,
 as described in the next subsection.


\subsubsection{ Link between the new 
Markov generator ${\bold G}^{[1]}$ and the new quantum Hamiltonian ${\bold H}^{[1]} $ with vanishing ground state energy }

The ground state $\Phi^{[1]}_0(x)$ of ${\bold H}^{[1]} $
that coincides with the ground state  $ {\breve \Phi}_1(x) $ of ${\breve {\bold H}}$
can be computed via the relation of Eq. \ref{Ahatphifromphi} for $n=1$ that
involves the first-excited-state $\Phi_1(x) = \Phi_0(x) L_1(x)=\frac{ e^{ - \frac{ U(x)}{2} } }{\sqrt { Z} } L_1(x) $ 
of the initial Hamiltonian ${\bold H}$
\begin{eqnarray}
\Phi^{[1]}_0(x) = {\breve \Phi}_1(x) &&= \frac{ {\bold Q}   \Phi_1 (x) }{\sqrt{E_1} } 
 = \frac{ 1 }{\sqrt{E_1} } \left[  e^{- \frac{U_I(x) }{2} }  \nabla  e^{ \frac{U(x) }{2} } \right] 
   \frac{ e^{ - \frac{ U(x)}{2} } }{\sqrt { Z} } L_1(x)
 \nonumber \\
 &&  =   \frac{ e^{- \frac{U_I(x) }{2} }  \nabla  L_1(x) }{\sqrt{ Z E_1} }  
 =   \frac{ e^{- \frac{U_I(x) }{2} }  {\breve L}_1(x) }{\sqrt{ Z E_1} } 
\label{Abrevephi1xQ}
\end{eqnarray}
where we have used ${\breve L}_1(x) =    \nabla L_1(x) $ of Eq. \ref{ChoiceTwoleftviaNabla}.

The identification with the parametrization $\Phi_0^{[1]}(x) = \frac{  e^{- \frac{ U^{[1]}(x)}{2} } }  {  \sqrt{ Z_1 } }$
of Eq. \ref{PHi0GSH1} with $Z_1=Z E_1$ leads to the following simple 
relations between the new potential $U^{[1]}(x) $ and the initial potential $U_I(x)$
 \begin{eqnarray}
 e^{- \frac{ U^{[1]}(x)}{2} } && = e^{- \frac{U_I(x) }{2} }  {\breve L}_1(x) = e^{- \frac{U_I(x) }{2} }  \nabla L_1(x)
 \nonumber \\
\text{ i.e. } \ \ \ \ \ \  U^{[1]}(x) && =  U_I(x) -2 \ln \big( {\breve L}_1(x)  \big)  = U_I(x) -2 \ln \big( \nabla L_1(x)  \big)
\label{AU1xbreve}
\end{eqnarray}

Plugging the rewriting ${\breve{\bold H}} =  {\bold H}^{[1]} + E_1$
of Eq. \ref{ApartnerHsusyrewriting} 
into the similarity transformation of Eq. \ref{AFPpartnersimilarity}
between ${\breve {\bold G} } $ and ${\breve {\bold H} } $ yields
\begin{eqnarray}
{\breve {\bold G}}  = -  e^{- \frac{U_I(x) }{2} } {\breve{\bold H}} e^{ \frac{U_I(x) }{2} }   =  - e^{- \frac{U_I(x) }{2} }   {\bold H}^{[1]}  e^{ \frac{U_I(x) }{2} }  -E_1
\label{AFPpartnersimilarityrewriting}
\end{eqnarray}
So the corresponding Markov generator ${\bold G}^{[1]}$ constructed via the Doob-transformation of Eq. \ref{ADoobRing} 
\begin{eqnarray}
 {\bold G}^{[1]} 
 && = {\breve L}_1(x) {\breve {\bold G} } \frac{1}{ {\breve L}_1(x) } + E_1
  =-  {\breve L}_1(x)   e^{- \frac{U_I(x) }{2} }  
  {\bold H}^{[1]}  \frac{1}{e^{- \frac{U_I(x) }{2} }  {\breve L}_1(x) } 
\nonumber \\
&&  =  -  e^{- \frac{U^{[1]}(x) }{2} } {\bold H}^{[1]} e^{ \frac{U^{[1]}(x) }{2} } 
=  - \Phi_0^{[1]}(x) {\bold H}^{[1]} \frac{1}{\Phi_0^{[1]}(x)}
  \label{ADoobRingsimilarity}
\end{eqnarray}
is related to the new quantum Hamiltonian $ {\bold H}^{[1]}$ via 
the similarity that involves its
ground state $\Phi_0^{[1]}(x) = \frac{  e^{- \frac{ U^{[1]}(x)}{2} } }  {  \sqrt{ Z_1 } }$,
i.e. it is the analog of the similarity of Eq. \ref{AgeneratorSimilarity}
concerning the initial Markov generator ${\bold G} $ and the initial Hamiltonian ${\bold H}$.

So the new Markov generator $  {\bold G}^{[1]} $ and its adjoint $ $
are parametrized by the two new potentials $U^{[1]}(x) $and $U_I^{[1]}(x)$
via the analog of Eqs \ref{AgeneratorUIU} and Eq. \ref{AFPpartneradjoint}
\begin{eqnarray}
 {\bold G}^{[1]}   =   \nabla   e^{- U_I^{[1]}(x) } \nabla  e^{ U^{[1]}(x) }  
 \nonumber \\
\left( {\bold G}^{[1]}  \right)^{\dagger} 
=  e^{U^{[1]}(x)}   \nabla e^{- U_I^{[1]}(x)} \nabla
\label{AgeneratorUIU1}
\end{eqnarray}


\begin{table}[h]
\centering
\begin{tabular}{|p{6cm}|p{5cm}|p{8cm}|}
\hline
\textbf{Physical meaning }
& \textbf{Factorized expression} 
& \textbf{Main property}  \\
\hline 
Current operator $ {\bold J}$ 
that produces the 
\newline current $J_t(x) = {\bold J} P_t(x)$ associated to the density $P_t(x)$
& ${\bold J}  =   - e^{- U_I(x)} \nabla e^{U(x)} $
& annihilates the steady state $ P^*(x)  = \frac{ e^{ -U(x)} }{Z} $ 
\newline
 (Detailed-Balance : the steady current vanishes \newline
 $  J_*(x) = {\bold J} P_*(x)=0$) 
\\
\hline
Markov Generator governing the dynamics 
\newline 
of the probability density $P_t(x)$
\newline 
$ \partial_t P_t( x) =  {\bold G} P_t(x)   $ 
& ${\bold G}  \equiv - \nabla  {\bold J}  
 =   \nabla   e^{- U_I(x) } \nabla  e^{ U(x) } $ 
& annihilates the steady state $ P^*(x)  = \frac{ e^{ -U(x)} }{Z} $ 

\\
\hline
Adjoint operator governing the dynamics 
\newline 
of averaged values of observables $O(x)$
\newline
$\partial_t {\mathbb E}_t [ O( x ) ] =  {\mathbb E}_t [ ({\bold G}^{\dagger} O ) ( x) ]  $
& ${\bold G}^{\dagger}={\bold J}^{\dagger}  \nabla
= e^{U(x)}   \nabla e^{- U_I(x)} \nabla
 $ 
& annihilates the constants $ {\bold G}^{\dagger} 1 =0 $ since $\nabla 1 =0 $
\newline
(Conservation of the probability normalization)
\\
\hline
Supersymmetric Quantum Hamiltonian   
\newline
obtained from the Markov Generator $ {\bold G} $ 
\newline via the similarity involving $U(x)$
&  ${\bold H} = - e^{ \frac{U(x) }{2} }{\bold G}    e^{ - \frac{U(x) }{2} } $
\newline
$= - e^{ \frac{U(x) }{2} }  \nabla  e^{- U_I(x) }  \nabla e^{ \frac{U(x) }{2} } 
\equiv {\bold Q}^{\dagger} {\bold Q}$ 
& annihilates the ground state $   \Phi_0(x) =   \sqrt{ P^*(x) } = \frac{ e^{ - \frac{ U(x)}{2} } }{\sqrt { Z} }$ 
\newline
corresponding to the square-root of the Markov  steady state $P^*(x) $
 \\
\hline
Annihilation operator
\newline
Creation operator
& ${\bold Q}  =   e^{- \frac{U_I(x) }{2} }  \nabla  e^{ \frac{U(x) }{2} }
 $ 
 \newline
  $ {\bold Q}^{\dagger}  =  -e^{ \frac{U(x) }{2} } \nabla e^{- \frac{U_I(x) }{2} } $
& annihilates the ground state $   \Phi_0(x) =   \sqrt{ P^*(x) } = \frac{ e^{ - \frac{ U(x)}{2} } }{\sqrt { Z} }$
  \\  
\hline
Supersymmetric partner Hamiltonian
& ${\breve{\bold H}} = {\bold Q} {\bold Q}^{\dagger} $
\newline
$ = -  e^{- \frac{U_I(x) }{2} }  \nabla  e^{ U(x) } \nabla e^{- \frac{U_I(x) }{2} } $ 
& 
Its ground state ${\breve \Phi}_1(x) = \frac{ {\bold Q}   \Phi_1 (x) }{\sqrt{E_1} } =   \frac{  e^{- \frac{ U^{[1]}(x)}{2} } }  {  \sqrt{ Z_1 } }$ has the energy $E_1>0$ 
of the first excited state $\Phi_1 (x) $ of ${\bold H}$
  \\  
\hline
New Supersymmetric Hamiltonian  
\newline
with vanishing ground state energy
&  ${\bold H}^{[1]} = {\breve{\bold H}} - E_1= \left( {\bold Q}^{[1]} \right)^{\dagger} {\bold Q}^{[1]} $
\newline
$ = -e^{ \frac{U^{[1]}(x) }{2} } \nabla e^{- U_I^{[1]}(x) }  \nabla  e^{ \frac{U^{[1]}(x) }{2} }$
& where ${\bold Q}^{[1]}  = e^{- \frac{U_I^{[1]}(x) }{2} }  \nabla  e^{ \frac{U^{[1]}(x) }{2} } $ annihilates 
\newline 
the ground state ${\breve \Phi}_1(x) =  \frac{  e^{- \frac{ U^{[1]}(x)}{2} } }  {  \sqrt{ Z_1 } }$  \\  
\hline
 Supersymmetric partner
 ${\breve {\bold G}} =      -  {\bold J}\nabla $ \newline
 of the Markov generator $ {\bold G}  = -\nabla  {\bold J} $ 
 \newline
 governing the dynamics $ \partial_t J_t( x ) =  {\breve {\bold G}} J_t(x) $ 
 of the current $J_t(x)= {\bold J} P_t(x)$
& $ {\breve {\bold G}} =      -  {\bold J}\nabla =
 e^{- U_I(x) } \nabla e^{ U(x) }   \nabla$
 \newline
 $=  -  e^{- \frac{U_I(x) }{2} } {\breve{\bold H}} e^{ \frac{U_I(x) }{2} }   $
 & related to the partner Hamiltonian ${\breve{\bold H}} = {\bold Q} {\bold Q}^{\dagger} $ \newline
 via the similarity involving $U_I(x)$ :
  \newline
  its lowest eigenvalue $E_1>0$ is associated 
  \newline to the left eigenvector $ {\breve L}_1(x) =  \nabla L_1(x) $\\
  \hline
The new Markov Generator obtained via the Doob transformation
involving $E_1$ and the left eigenvector $ {\breve L}_1(x)$
& ${\bold G}^{[1]}  \equiv {\breve L}_1(x) {\breve {\bold G} } \frac{1}{ {\breve L}_1(x) } + E_1 $
\newline
$=  -  e^{- \frac{U^{[1]}(x) }{2} } {\bold H}^{[1]} e^{ \frac{U^{[1]}(x) }{2} } $
& annihilates the new steady state $ P_*^{[1]}(x)   = {\breve \Phi}_1^2 (x)$ 
\newline
and is related to $ {\bold H}^{[1]} $ via the similarity involving $U^{[1]}(x) $  \\
  \hline 
\end{tabular}
\caption{Table summarizing the various operators introduced in section \ref{sec_general}. Their
factorized expressions in terms of the divergence operator $\nabla$
 (defined either in continuous space or in discrete space)
and in terms of the two potentials $U(x)$ and $U_I(x)$ make obvious their main properties.
 The two other potentials  $ U^{[1]}(x)$ and $U_I^{[1]}(x) $ appear when discussing
  the supersymmetric partners (see section \ref{sec_general} for detailed explanations).
} 
\label{table}
\end{table}


\subsubsection{ Discussion }

\label{sub_generalSusyDoobEquiv}

In this section, we have described how the factorized expressions of various operators 
in terms of the divergence operator $\nabla$
 (defined either in continuous space or in discrete space)
and in terms of the two potentials $U(x)$ and $U_I(x)$ (see the summarizing Table 
 \ref{table}) are useful to better understand their properties.
 
Then we have explained the equivalence between two points of view 
to find the two new potentials $U^{[1]}(x) $and $U_I^{[1]}(x)$ in terms of the two initial potentials $U(x) $ and $U_I(x)$ :

(i) In the quantum perspective, the new Hamiltonian ${\bold H}^{[1]}
= -e^{ \frac{U^{[1]}(x) }{2} } \nabla e^{- U_I^{[1]}(x) }  \nabla  e^{ \frac{U^{[1]}(x) }{2} }$
of Eq. \ref{Av1fromu1}
is constructed via the supersymmetric
 recursion $ {\bold H}^{[1]}= {\breve{\bold H}} - E_1$  of Eq. \ref{ApartnerHsusyrewriting} 
 from the partner ${\breve{\bold H}} = {\bold Q} {\bold Q}^{\dagger}=-  e^{- \frac{U_I(x) }{2} }  \nabla  e^{ U(x) } \nabla e^{- \frac{U_I(x) }{2} }$ of Eq. \ref{AquantumHpartner}
of the initial Hamiltonian $ {\bold H} = {\bold Q}^{\dagger}  {\bold Q}=  - e^{ \frac{U(x) }{2} }  \nabla  e^{- U_I(x) }  \nabla e^{ \frac{U(x) }{2} }$ of Eq. \ref{AquantumHsusy}.

(ii) In the Markov perspective, the new Markov generator ${\bold G}^{[1]}=  \nabla   e^{- U_I^{[1]}(x) } \nabla  e^{ U^{[1]}(x) }$ 
is constructed via the Doob transformation of Eq. \ref{ADoobRing} that involves 
the left eigenvector ${\breve L}_1(x) =    \nabla L_1(x) $ of Eq. \ref{ChoiceTwoleftviaNabla}
that is useful to relate the new potential $U^{[1]}(x) $ and the initial potential $U_I(x) $ via Eq. \ref{AU1xbreve}.

Now to go forward and write the explicit forms of the supersymmetric
 recursion for quantum Hamiltonians of Eq. \ref{ApartnerHsusyrewriting}
or of the Doob transformation of Eq. \ref{ADoobRing}, 
one needs to be able to exchange the order between the application of the operator $\nabla$ and the multiplication by a function,
so one needs to 
use the specific forms of the operator $\nabla$ of Eq. \ref{Adefnabla}
for continuous space (see the following section \ref{sec_FPx})
and for discrete space (see Appendix \ref{app_Jump}).
We will see that in both cases, the introduction of the first-excited left eigenvectors 
$L_1(x)$ and ${\breve L}_1(x) =    \nabla L_1(x) $ are very useful to better understand many properties.


\section{ Application to one-dimensional Fokker-Planck dynamics }

\label{sec_FPx}

In this section, we focus on one-dimensional Fokker-Planck dynamics in continuous space
where the operator $\nabla$ introduced in Eq. \ref{Adefnabla} is 
the derivative operator in continuous space
 \begin{eqnarray}
\nabla \equiv \frac{\partial}{\partial x}  
 \label{nabladerivative}
\end{eqnarray}
in order to write the analytical counterparts of the algebraic properties described in the previous section.


\subsection{ Specific properties of the derivative operator $\nabla \equiv \frac{\partial}{\partial x} $ in continuous space  }

The additional property of the derivative operator $\nabla \equiv \frac{\partial}{\partial x} $
with respect to Eq. \ref{AnablaProp}
is the Leibniz rule for the derivative of a product of two functions $O(x)$ and $\Omega(x)$
\begin{eqnarray}
 \frac{\partial}{\partial x} \bigg( O(x) \Omega(x) \bigg) 
 = O(x)   \frac{\partial \Omega(x)}{\partial x} +  \frac{\partial O(x)}{\partial x}\Omega(x)
 \label{Leibniz}
\end{eqnarray}
At the operator level, this translates into the rule
for exchanging the order of the derivative and the multiplication by a function $O(x)$
\begin{eqnarray}
 \frac{\partial}{\partial x} O(x) = O(x)   \frac{\partial}{\partial x} + O'(x) 
 \label{commutderivativeO}
\end{eqnarray}
that can be used to move the operator $\nabla \equiv \frac{\partial}{\partial x}  $
in the various expressions written in the previous section.


\subsection{ Explicit forms of the Fokker-Planck generator ${\bold G}$ and its adjoint ${\bold G}^{\dagger} $
as differential operators}

\subsubsection{ Link between the two parametrizations of the current operator ${\bold J}=- e^{- U_I(x)}  \frac{\partial}{\partial x} e^{U(x)} =F(x)   - D(x) \frac{\partial}{\partial x}   $   
}

The application of the rule of Eq. \ref{commutderivativeO} with $O(x)=e^{U(x)}$ to the current operator of Eq. \ref{AJUIU} 
\begin{eqnarray}
{\bold J} && =   - e^{- U_I(x)}  \frac{\partial}{\partial x} e^{U(x)} 
= - e^{- U_I(x)} \left(U'(x) e^{U(x)} + e^{U(x)} \frac{\partial}{\partial x}   \right)
\nonumber \\
&& = - e^{ U(x) - U_I(x)} \left( U'(x) + \frac{\partial}{\partial x}    \right)
 \label{AJUIUFP}
\end{eqnarray}
yields that the identification 
with the standard parametrization of the current operator ${\bold J} $
in terms of the Fokker-Planck force $F(x)$ and the positive diffusion coefficient $D(x) >0$
\begin{eqnarray}
 {\bold J} && \equiv F(x)   - D(x) \frac{\partial}{\partial x}    
\label{JdiffFD}
\end{eqnarray}
leads to the following correspondence
\begin{eqnarray}
D(x) && \equiv  e^{ U(x) - U_I(x)} 
\nonumber \\
  F(x)  && \equiv - D(x) U'(x) 
 \label{DFUIU}
\end{eqnarray}

The potential $U_I(x)$ can be thus rewritten in terms of $U(x)$ and the diffusion coefficient $D(x)$
\begin{eqnarray}
U_I(x) && = U(x) - \ln D(x) 
 \label{UIfromU}
\end{eqnarray}
so that the steady state of Eq. \ref{APstarx} involving $U(x)$ can be alternatively rewritten in terms of $U_I(x)$ and $D(x)$
\begin{eqnarray}
  P_*(x)  = \frac{ e^{ -U(x)} }{Z}   = \frac{ e^{ -U_I(x)} }{D(x) Z}
 \label{steadyeq}
\end{eqnarray}
while the partition function $Z$ is an integral over the domain $]x_L,x_R[$ where the diffusion takes place
\begin{eqnarray}
Z=  \int_{x_L}^{x_R} dx  e^{ -U(x)}    =\int_{x_L}^{x_R} dx \frac{ e^{ -U_I(x)} }{D(x) }
 \label{steadyeqnormaZ}
\end{eqnarray}

\subsubsection{ Vanishing-current boundary conditions and its consequences for the eigenvectors }

To have a genuine Markov operator ${\bold G}$
that conserves the probability normalization on the interval $]x_L,x_R[$ where the diffusion takes place
\begin{eqnarray}
0= \partial_t \int_{x_L}^{x_R} dx P_t(x)    && =   - \int_{x_L}^{x_R} dx  \partial_{x}   J_t(x)
= - \left[J_t(x) \right]_{x=x_L}^{x=x_R} = J_t(x_L)-J_t(x_R)
\label{conservnorma}
\end{eqnarray}
one needs to impose the vanishing of the current $J_t(x)={\bold J} P_t(x)$ at the boundaries $x_L$ and $x_R$
\begin{eqnarray}
0=J_t(x)  = {\bold J} P_t(x) =  - e^{- U_I(x)}  \frac{\partial}{\partial x} \left( e^{U(x)}  P_t(x) \right) \ \ \text{for $x=x_L$ and for $x=x_R$}
\label{jboundaries}
\end{eqnarray}
As a consequence, the right eigenvectors $R_n(x)$ 
of the Markov generator ${\bold G}= - \frac{\partial}{\partial x } {\bold J}$
inherit these vanishing-current boundary conditions
\begin{eqnarray}
 0  = {\bold J} R_n(x) =  - e^{- U_I(x)}  \frac{\partial}{\partial x} \left( e^{U(x)}  R_n(x) \right) \ \ \text{for $x=x_L$ and for $x=x_R$} 
 \label{jnright}
\end{eqnarray}
that can be translated for the left eigenvector $L_n(x)= \frac{R_n(x)}{P_*(x)} = Z e^{U(x)} R_n(x)$ into
the simpler boundary conditions
\begin{eqnarray}
 0 =  - e^{- U_I(x)}  \frac{L_n'(x)}{Z} 
 = - D(x) e^{- U(x)}  \frac{L_n'(x)}{Z} = - D(x) P_*(x) L_n'(x)  \ \ \text{for $x=x_L$ and for $x=x_R$} 
 \label{jnleft}
\end{eqnarray}


\subsubsection{ Explicit forms of the adjoints ${\bold G}^{\dagger} $ and ${\breve {\bold G}}^{\dagger}  $
with the corresponding eigenvalue equations for $ L_n(x) $ and $ {\breve L}_n(x) $}

The application of the rule of Eq. \ref{commutderivativeO} with $O(x)=e^{-U_I(x)}$ to the adjoint of Eq. \ref{AJUIUdagger}
\begin{eqnarray}
{\bold J}^{\dagger} && =  e^{U(x)}  \frac{\partial}{\partial x} e^{- U_I(x)}
=  e^{U(x)}\left(e^{-U_I(x)} \frac{\partial}{\partial x}  - U_I'(x) e^{-U_I(x)} \right)
\nonumber \\
&& =  e^{U(x)-U_I(x)}\left( \frac{\partial}{\partial x}  - U_I'(x)  \right)
= D(x) \left( \frac{\partial}{\partial x}  - U_I'(x)  \right) \equiv D(x)  \frac{\partial}{\partial x} +F_I(x)
 \label{AJUIUdaggerFP}
\end{eqnarray}
shows that 
it is also useful to introduce the Ito force $F_I(x) $ associated with the potential $U_I(x)$
\begin{eqnarray}
    F_I(x) && \equiv - D(x)U_I'(x) = -D(x) \bigg( U'(x)-  \frac{D'(x)}{D(x)}  \bigg) = - D(x) U'(x)  + D'(x) 
     = F(x)+ D' (x)
\label{FI}
\end{eqnarray}
even if it can be computed in terms of the Fokker-Planck force $F(x)$ and the the diffusion coefficient $D(x)$
of Eq. \ref{DFUIU}.

Then the adjoint ${\bold G}^{\dagger} $ of Eq. \ref{AGadjoint} that governs the dynamics of observables
\begin{eqnarray}
  {\bold G}^{\dagger} = {\bold J} ^{\dagger} \frac{\partial}{\partial x}
    &&  =     D(x)  \frac{\partial^2 }{\partial x^2} + F_I(x) \frac{\partial }{\partial x}  
\label{adjointIto}
\end{eqnarray}
is simpler with all the derivatives on the right.
In particular, the eigenvalue equations for the left eigenvectors $L_n(x)$ 
\begin{eqnarray}
- E_n L_n(x) && = {\bold G}^{\dagger}  L_n(x) = D(x)  L_n''(x) + F_I(x) L_n'(x)  
\label{EigenLeftFP}
\end{eqnarray}
 display the factorization of Eq. \ref{ChoiceTwoleftviaNabla}
\begin{eqnarray}
 {\breve L}_n(x) &&=   L_n'(x)
 \nonumber \\
 -E_n L_n(x) &&  =   {\bold J}^{\dagger } {\breve L}_n(x) = D(x)   {\breve L}_n'(x) + F_I(x)  {\breve L}_n(x)
\label{ChoiceTwoleftviaNablaFP}
\end{eqnarray}
when one introduces the left eigenvectors $ {\breve L}_n(x) $ of the partner ${\breve {\bold G}}  $ satisfying
\begin{eqnarray}
 -E_n {\breve L}_n(x)  = {\breve {\bold G}}^{\dagger} {\breve L}_n(x) 
 && = \frac{\partial}{\partial x}  {\bold J}^{\dagger } {\breve L}_n(x) 
\label{EigenLeftbreveFP}
\end{eqnarray}
with
\begin{eqnarray}
 {\breve {\bold G} }^{\dagger} = \frac{\partial}{\partial x}  {\bold J}^{\dagger } 
&& = \frac{\partial}{\partial x} \bigg( F_I(x)   + D(x) \frac{\partial}{\partial x}  \bigg)
 = \frac{\partial}{\partial x} \left[D(x)  \frac{\partial}{\partial x}+ F_I(x)   \right] 
 \nonumber \\
 && = D(x)  \frac{\partial^2}{\partial x^2} + \bigg( D'(x) + F_I(x) \bigg) \frac{\partial}{\partial x} + F_I'(x)
  \label{DoobRingDaggerenx}
\end{eqnarray}

The comparison with the adjoint ${\bold G}^{\dagger} $ of Eq. \ref{adjointIto} that conserves probability 
with $L_0(x)=1$ as left-eigenvector associated with the vanishing eigenvalue $E_0=0$ (i.e. ${\bold G}^{\dagger}L_0(x)={\bold G}^{\dagger} 1=0 $)
shows that the last term ${\breve {\bold G} }^{\dagger} $ in  Eq. \ref{DoobRingDaggerenx}
containing no derivative can be interpreted as a killing rate $K(x)= - F_I'(x)$. 


\subsection{ Explicit forms of the quantum Hamiltonians
${\bold H} = {\bold Q}^{\dagger} {\bold Q}$ and ${\breve{\bold H}} =  {\bold Q} {\bold Q}^{\dagger} $ 
and ${\bold H}^{[1]} = {\breve{\bold H}} - E_1=\left( {\bold Q}^{[1]} \right)^{\dagger} {\bold Q}^{[1]} $}

\subsubsection{ Explicit form of the quantum Hamiltonian ${\bold H} = {\bold Q}^{\dagger} {\bold Q}$ }

Using $U_I(x)  = U(x) - \ln D(x) $
 of Eq. \ref{UIfromU}, the operators ${\bold Q} $ of Eq. \ref{Aqsusy}
and ${\bold Q}^{\dagger}  $ of Eq. \ref{Aqdaggersusy}
become the first-order differential operators 
\begin{eqnarray}
{\bold Q}  && =  e^{- \frac{U_I(x) }{2} }  \frac{\partial}{\partial x}  e^{ \frac{U(x) }{2} }
=  \sqrt{ D(x) } e^{- \frac{U(x) }{2} }  \frac{\partial}{\partial x}  e^{ \frac{U(x) }{2} }
=  \sqrt{ D(x) }  \left( \frac{\partial}{\partial x}  +\frac{ U'(x)}{2 } \right) 
\nonumber \\
{\bold Q}^{\dagger}  &&= -e^{ \frac{U(x) }{2} } \frac{\partial}{\partial x} e^{- \frac{U_I(x) }{2} } 
=  -e^{ \frac{U(x) }{2} }
 \frac{\partial}{\partial x} e^{- \frac{U(x) }{2} } \sqrt{ D(x) }
= \left(  -  \frac{\partial}{\partial x}  +\frac{ U'(x)}{2 } \right)\sqrt{ D(x) }
\label{qqdaggerdiff}
\end{eqnarray}

Then the quantum Hamiltonian of Eq. \ref{AquantumHsusy}
\begin{eqnarray}
{\bold H} = - e^{ \frac{U(x) }{2} }  \frac{ \partial  }{\partial x}  e^{- U_I(x) }  \frac{ \partial  }{\partial x} e^{ \frac{U(x) }{2} }
= {\bold Q}^{\dagger} {\bold Q}=  - \frac{ \partial  }{\partial x} D(x) \frac{ \partial  }{\partial x} +V(x)
\label{hsusy}
\end{eqnarray}
involves the diffusion coefficient $D(x)$ between the two derivatives of the kinetic term $\frac{ \partial  }{\partial x} D(x) \frac{ \partial  }{\partial x} $, while the scalar potential $V(x)$
reads
\begin{eqnarray}
V(x) && \equiv D(x)  \frac{ [U'(x)]^2 }{4 }  -D(x) \frac{U''(x)}{2} -D'(x) \frac{U'(x)}{2}
\nonumber \\
&& = \frac{ F^2(x) }{4 D(x) } + \frac{F'(x)}{2} 
\label{vfromu}
\end{eqnarray}
and is thus simpler in terms of the Fokker-Planck force $F(x)=-D(x) U'(x)$ of Eq. \ref{DFUIU}.


\subsubsection{ Explicit form of the quantum partner ${\breve{\bold H}} =  {\bold Q} {\bold Q}^{\dagger} $ }

The supersymmetric partner ${\breve{\bold H}} = {\bold Q} {\bold Q}^{\dagger} $ of Eq. \ref{AquantumHpartner}
\begin{eqnarray}
{\breve{\bold H}} =  -  e^{- \frac{U_I(x) }{2} }  \nabla  e^{ U(x) } \nabla e^{- \frac{U_I(x) }{2} } = {\bold Q} {\bold Q}^{\dagger} 
  = - \frac{ \partial  }{\partial x} D(x) \frac{ \partial  }{\partial x} +\breve{V}(x)
\label{hsusypartner}
\end{eqnarray}
involves the same kinetic term $\frac{ \partial  }{\partial x} D(x) \frac{ \partial  }{\partial x} $ as the Hamiltonian ${\bold H}$ of Eq. \ref{hsusy},
while the partner scalar potential $\breve{V}(x) $ given by
\begin{eqnarray}
\breve{V }(x) && \equiv D(x)  \frac{ [U'(x)]^2 }{4 }  +D(x) \frac{U''(x)}{2}
 + \frac{ [D'(x)]^2 }{4 D(x) }-\frac{D''(x)}{2}
\nonumber \\
&& = \frac{ F^2(x) }{4 D(x) } - \frac{F'(x)}{2} +\frac{F(x) D'(x) }{2D(x)}+ \frac{ [D'(x)]^2 }{4 D(x) }-\frac{D''(x)}{2}
\nonumber \\
&& = \frac{ F^2_I(x) }{4 D(x) } - \frac{ F_I'(x)  }{2} 
\label{vpartner}
\end{eqnarray}
turns out to be simpler on the last line in terms of the Ito force $F_I(x)= F(x)+D'(x)= -D(x) U'(x)+D'(x)$
of Eq. \ref{FI}.


\subsubsection{ Construction of the new Hamiltonian  ${\bold H}^{[1]} \equiv \left( {\bold Q}^{[1]} \right)^{\dagger} {\bold Q}^{[1]}= {\breve{\bold H}} - E_1= {\bold Q} {\bold Q}^{\dagger}-E_1 $   }

Plugging the explicit form of ${\breve{\bold H}} $ of Eq. \ref{hsusypartner}
into the new Hamiltonian $ {\bold H}^{[1]} $ of Eq. \ref{ApartnerHsusyrewriting}
\begin{eqnarray}
 {\bold H}^{[1]} = {\breve{\bold H}} - E_1
 && =  - \frac{ \partial  }{\partial x} D(x) \frac{ \partial  }{\partial x} +\breve{V}(x)- E_1
 \nonumber \\
 && =  - \frac{ \partial  }{\partial x} D(x) \frac{ \partial  }{\partial x} +V^{[1]}(x)
\label{H1QMsameD}
\end{eqnarray}
yields that $ {\bold H}^{[1]}  $ involves the same kinetic term $\frac{ \partial  }{\partial x} D(x) \frac{ \partial  }{\partial x} $
with the same diffusion coefficient $D(x)$,
while the new scalar potential $V^{[1]}(x) =\breve{V}(x)- E_1$ is simply given by the partner potential $\breve{V}(x) $
shifted by $E_1$.

 Using the correspondence of Eq. \ref{DFUIU} for the new model involving the two potentials $[U^{[1]}(x) ; U_I^{[1]}(x)] $
\begin{eqnarray}
D(x) && =  e^{ U^{[1]}(x) - U_I^{[1]}(x)} 
\nonumber \\
 F^{[1]}(x) && =-D(x) \frac{d U^{[1]}(x)}{dx} 
 \label{DFUIU1}
\end{eqnarray}
one obtains the following conclusions :

(i) The potential $U_I^{[1]}(x) $ is related to $U^{[1]}$ that parametrizes the ground state of $ {\bold H}^{[1]} $
via the relation analogous to Eq. \ref{UIfromU} 
\begin{eqnarray}
U_I^{[1]}(x) && = U^{[1]}(x) - \ln D(x) 
 \label{UI1fromU1}
\end{eqnarray}
so that the operator ${\bold Q}^{[1]}  $ of Eq. \ref{Aq1q1daggersusy} is the analog of Eq. \ref{qqdaggerdiff}
\begin{eqnarray}
{\bold Q}^{[1]}  && \equiv  \sqrt{ D(x) } e^{- \frac{U^{[1]}(x) }{2} }  \frac{\partial}{\partial x}  e^{ \frac{U^{[1]}(x) }{2} }
=  \sqrt{ D(x) }  \left( \frac{\partial}{\partial x}  +\frac{ \frac{d U^{[1]}(x)}{dx}}{2 } \right) 
\nonumber \\
\left( {\bold Q}^{[1]} \right)^{\dagger}  &&\equiv  -e^{ \frac{U^{[1]}(x) }{2} }
 \frac{\partial}{\partial x} e^{- \frac{U^{[1]}(x) }{2} } \sqrt{ D(x) }
= \left(  -  \frac{\partial}{\partial x}  +\frac{ \frac{d U^{[1]}(x)}{dx}}{2 } \right)\sqrt{ D(x) }
\label{q1q1daggersusy}
\end{eqnarray}

(ii)  The corresponding new scalar potential $V^{[1]}(x)$ computed in terms of the Fokker-Planck force $ F^{[1]}(x) $
 via Eq. \ref{vfromu}
\begin{eqnarray}
V^{[1]}(x) &&  = \frac{ \left[ F^{[1]}(x) \right]^2 }{4 D(x) } + \frac{\frac{d F^{[1]}(x)}{dx}}{2} 
\label{v1fromu1}
\end{eqnarray}
should be equal to the difference between the partner potential $\breve{V}(x) $ of Eq. \ref{vpartner}
and $E_1$ 
\begin{eqnarray}
V^{[1]}(x) && = \breve{V}(x)- E_1
 \nonumber \\
\text{ i.e.} \ \ \ \ \ \ \ \
\frac{ \left[ F^{[1]}(x) \right]^2 }{4 D(x) } + \frac{\frac{d F^{[1]}(x)}{dx}}{2} 
&& = \frac{ F^2_I(x) }{4 D(x) } - \frac{ \frac{d F_I(x)}{dx}  }{2} - E_1
\label{Riccati}
\end{eqnarray}
When the diffusion coefficient $D(x)$ and the Ito force $F_I(x)$ of the initial model  are given,
one then needs to solve this non-linear first-order differential equation
to obtain the new Fokker-Planck force $F^{[1]}(x)$. However, the solution
of such Riccati equation is not known for arbitrary parameters $[D(x);F_I(x)]$.


\subsection{ Construction of the new Fokker-Planck generator ${\bold G}^{[1]}$
via the Doob-transformation of the partner ${\breve {\bold G} } $  }

Let us now describe the construction of the new Fokker-Planck generator ${\bold G}^{[1]}$
via the Doob-transformation of Eq. \ref{ADoobRing}
that involves the left eigenvector ${\breve L}_1(x) $.

The explicit action of the adjoint $ \left( {\bold G}^{[1]}  \right)^{\dagger}$ of Eq. \ref{ADoobRingDagger}
on an arbitrary observable $O(x)$
\begin{eqnarray}
\left( {\bold G}^{[1]}  \right)^{\dagger} O(x) && 
= \frac{1}{ {\breve L}_1(x) } {\breve {\bold G} }^{\dagger} \left[ {\breve L}_1(x) O(x) \right]+ E_1 O(x) 
  \label{DoobRingDaggerOdef}
\end{eqnarray}
can be evaluated using Eq. \ref{DoobRingDaggerenx} for $ {\breve {\bold G} }^{\dagger}   $.
Using the eigenvalue equation  ${\breve {\bold G} }^{\dagger} {\breve L}_1(x)= - E_1 {\breve L}_1(x)$,
one obtains that Eq. \ref{DoobRingDaggerOdef}
reduces to
\begin{eqnarray}
\left( {\bold G}^{[1]}  \right)^{\dagger}  O(x)  = D(x) O''(x) +  F_I^{[1]}  (x) O'(x)
  \label{DoobRingDaggerOx}
\end{eqnarray}
where the term containing the second derivative $O''(x)$ involves the same diffusion coefficient $D(x)$ as 
the initial adjoint ${\bold G}^{\dagger} $ of Eq. \ref{adjointIto}, while 
the term containing the first derivative $O'(x)$ involves the new Ito force $F_I^{[1]} (x) $
given by
\begin{eqnarray}
  F_I^{[1]}  (x) = F_I(x) +D'(x)+ 2 D(x) \frac{ {\breve L}_1'(x)}{{\breve L}_1(x) }  
  \label{DoobRingFI}
\end{eqnarray}
Equivalently, the new Fokker-Planck force $F^{[1]}  (x) \equiv F_I^{[1]}  (x) -D'(x) $ reads
in terms of the initial Ito force $F_I(x) $, of the diffusion coefficient $D(x)$
 and of the left eigenvector ${\breve L}_1(x) $ of ${\breve {\bold G} } $
\begin{eqnarray}
F^{[1]}  (x) \equiv F_I^{[1]}  (x) -D'(x) && = F_I(x) + 2 D(x) \frac{ {\breve L}_1'(x)}{{\breve L}_1(x) }  
  \label{DoobRingFbreve}
\end{eqnarray}

Plugging this expression for $F^{[1]}  (x) $
into the Riccati Eq. \ref{Riccati}
\begin{eqnarray}
- E_1 && =\frac{ \left[ F^{[1]}(x) \right]^2- F_I^2(x) }{4 D(x) } + \frac{\frac{d F^{[1]}(x)}{dx}}{2} + \frac{ \frac{d F_I(x)}{dx}  }{2} 
\nonumber \\
&& = \frac{  \left[ F^{[1]}(x) - F_I(x) \right] \left[ F^{[1]}(x) + F_I(x) \right] }{4 D(x) } + \frac{d }{dx} \frac{\left[ F^{[1]}(x) + F_I(x) \right]}{2} 
\nonumber \\
&& = \frac{  \left[ 2 D(x) \frac{ {\breve L}_1'(x)}{{\breve L}_1(x) }  \right] \left[ 2F_I(x) + 2 D(x) \frac{ {\breve L}_1'(x)}{{\breve L}_1(x) } \right] }{4 D(x) } + \frac{d }{dx} \frac{\left[ 2F_I(x) + 2 D(x) \frac{ {\breve L}_1'(x)}{{\breve L}_1(x) } \right]}{2} 
\nonumber \\
&& =   \frac{ {\breve L}_1'(x)}{{\breve L}_1(x) }  \left[ F_I(x) +  D(x) \frac{ {\breve L}_1'(x)}{{\breve L}_1(x) } \right]
+ F_I'(x) +  D'(x) \frac{ {\breve L}_1'(x)}{{\breve L}_1(x) } 
+   D(x) \frac{ {\breve L}_1''(x)}{{\breve L}_1(x) } 
-  D(x) \left( \frac{ {\breve L}_1'(x)}{{\breve L}_1(x) } \right)^2
\nonumber \\
&& = D(x) \frac{ {\breve L}_1''(x)}{{\breve L}_1(x) } 
+  \bigg[  F_I(x) +D'(x)  \bigg] \frac{ {\breve L}_1'(x)}{{\breve L}_1(x) }  
+ F_I'(x)     
\label{H1QMsameDplugDoob}
\end{eqnarray}
leads to the eigenvalue Equation of Eq \ref{EigenLeftbreveFP} with Eq. \ref{DoobRingDaggerenx}
for ${\breve L}_1(x) $
\begin{eqnarray}
 -E_1 {\breve L}_1(x)  = {\breve {\bold G}}^{\dagger} {\breve L}_1(x) 
  = D(x)   {\breve L}_1''(x) + \bigg( D'(x) + F_I(x) \bigg) {\breve L}_1'(x) + F_I'(x){\breve L}_1(x)
\label{EigenLeftbreveFPL1}
\end{eqnarray}

In summary, the left eigenvector ${\breve L}_1(x)=L_1'(x) $ satisfying this second-order linear eigenvalue equation
is useful to write the force $F^{[1]}  (x) $ of Eq. \ref{DoobRingFbreve}
satisfying the Riccati non-linear first-order differential Eq. \ref{Riccati}.

The relation of Eq. \ref{DoobRingFbreve} between the forces $F^{[1]}(x)  =-D(x) \frac{d U^{[1]}(x)}{dx} $ and $F_I(x)= -D(x)  U_I'(x)$ 
is the counterpart of the relation of Eq. \ref{AU1xbreve} between the potentials $U^{[1]}(x) $ and $U_I(x) $
\begin{eqnarray}
U^{[1]}(x) \equiv  U_I(x) -2 \ln \big( {\breve L}_1(x)  \big) = U_I(x) -2 \ln \big( L_1'(x)  \big) 
\label{AU1xbreveFP}
\end{eqnarray}


\begin{table}[h]
\centering
\begin{tabular}{|p{4cm}|p{4.5cm}|p{4.5cm}|p{8cm}|}
\hline
\textbf{Operator }
& \textbf{Factorized expression} 
& \textbf{Expanded expression}
& \textbf{Properties}  \\
\hline 
Current operator $ {\bold J}$ 
producing the 
 current $J_t(x) = {\bold J} P_t(x)$ associated to the density $P_t(x)$
& ${\bold J}  =   - e^{- U_I(x)} \frac{\partial}{\partial x} e^{U(x)} $
\newline
$=- e^{ U(x) - U_I(x)} \left( U'(x) + \frac{\partial}{\partial x}    \right)$
& $ {\bold J} = F(x)   - D(x) \frac{\partial}{\partial x}     $
& Correspondence: 
\newline $D(x) =  e^{ U(x) - U_I(x)} $
\newline
  $F(x)  = - D(x) U'(x) $
\\
\hline
Adjoint ${\bold J}^{\dagger} $ of the current operator 
& ${\bold J}^{\dagger}  =  e^{U(x)}  \frac{\partial}{\partial x} e^{- U_I(x)} $
\newline
$=e^{U(x)-U_I(x)}\left(  - U_I'(x) + \frac{\partial}{\partial x}   \right)$
& ${\bold J}^{\dagger}= F(x)+D(x)  \frac{\partial}{\partial x} $
\newline $= (F(x)+D'(x)) + D(x) \frac{\partial}{\partial x}$
\newline $= F_I(x) +D(x)  \frac{\partial}{\partial x} $
& The Ito force $F_I(x)= - D(x)U_I'(x) $ 
\newline is related to the Fokker-Planck force $F(x)$ :
\newline $  via F_I(x)   = F(x)+ D' (x)$
\\
\hline
Adjoint operator governing the dynamics 
of averaged values of observables $O(x)$
\newline
$\partial_t {\mathbb E}_t [ O( x ) ] =  {\mathbb E}_t [ ({\bold G}^{\dagger} O ) ( x) ]  $
& ${\bold G}^{\dagger}={\bold J}^{\dagger}  \frac{\partial }{\partial x} $
\newline $
= e^{U(x)}   \frac{\partial }{\partial x} e^{- U_I(x)} \frac{\partial }{\partial x}
 $ 
& $ {\bold G}^{\dagger} =     D(x)  \frac{\partial^2 }{\partial x^2} + F_I(x) \frac{\partial }{\partial x} $
& Eigenvalue Eqs for the left eigenvectors: 
\newline
$- E_n L_n(x)  = {\bold G}^{\dagger}  L_n(x) = {\bold J}^{\dagger}  \frac{\partial }{\partial x}   L_n(x) $ 
\newline 
i.e. $ - E_n L_n(x)= D(x)  L_n''(x) + F_I(x) L_n'(x)$
\\
\hline
Annihilation operator
\newline
Creation operator
& ${\bold Q}  =   e^{- \frac{U_I(x) }{2} }  \frac{\partial }{\partial x}  e^{ \frac{U(x) }{2} } $ 
 \newline
  $ {\bold Q}^{\dagger}  =  -e^{ \frac{U(x) }{2} } \frac{\partial }{\partial x} e^{- \frac{U_I(x) }{2} } $
& ${\bold Q}  =  \sqrt{ D(x) }  \left( \frac{\partial}{\partial x}  +\frac{ U'(x)}{2 } \right)  $ 
 \newline
  $ {\bold Q}^{\dagger}  = \left(  -  \frac{\partial}{\partial x}  +\frac{ U'(x)}{2 } \right)\sqrt{ D(x) } $
& simpler in terms of $D(x)$ and $U(x)$
  \\  
  \hline
Supersymmetric Quantum Hamiltonian $ {\bold H}$  
obtained from the Markov Generator $ {\bold G} $ 
 via the similarity involving $U(x)$
&  ${\bold H} = - e^{ \frac{U(x) }{2} }{\bold G}    e^{ - \frac{U(x) }{2} } $
\newline
$= - e^{ \frac{U(x) }{2} }  \frac{\partial }{\partial x}  e^{- U_I(x) }  \frac{\partial }{\partial x} e^{ \frac{U(x) }{2} } $
\newline $= {\bold Q}^{\dagger} {\bold Q}$ 
& $ {\bold H} =  - \frac{ \partial  }{\partial x} D(x) \frac{ \partial  }{\partial x} +V(x)$
\newline with the scalar potential
\newline
$ V(x)= \frac{ F^2(x) }{4 D(x) } + \frac{F'(x)}{2} $
&
$V(x) $ simpler in terms of 
\newline the Fokker-Planck force $F(x)$
 \\
\hline
Supersymmetric partner Hamiltonian
& ${\breve{\bold H}} = {\bold Q} {\bold Q}^{\dagger} $
\newline
$ = -  e^{- \frac{U_I(x) }{2} }  \frac{\partial }{\partial x}  e^{ U(x) } \frac{\partial }{\partial x} e^{- \frac{U_I(x) }{2} } $ 
& 
${\breve{\bold H}}   = - \frac{ \partial  }{\partial x} D(x) \frac{ \partial  }{\partial x} +\breve{V}(x)$
  \newline
  with the scalar potential
  \newline
  $ \breve{V }(x)  = \frac{ F^2_I(x) }{4 D(x) } - \frac{ F_I'(x)  }{2} $
&
$\breve{V }(x) $ simpler in terms of 
\newline 
the Ito force $F_I(x)$
  \\  
\hline
New Supersymmetric Hamiltonian  
with vanishing ground state energy
&  ${\bold H}^{[1]} = {\breve{\bold H}} - E_1= \left( {\bold Q}^{[1]} \right)^{\dagger} {\bold Q}^{[1]} $
\newline
$ = -e^{ \frac{U^{[1]}(x) }{2} } \frac{\partial }{\partial x} e^{- U_I^{[1]}(x) }  \frac{\partial }{\partial x}  e^{ \frac{U^{[1]}(x) }{2} }$
& 
 $ {\bold H}^{[1]}  =  - \frac{ \partial  }{\partial x} D(x) \frac{ \partial  }{\partial x} +V^{[1]}(x)$
\newline 
 with the scalar potential
\newline
$V^{[1]}(x) =\breve{V}(x)- E_1 $
\newline
$=\frac{ \left[ F^{[1]}(x) \right]^2 }{4 D(x) } + \frac{\frac{d F^{[1]}(x)}{dx}}{2} $
& Riccati Equation 
\newline
for the new Fokker-Planck force $F^{[1]}(x)$ :
\newline
$\frac{ \left[ F^{[1]}(x) \right]^2 }{4 D(x) } + \frac{\frac{d F^{[1]}(x)}{dx}}{2} 
 = \frac{ F^2_I(x) }{4 D(x) } - \frac{ \frac{d F_I(x)}{dx}  }{2} - E_1 $
  \\  
\hline
 Adjoint $ {\breve {\bold G} }^{\dagger}  $ of the supersymmetric partner
 ${\breve {\bold G}} =      -  {\bold J}\frac{\partial }{\partial x} $ \newline
 of the Markov generator $ {\bold G}  = -\frac{\partial }{\partial x}  {\bold J} $ 
& $ {\breve {\bold G} }^{\dagger} = \frac{\partial}{\partial x}  {\bold J}^{\dagger }  $
\newline
$ =\frac{\partial}{\partial x} e^{U(x)}  \frac{\partial}{\partial x} e^{- U_I(x)}
 $
 &  $ {\breve {\bold G} }^{\dagger} = \frac{\partial}{\partial x} \bigg( F_I(x)   + D(x) \frac{\partial}{\partial x}  \bigg)$
 \newline
 $ = F_I'(x)  + (D'(x) + F_I(x)) \frac{\partial}{\partial x}  $
 \newline
$ +D(x)  \frac{\partial^2}{\partial x^2}$
 &Eigenvalue Eqs for the left eigenvectors: 
\newline
$-E_n {\breve L}_n(x)  = {\breve {\bold G}}^{\dagger} {\breve L}_n(x) 
  = \frac{\partial}{\partial x}  {\bold J}^{\dagger } {\breve L}_n(x)  $ 
\newline 
so  $  {\breve L}_n(x) =   L_n'(x)$ and
 \newline  the lowest eigenvalue $E_1>0$ is associated 
  \newline to the left eigenvector $ {\breve L}_1(x) =   L_1'(x) $
  \\
  \hline
Adjoint $ \left( {\bold G}^{[1]}  \right)^{\dagger}$ obtained via the Doob transformation
involving $E_1$ and the left eigenvector $ {\breve L}_1(x)$
& $ \left( {\bold G}^{[1]}  \right)^{\dagger}  
= \frac{1}{ {\breve L}_1(x) } {\breve {\bold G} }^{\dagger} {\breve L}_1(x) + E_1 $
& $ \left( {\bold G}^{[1]}  \right)^{\dagger}  = D(x) \frac{\partial^2}{\partial x^2} +  F_I^{[1]}  (x)  \frac{\partial}{\partial x} $
& The new Ito force 
\newline
$  F_I^{[1]}  (x) = F_I(x) +D'(x)+ 2 D(x) \frac{ {\breve L}_1'(x)}{{\breve L}_1(x) }  $
\newline
and the new Fokker-Planck force
\newline
$F^{[1]}  (x) = F_I^{[1]}  (x) -D'(x)  $
\newline 
$= F_I(x) + 2 D(x) \frac{ {\breve L}_1'(x)}{{\breve L}_1(x) } $
   \\
  \hline 
\end{tabular}
\caption{Table summarizing the notations for the various operators introduced in section \ref{sec_FPx}
when the divergence operator is the derivative $\frac{\partial }{\partial x}=  \frac{\partial}{\partial x}  $. 
Their factorized expressions in terms of the two potentials $U(x)$ and $U_I(x)$ 
and their usual expanded expressions in terms of the diffusion coefficient $D(x)$
and the Fokker-Planck force $F(x)$ or the Ito force $F_I(x)$ are compared 
 (see section \ref{sec_FPx} for detailed explanations).
} 
\label{tableFP}
\end{table}

\subsection{ Discussion }

In this section, we have explained how the algebraic framework described in section \ref{sec_general}
can be applied to Fokker-Planck differential generators in continuous space,
and we have described the link with the usual notations involving the diffusion coefficient $D(x)$ and
the Fokker-Planck force $F(x)$ and the Ito force $F_I(x)$ (see the summarizing Table \ref{tableFP}).
Then we have emphasized the major role of the left eigenvector $L_1(x)$ and of its derivatives
to construct the new Fokker-Planck force $F^{[1]}  (x) $.
In the next section, we analyze how the left eigenvector $L_1(x)$ can be used  
as the central object to construct quasi-exactly-solvable Fokker-Planck generators with two explicit levels.


\section{ Constructing Fokker-Planck models with explicit eigenstates for $E_0=0$ and $E_1>0$}

\label{sec_2Eigen}

In this section, we describe how the analysis of the previous section 
can be used to construct one-dimensional
models with explicit eigenstates associated with the $N=2$ energies $E_0=0$ and $E_1>0$
when the diffusion coefficient $D(x)$ is given, with the two perspectives: 

(i) for the quantum Hamiltonian ${\bold H}$ of Eq. \ref{hsusy}
with its two eigenstates $\Phi_0(x)$ and $\Phi_1(x)$

(ii) for the Fokker-Planck generator ${\bold G}= - \frac{\partial}{\partial x} {\bold J}$ with the steady state $P_*(x)=R_0(x)=\Phi^2_0(x)$
and the first excited left eigenvector $L_1(x)=\frac{\Phi_1(x)}{\Phi_0(x)}$, 
while the right eigenvector is given by $ R_1(x)=\Phi_1(x)\Phi_0(x)= L_1(x) P_*(x)$.

Let us stress that if one chooses the steady state $P_*(x) $ of the Fokker-Planck generator ${\bold G}$,
  or equivalently the ground state $\Phi_0(x)$ of the quantum Hamiltonian ${\bold H}$,
  while the diffusion coefficient $D(x)$ is given,
then the force $F(x)$ of the Fokker-Planck generator ${\bold G}$ 
and the scalar potential $V(x)$ of the quantum Hamiltonian ${\bold H}$ are fixed,
and in general one cannot solve explicitly the eigenvalue equations associated with the first excited energy $E_1>0$
and thus one cannot write explicitly the first excited states. 
So the main idea of the present section is to choose instead the first excited left eigenvector $L_1(x)$ and its energy $E_1$,
and then to reconstruct the steady state $P_*(x)$ or equivalently the quantum ground state $\Phi_0(x)$
from $L_1(x)$ and $E_1$.


\subsection{ Using the first excited left eigenvector $L_1(x)$ to construct the Ito force $F_I(x)$ and the steady state $P_*(x)$}

Since the first excited left eigenvector $L_1(x)$ plays the central role in this section,
it is useful to summarize the various conditions that $L_1(x)$ should satisfy 
and to write the various consequences for the other observables step by step :

$\bullet$ While the ground state energy $E_0=0$ is associated with eigenstates that have no nodes,
namely the trivial left eigenvector $L_0(x)=1$, the steady state $R_0(x)=P_*(x)$
and the quantum ground state $\Phi_0(x)=\sqrt{P_*(x)}$,
the first excited energy $E_1$ is associated with eigenstates that have exactly one node,
so it will be convenient to introduce the single root $x_1$
of the first excited left eigenvector $L_1(x)$, that will also be the single root of the first excited right eigenvector 
$R_1(x)=L_1(x) P_*(x)$, and of 
 the first excited quantum eigenstate $\Phi_1(x)= L_1(x) \Phi_0(x)$ :
\begin{eqnarray}
 \text{  $x_1$ unique root of the first excited states $L_1(x_1)=0 = R_1(x_1)=\Phi_1(x_1)$}
  \label{x1uniqueroot}
\end{eqnarray}

$\bullet$ The left eigenvector  ${\breve L}_1(x) =  L_1'(x) $ of Eq. \ref{ChoiceTwoleftviaNablaFP}
is associated with the ground state energy $E_1$ of the partner ${\breve{\bold G}}$ and thus has no node,
so the derivative $L_1'(x) $ of $L_1(x)$ cannot vanish and can be chosen to be always positive
\begin{eqnarray}
L_1'(x) >0
  \label{L1primepositive}
\end{eqnarray}

$\bullet$ The first excited left eigenvector $L_1(x)$ should satisfy the eigenvalue Eq. \ref{EigenLeftFP} 
associated with the eigenvalue $E_1>0$
\begin{eqnarray}
 - E_1 L_1(x)  =  {\bold G}^{\dagger} L_1(x) =  F_I(x) L_1'(x)  + D(x)  L_1''(x)
 \label{EigenL1}
\end{eqnarray}
that can be used to write the Ito force $F_I(x)$ in terms of $L_1(x)$ and its first two derivatives $L_1'(x) $ and $L_1''(x) $
when the diffusion coefficient $D(x)$ is given
\begin{eqnarray}
F_I(x)  =  -   E_1 \frac{ L_1(x)   }{L_1' (x) } -  D(x) \frac{ L_1''(x) }{L_1' (x) }
\label{FIEigenl1}
\end{eqnarray}

$\bullet$ The corresponding derivative $U_I'(x) $ of the Ito potential $U_I(x)$ computed from $F_I(x)$ via Eq. \ref{FI} 
\begin{eqnarray}
 U_I'(x) =  - \frac{F_I(x)}{D(x)} &&  =     E_1 \frac{ L_1(x)   }{ D(x) L_1' (x) } +   \frac{ L_1''(x) }{L_1' (x) }
 \nonumber \\
 && =  E_1 \frac{ L_1(x)   }{ D(x) L_1' (x) } +  \frac{d}{dx} \ln \big( L_1'(x) \big)
 \label{UIprimefromL1}
\end{eqnarray}
leads to the steady state of Eq. \ref{steadyeq} 
\begin{eqnarray}
  P_*(x)  && =  \frac{ e^{ -U_I(x)} }{D(x) Z}
  = \frac{ e^{ - \ln \big( L_1'(x) \big) - E_1 \int^x dX  \frac{ L_1(X)   }{ D(X) L_1' (X) } } }{D(x) Z}
  \nonumber \\
  && =  \frac{ e^{  - E_1\int^x dX  \frac{ L_1(X)   }{ D(X) L_1' (X) } } }
  {D(x) L_1'(x)  Z}
 \label{steadyeqfroml1}
\end{eqnarray}
where the partition function of Eq. \ref{steadyeqnormaZ} should be finite
\begin{eqnarray}
Z=  \int_{x_L}^{x_R} dx \frac{ e^{ -U_I(x)} }{D(x) }=\int_{x_L}^{x_R} dx  \frac{ e^{  - E_1\int^x dX  \frac{ L_1(X)   }{ D(X) L_1' (X) } } }
  {D(x) L_1'(x)  } < +\infty
 \label{steadyeqnormaZQES}
\end{eqnarray}

$\bullet$ The right eigenvector $R_1(x)$ given by the product of the steady state $P_*(x)$ with $L_1(x)$
\begin{eqnarray}
R_1(x) =L_1(x)  P_*(x)  && =  \frac{ L_1(x) } {D(x) L_1'(x)  Z} e^{  - E_1\int^x dX  \frac{ L_1(X)   }{ D(X) L_1' (X) } } 
  \nonumber \\
  && = - \frac{d}{dx} \left(   \frac{ 1 } { E_1  Z} e^{  - E_1\int^x dX  \frac{ L_1(X)   }{ D(X) L_1' (X) } } \right)
 \label{prodL1Pstar}
\end{eqnarray}
is a total derivative and thus satisfies the various orthogonality conditions of Eqs \ref{Aortholr} for $n=0$ and $n'=1$
\begin{eqnarray}
0 && = \int_{x_L}^{x_R} dx R_1(x) =  \int_{x_L}^{x_R} dx L_0(x)  R_1(x) = \langle L_0 \vert R_1 \rangle
\nonumber \\
&& = \int_{x_L}^{x_R} dx L_1(x)  P_*(x)  = \langle L_1 \vert R_0 \rangle = \langle \Phi_1 \vert \Phi_0 \rangle
 \label{orthoprodL1Pstar}
\end{eqnarray}
that also corresponds to the orthogonality of Eq. \ref{Aorthophin} for the two quantum eigenstates $\Phi_0$ and $\Phi_1$.

$\bullet$ The remaining normalization of Eqs \ref{Aortholr} for $n=1=n'$, that is equivalent to the 
normalization of Eq. \ref{Aorthophin} for the quantum eigenstate $\Phi_1$,
can be also rewritten using Eq. \ref{prodL1Pstar} and an integration by parts if one prefers
\begin{eqnarray}
1 &&= \langle \Phi_1 \vert \Phi_1 \rangle = \langle L_1 \vert R_1 \rangle =  \int_{x_L}^{x_R} dx L_1^2(x)  P_*(x)
=  -  \int_{x_L}^{x_R} dx L_1(x) \frac{d}{dx} \left(   \frac{ 1 } { E_1  Z} e^{  - E_1\int^x dX  \frac{ L_1(X)   }{ D(X) L_1' (X) } } \right)
\nonumber \\
&& = \int_{x_L}^{x_R} dx L_1'(x)  \left(   \frac{ 1 } { E_1  Z} e^{  - E_1\int^x dX  \frac{ L_1(X)   }{ D(X) L_1' (X) } } \right)
= \frac{1}{E_1} \int_{x_L}^{x_R} dx [L_1'(x)]^2  D(x)  P_*(x) 
 \label{ortholrL1square}
\end{eqnarray}
This equation determines the global normalization of $ L_1(x)$ or $ L_1'(x)$ if $E_1$ is given.

$\bullet$ The boundary conditions of Eq. \ref{jnleft} then read for the left eigenvector $L_1(x)$
using Eq. \ref{steadyeqfroml1}
\begin{eqnarray}
\text{for $x=x_L$ and for $x=x_R$} : \ \ 
  0  = - D(x) P_*(x) L_1'(x)  
  = -   \frac{ e^{  - E_1\int^x dX  \frac{ L_1(X)   }{ D(X) L_1' (X) } } }
  {  Z}
 \label{jnleft1}
\end{eqnarray}

\subsubsection*{Summary}

In summary, the energy $E_1>0$ and the first left eigenvector $L_1(x)$
 that should have a positive derivative $L_1'(x)>0$ 
and a single root $L_1(x_1)=0$ can be chosen to construct Fokker-Planck generators ${\bold G}$
whose positive diffusion coefficient $D(x)$ is given :
one then needs to check the normalizations of Eqs \ref{steadyeqnormaZQES} and \ref{ortholrL1square}
as well as the boundary conditions of Eq. \ref{jnleft1} in order to obtain genuine Fokker-Planck generators.

In the next section, it is interesting to consider the further consequences for 
the supersymmetric partners ${\breve {\bold G}}$ and  ${\breve  {\bold H}}$.


\subsection{ Consequences for the eigenstates of the partners ${\breve {\bold G}}$, and  ${\breve  {\bold H}}$
associated with their ground state energy $E_1$ }

Here the goal is to write the explicit forms of the left eigenvector $ {\breve L}_1(x) $ 
and of the right eigenvector ${\breve R}_1(x) $ associated with the lowest energy $E_1$ of 
the partner ${\breve {\bold G}}$ in terms of $L_1(x)$ :

$\bullet $ The left eigenvector $ {\breve L}_1(x)$ is directly related to the derivative $L_1'(x) $  of $L_1(x)$ via Eq. \ref{ChoiceTwoleftviaNablaFP}
\begin{eqnarray}
 {\breve L}_1(x) =     L_1'(x)  
\label{hatL1}
\end{eqnarray}

$\bullet $ The right eigenvector ${\breve R}_1(x)$ of Eq. \ref{ARightEigenPartner}
reads using $P_*(x) = \frac{ e^{ -U_I(x)} }{D(x) Z}$
 of Eqs \ref{steadyeq} and \ref{steadyeqfroml1}
\begin{eqnarray}
{\breve R}_1(x) && =\frac{e^{-U_I(x)}}{Z E_1}   {\breve L}_1 ( x )  = \frac{ D(x) L_1'(x) P_*(x) }{E_1}
=  \frac{ e^{  - E_1\int^x dX  \frac{ L_1(X)   }{ D(X) L_1' (X) } } }  {  Z E_1}
\label{RightEigenPartner1FP}
\end{eqnarray}

$\bullet$ The product of the two previous equations produces the steady state $P_*^{[1]}(x) =  {\breve R}_1(x) {\breve L}_1(x)  = {\breve \Phi}_1^2 $ of Eq. \ref{AP1star} of the Fokker-Planck generator ${\bold G}^{[1]}$
\begin{eqnarray}
P_*^{[1]}(x)= {\breve \Phi}^2_1(x) =  {\breve L}_1(x) {\breve R}_1(x) && =  \frac{ D(x) \left[ L_1'(x) \right]^2 P_*(x) }{E_1}
= L_1'(x)   \frac{ e^{  - E_1\int^x dX  \frac{ L_1(X)   }{ D(X) L_1' (X) } } }  {  Z E_1}
\label{hatphi1square}
\end{eqnarray}
that satisfies automatically the appropriate normalization 
as a consequence of Eq. \ref{ortholrL1square}.

$\bullet$ The identification of Eq. \ref{hatphi1square}
with the parametrization ${\breve \Phi}_1(x) =  \frac{  e^{- \frac{ U^{[1]}(x)}{2} } }  {  \sqrt{ Z E_1 } }  $ of  
Eq. \ref{PHi0GSH1} with $Z_1=Z E_1$
\begin{eqnarray}
P_*^{[1]}(x)={\breve \Phi}_1^2(x) =    \frac{  e^{-  U^{[1]}(x) } }  {  Z E_1  }
\label{P1starU1}
\end{eqnarray}
leads to the potential 
\begin{eqnarray}
U^{[1]}(x) && = - \ln \left(   L_1'(x) \right)   + E_1\int^x dX  \frac{ L_1(X)   }{ D(X) L_1' (X) }  
\nonumber \\
\frac{ d U^{[1]}(x) }{dx} && = - \frac{ L_1''(x)}{   L_1'(x) }  + E_1  \frac{ L_1(x)   }{ D(x) L_1' (x) }  
\label{U1etU1prime}
\end{eqnarray}
The corresponding Fokker-Planck force
\begin{eqnarray}
F^{[1]}(x) = - D(x) \frac{ d U^{[1]}(x) }{dx}  = D(x) \frac{ L_1''(x)}{   L_1'(x) }  - E_1  \frac{ L_1(x)   }{  L_1' (x) }  
\label{F1}
\end{eqnarray}
can be compared to the initial Ito force $F_I(x)$ of Eq. \ref{FIEigenl1}
to obtain that it can be convenient to introduce
their half-sum $ F_+(x) $ and their half-difference $F_-(x) $ that have the simple expressions
\begin{eqnarray}
F_+(x) && \equiv \frac{F^{[1]}(x) +  F_I(x) }{2} =- E_1  \frac{ L_1(x)   }{  L_1' (x) } 
\nonumber \\
F_-(x) && \equiv \frac{ F^{[1]}(x) - F_I(x) }{2  } =D(x) \frac{ L_1''(x)}{   L_1'(x) } 
\label{Fpml1}
\end{eqnarray}

Note that these half-sum $ F_+(x) $ and half-difference $F_-(x) $ also appear naturally
in the Riccati Eq. \ref{Riccati} concerning the quantum scalar potentials 
that can be rewritten as
\begin{eqnarray}
0  = - \breve{V }(x) + V^{[1]}(x)+E_1 && = \frac{ \left[ F^{[1]}(x) \right]^2 - F^2_I(x) }{4 D(x) } + \frac{\frac{d F^{[1]}(x)}{dx} +\frac{d F_I(x)}{dx}  }{2}   + E_1
\nonumber \\
&& =   \frac{ F_-(x) F_+(x) }{D(x) } + F_+'(x) + E_1
\label{susyrecursionFpm}
\end{eqnarray}
This differential equation can be indeed obtained via the direct computation of
 the derivative $F_+'(x)$ of $F_+(x)$ of Eq. \ref{Fpml1}
\begin{eqnarray}
F_+'(x) && = - E_1 \left( \frac{ L_1'(x)   }{  L_1' (x) }  - \frac{ L_1(x) L_1''(x) }{ \big( L_1'(x) \big)^2 } \right)
= -E_1 + \left( E_1  \frac{ L_1(x)   }{  L_1' (x) } \right) \left( \frac{ L_1''(x)}{   L_1'(x) }\right) 
\nonumber \\
 && = -E_1 - F_+(x) \frac{F_-(x)}{ D(x)}
\label{Fpl1deri}
\end{eqnarray}

$\bullet $ Finally, let us mention that the root $x_1$ of Eq. \ref{x1uniqueroot}
where the first-left eigenvector vanishes $L_1(x_1)=0$
is also a root for $F_+(x)$ of Eq. \ref{Fpml1},
while the derivative $F_+'(x_1) $ is directly related to $E_1$ as a consequence of Eq. \ref{Fpl1deri}
\begin{eqnarray}
F_+(x_1) &&  =- E_1  \frac{ L_1(x_1)   }{  L_1' (x_1) } =0
\nonumber \\
F_+'(x_1)  && = -E_1 - F_+(x_1) \frac{F_-(x_1)}{ D(x_1)} = - E_1
\label{F1F1derix0}
\end{eqnarray}


\subsection{ Link with previous works on Quasi-Exactly-Solvable Quantum Hamiltonians with $N=2$ explicit eigenstates  }

As mentioned in the Introduction, the analysis of Quasi-Exactly-Solvable Quantum Hamiltonians with $N=2$ explicit 
eigenstates has attracted a lot of interest \cite{Tkachuk98,Tkachuk98bis,Tkachuk99,Tkachuk01,N2Examples,Quesne}
and it is thus useful to discuss the links with the previous subsections. 
In quantum mechanics, the analysis is based on the supersymmetric recursion of 
Eq. \ref{susyrecursionFpm} concerning the quantum scalar potentials that
can be rewritten as Eq. \ref{susyrecursionFpm}
in terms of the half-sum $ F_+(x)\equiv \frac{F^{[1]}(x) +  F_I(x) }{2} $ and the half-difference $F_-(x)\equiv \frac{ F^{[1]}(x) - F_I(x) }{2  } $
from which the initial Ito force $F_I(x)$ and the new force $F^{[1]}(x)  $ can be reconstructed via
\begin{eqnarray}
F_I(x) && = F_+(x) -F_-(x)  
\nonumber \\
F^{[1]}(x) && = F_+(x) + F_-(x)  
\label{reconstructFromFpm}
\end{eqnarray}

So the discussion can be summarized as follows (see the various presentations and examples in \cite{Tkachuk98,Tkachuk98bis,Tkachuk99,Tkachuk01,N2Examples,Quesne}): 

(i) in the first method, one uses Eq. \ref{susyrecursionFpm} to compute $F_-(x)$ in terms of $F_+(x)$ via
\begin{eqnarray}
F_-(x) = -  \frac{ D(x) }{ F_+(x)  }\left[ F_+'(x) +  E_1       \right] =  -  \frac{ D(x) }{ F_+(x)  }\left[ F_+'(x) - F_+'(x_1)       \right]
\label{vpartneridpmsolm}
\end{eqnarray}
where the conditions of Eq. \ref{F1F1derix0} are important to avoid any singularity at $x_1$ for $F_-(x)$.
So the first method to construct QES-models with $N=2$ 
amounts to choose an appropriate $F_+(x)$, to compute $F_-(x)$ via Eq. \ref{vpartneridpmsolm}
and then to reconstruct everything from them.

(ii) in the second method, one uses Eq. \ref{susyrecursionFpm} to compute $F_+(x)$ in terms of $F_-(x)$ : 
this amounts
to solve the differential equation via the method of variations of constants, i.e. 
one makes the change of variable from $F_+(x)$ to the new function $C(x)$
\begin{eqnarray}
F_+(x) &&  = C(x) e^{ - \int^x dX \frac{F_-(X)}{ D(X)} }
\nonumber \\
F_+'(x) &&  = \left[ C'(x)  - C(x) \frac{F_-(x)}{ D(x)}\right] e^{ - \int^x dX \frac{F_-(X)}{ D(X)} } 
= C'(x)   e^{ - \int^x dX \frac{F_-(X)}{ D(X)} }- F_+(x) \frac{F_-(x)}{ D(x)}
\label{VariationCst}
\end{eqnarray}
and one obtains the solution
\begin{eqnarray}
C'(x)   &&  = -E_1  e^{  \int^x dX \frac{F_-(X)}{ D(X)} }
\nonumber \\
C(x)   &&  = -E_1 \int^x dX' e^{  \int^{X'} dX \frac{F_-(X)}{ D(X)} }
\nonumber \\
F_+(x)   &&  = -E_1 \frac{ \int^x dX' e^{  \int^{X'} dX \frac{F_-(X)}{ D(X)} } }{e^{  \int^x dX \frac{F_-(X)}{ D(X)} }}
\label{VariationCstSolu}
\end{eqnarray}
So the second method to construct QES-models with $N=2$ 
amounts to choose an appropriate $F_-(x)$, to compute $F_+(x)$ via Eq. \ref{VariationCstSolu}
and then to reconstruct everything from them.

From the perspective of the present paper, both methods based either on the choice of $F_+(x)$ or on the choice of $F_-(x)$
can be reformulated as choices of the left eigenvector $L_1(x)$ as a consequence of the simple relations of Eq. \ref{Fpml1},
so all these methods can be considered as equivalent, but we feel that the choice of $L_1(x)$ in the Markov perspective
is clearer both physically and technically and thus more intuitive.


\subsection{ Discussion }

In this section, we have described why the choice of the first excited left eigenvectors $L_1(x)$
is simpler both technically and intuitively to construct Markov generators 
with a given diffusion coefficient $D(x)$ that are Quasi-Exactly-Solvable 
with $N=2$ explicit levels. 
In the remaining sections of the main text, we discuss how changes of variables
can be used to simplify the analysis.


\section{ Properties of changes of variables $x \to {\mathring x}$ in Fokker-Planck dynamics }

\label{sec_ChangeVariables}

In this section, we recall the properties of a change of variables $x \to {\mathring x}$ for the various properties of the Fokker-Planck dynamics, before stressing the role of the first left eigenvector $L_1(x)$ and its derivatives.

\subsection{ Reminder on changes of variables $x \to {\mathring x}$ in Fokker-Planck dynamics }

\subsubsection{ Reminder on the Stratonovich force $F_S(x)= F(x)+  \frac{D' (x)}{2}  = F_I(x)-  \frac{D' (x)}{2} $}

Besides the Fokker-Planck force $F(x)= - D(x) U'(x) $ of Eq. \ref{DFUIU} and the Ito force $F_I(x)=F(x)+ D' (x)$
of Eq. \ref{FI},
it is often useful to introduce also the Stratonovich force $ F_S(x) $ related to both previous forces via
\begin{eqnarray}
   F_S(x) && = F(x)+  \frac{D' (x)}{2} 
   \nonumber \\
    && = F_I(x)-  \frac{D' (x)}{2} 
\label{forceStrato}
\end{eqnarray}
with the corresponding potential $U_S(x)$ related to the potentials $U(x)$ and $U_I(x)  = U(x) - \ln D(x) $ of Eq. \ref{UIfromU}
\begin{eqnarray}
U_S(x) && = U(x) - \frac{\ln D(x) }{2} =  U_I(x) + \frac{\ln D(x) }{2}
\nonumber \\
U_S'(x) && = - \frac{ F_S(x) }{D(x) }
 \label{USfromU}
\end{eqnarray}
so that the steady state of Eq. \ref{steadyeq} can be also rewritten in terms of $U_S(x)$ and $D(x)$ as
\begin{eqnarray}
  P_*(x)  = \frac{ e^{ -U(x)} }{Z}   = \frac{ e^{ -U_I(x)} }{Z D(x) } = \frac{ e^{ -U_S(x)} }{Z \sqrt{D(x) } }
 \label{steadyeqUs}
\end{eqnarray}

The adjoint 
${\bold J}^{\dagger} $ of Eq. \ref{AJUIUdaggerFP}
becomes in terms of $U_S(x)$ or in terms of $F_S(x)$
\begin{eqnarray}
{\bold J}^{\dagger} && = e^{U(x)}  \frac{\partial}{\partial x} e^{- U_I(x)} =  e^{U_S(x)} \sqrt{D(x) } \frac{\partial}{\partial x} \sqrt{D(x) } e^{- U_S(x)}  
\nonumber \\
&& =   F_I(x) + D(x)  \frac{\partial}{\partial x} =  F_S(x)  + \sqrt{D(x) } \frac{\partial }{\partial x}  \sqrt{D(x) }   
 \label{AJUIUdaggerFPS}
\end{eqnarray}
and can be plugged into the adjoint ${\bold G}^{\dagger} $ to obtain 
\begin{eqnarray}
  {\bold G}^{\dagger} = {\bold J} ^{\dagger} \frac{\partial}{\partial x}
    &&   =   e^{U(x)}  \frac{\partial}{\partial x} e^{- U_I(x)} =  e^{U_S(x)} 
    \left(\sqrt{D(x) } \frac{\partial}{\partial x} \right)  e^{- U_S(x)}  \left( \sqrt{D(x) } \frac{\partial }{\partial x} \right) 
  \nonumber \\
  && =    F_S(x) \frac{\partial }{\partial x}  
  + \left( \sqrt{D(x) } \frac{\partial }{\partial x} \right)   \left( \sqrt{D(x) } \frac{\partial }{\partial x} \right)
\label{adjoint}
\end{eqnarray}
that clearly show the respective advantages of each perspective :

(i) The Ito force $ F_I( x) $ is more convenient when one wishes 
to write the differential operator $ {\bold G}^{\dagger} $ with all the derivatives on the right,
as already stressed after Eq. \ref{adjointIto}

(ii)  The Stratonovich force $F_S(x) $ is simpler 
when one wishes to make changes of variables, as recalled in more detail in 
the following subsection.


\subsubsection{ Transformation rules for the various properties of the Fokker-Planck dynamics
for a change of variables $x \to {\mathring x}$} 

\label{subsec_fullRulesChangevar}

$\bullet$ For all the observables $O(x)$, and thus in particular for the left eigenvectors $L_n(x)$,
the change of variables reduces to a change in the argument
\begin{eqnarray}
 {\mathring L}_n({\mathring x}) = L_n(x) \bigg\vert_{x=x({\mathring x})}
\label{Lnchange}
\end{eqnarray}

$\bullet$ The corresponding change of the adjoint of Eq. \ref{adjoint} written in terms of the Stratonovich forces
\begin{eqnarray}
  {\bold G}^{\dagger}    &&  =   \frac{ F_S(x) }{ \sqrt{D(x) }} \left( \sqrt{D(x) } \frac{\partial }{\partial x} \right)  
  + \left( \sqrt{D(x) } \frac{\partial }{\partial x} \right)   \left( \sqrt{D(x) } \frac{\partial }{\partial x} \right)
  \nonumber \\
  && =    \frac{ {\mathring F}_S(({\mathring x})) }{ \sqrt{{\mathring D}({\mathring x}) }} \left( \sqrt{{\mathring D}({\mathring x}) } \frac{\partial }{\partial {\mathring x}} \right)
  + \left( \sqrt{{\mathring D}({\mathring x}) } \frac{\partial }{\partial {\mathring x}} \right) 
    \left( \sqrt{{\mathring D}({\mathring x}) } \frac{\partial }{\partial {\mathring x}} \right) \equiv {\mathring {\bold G}}^{\dagger}
\label{adjointchange}
\end{eqnarray}
leads to the relation between the two partial derivatives  
that involves the two diffusion coefficients $ D(x) $ and ${\mathring D}({\mathring x}) $.
\begin{eqnarray}
 \sqrt{{\mathring D}({\mathring x}) } \frac{\partial }{\partial {\mathring x}} && = \sqrt{D(x) } \frac{\partial }{\partial x}
\label{partialderichange}
\end{eqnarray}
while the relation between the two Stratonovich forces read
\begin{eqnarray}
 \frac{ {\mathring F}_S({\mathring x}) }{ \sqrt{{\mathring D}({\mathring x}) }} =  \frac{ F_S(x) }{ \sqrt{D(x) }}  \bigg\vert_{x=x({\mathring x})}
\label{stratochange}
\end{eqnarray}

$\bullet$ For all probability densities such as the time-dependent probability density $P_t(x)$,
  the change of variables 
\begin{eqnarray}
{\mathring P}_t({\mathring x}) d{\mathring x}=   P_t(x) dx
\label{ProbaChange}
\end{eqnarray}
involves an additional prefactor coming from the change of measure determined by Eq. \ref{partialderichange}
that involves the two diffusion coefficients $ D(x) $ and ${\mathring D}({\mathring x}) $
\begin{eqnarray}
 \frac{ dx}{d{\mathring x}}  = \sqrt{  \frac{  D(x)  }{{\mathring D}({\mathring x}) } }
\label{ChangeMeasure}
\end{eqnarray}
In particular, this additional prefactor coming from the change of measure is present
between the right eigenvectors 
\begin{eqnarray}
{\mathring R}_n({\mathring x}) = \frac{ dx}{d{\mathring x}} R_n(x) 
= \sqrt{  \frac{  D(x)  }{{\mathring D}({\mathring x}) } }  R_n(x) \bigg\vert_{x=x({\mathring x})}
\label{Rnchange}
\end{eqnarray}
with the special case of the steady state $R_0(x)=P_*(x)$ 
\begin{eqnarray}
{\mathring P}_*({\mathring x}) = \frac{ dx}{d{\mathring x}} P_*(x) 
= \sqrt{  \frac{  D(x)  }{{\mathring D}({\mathring x}) } } P_*(x) \bigg\vert_{x=x({\mathring x})}
\label{Pstarchange}
\end{eqnarray}

$\bullet$ Finally for the quantum eigenstates $\Phi_n(x)=\sqrt{ L_n(x) R_n(x)}$,
the change can be obtained by putting together Eqs \ref{Lnchange} and \ref{Rnchange}
\begin{eqnarray}
 {\mathring {\Phi}}_n({\mathring x})=    \sqrt{{\mathring L}_n({\mathring x}) {\mathring R}_n({\mathring x})  } 
  =  \sqrt{ L_n(x) \sqrt{  \frac{  D(x)  }{{\mathring D}({\mathring x}) } }  R_n(x)  } 
  \bigg\vert_{x=x({\mathring x})} 
 = \left( \frac{ D(x)}{{\mathring D}({\mathring x}) } \right)^{\frac{1}{4}} \Phi_n(x) \bigg\vert_{x=x({\mathring x})} 
 \label{phinchange}
\end{eqnarray}

In conclusion, the change of variables for the left eigenvectors $L_n(x)$ of Eq. \ref{Lnchange}
is simpler than for the right eigenvectors $R_n(x)$ of Eq. \ref{Rnchange}
and than for the quantum eigenstates $\Phi_n(x)$ of Eq. \ref{phinchange}


\subsection{ Transformation rules from the point of view of the first left eigenvector $L_1(x)$ and the diffusion coefficient $D(x)$}

In section \ref{sec_2Eigen}, we have described how all properties defining the model could be rewritten in terms of 
the diffusion coefficient $D(x)$, the first left eigenvector $L_1(x)$ and its first two derivatives $L_1'(x)$ and $L_1''(x)$.
As a consequence, one may also write directly all the properties of the Fokker-Planck dynamics in the new variable ${\mathring x} $
in terms of the new diffusion coefficient ${\mathring D}({\mathring x})$ 
and in terms of the new first left eigenvector ${\mathring L}_1({\mathring x})$ with its first two derivatives 
${\mathring L}_1'({\mathring x})$ and ${\mathring L}_1''({\mathring x})$.
It is thus useful to stress here how their transformation rules are related.

For $n=1$, the rule of Eq. \ref{Lnchange}
for the transformation of the first left eigenvector
\begin{eqnarray}
 {\mathring L}_1({\mathring x}) = L_1(x) \bigg\vert_{x=x({\mathring x})}
\label{L1change}
\end{eqnarray}
yields that the first derivatives of these two left eigenvectors 
are related via the change of measure of Eq. \ref{ChangeMeasure}
that involves the two diffusion coefficients
\begin{eqnarray}
 {\mathring L}_1'({\mathring x}) = \frac{d {\mathring L}_1({\mathring x})}{d {\mathring x}}= \frac{ dx}{d{\mathring x}}  L_1'(x) \bigg\vert_{x=x({\mathring x})}
 =  \sqrt{  \frac{  D(x)  }{{\mathring D}({\mathring x}) } }  L_1'(x) \bigg\vert_{x=x({\mathring x})}
\label{L1changederi}
\end{eqnarray}
so that the change between the two diffusion coefficients is directly related to the first two derivatives $L_1'(x) $ and ${\mathring L}_1'({\mathring x}) $
\begin{eqnarray}
{\mathring D}({\mathring x}) [ {\mathring L}_1'({\mathring x}) ]^2=    D(x)  [ L_1'(x)]^2 \bigg\vert_{x=x({\mathring x})}
\label{DchangefrimL1}
\end{eqnarray}

This property shows again the essential role of the first-excited left eigenvectors $ {\breve L}_1(x)=L_1'(x)$ 
even for the change of measure controlled by the diffusion coefficients.

In practice, the expressions of all properties in terms of $D(x)$ and $L_1(x)$ and its derivatives
as given in section \ref{sec_2Eigen} can be written directly for the Fokker-planck model in the new variables ${\mathring x} $
in terms of the new diffusion coefficient ${\mathring D}({\mathring x}) $ and $ {\mathring L}_1({\mathring x}) $ with its first two derivatives,
so that one does not need to redo the whole discussion of subsection \ref{subsec_fullRulesChangevar}
in the specific examples that will be considered in the two next sections.


\subsection{ Discussion} 

After the above description of arbitrary changes of variables $x \to {\mathring x}$,
let us now discuss what type of new variables ${\mathring x} $ can lead to simplifications :

(i) A standard choice is the new variable $z$ associated with the constant diffusion coefficient 
$d(z)=1$ that will be discussed in section \ref{sec_z}, that leads to the standard quantum Hamiltonians 
with the kinetic term $\frac{\partial^2}{\partial z^2}$ that are the most often discussed in the literature on
supersymmetric quantum mechanics. Then the only remaining parameter of the model
is the Fokker-Planck force $f(z)$ that coincides with the two others $f(z)=f_I(z)=f_S(z) $ as a consequence of $d'(z)=0$.
Note that we will use small letters for all properties of the Fokker-Planck model in the variable $z$ to stress its specific properties
with respect to the general case of the variable $x$..

(ii)  The analysis of the previous sections where the left eigenvector $L_1(x)$ plays a major role
suggests to choose to use $L_1(x)$ itself to define the new variable $y-y_1=L_1(x)= {\cal L}_1(y)$,
 where the first left eigenvector $ {\cal L}_1(y) $ is thus linear,
 that will be discussed in the next section \ref{sec_y}. Then the only remaining parameter of the model
is the diffusion coefficient ${\cal D}(y)$. Note that we will use calligraphic
 letters for all properties of the Fokker-Planck model in the variable $y$ to stress its specific properties
with respect to all the other cases.


\section{ Variable $y-y_1=L_1(x)= {\cal L}_1(y)$ where the first left eigenvector $ {\cal L}_1(y) $ is linear } 

\label{sec_y}

In this section, we describe the simplifications of the Fokker-Planck dynamics 
in the new variable $y-y_1=L_1(x)$ based on the first left eigenvector $L_1(x)$.

\subsection{ Change of variables towards the variable $y-y_1=L_1(x)={\cal L}_1(y)$ 
with the corresponding linear Ito force $ {\cal F}_I(y)=-E_1 (y-y_1) $  
} 

Since the first excited left eigenvector $L_1(x)$ plays a major role in the previous sections
and has a positive
derivative $L_1'(x)>0$, it can be used to define the change of variables 
\begin{eqnarray}
y-y_1 && =L_1(x)  \ \ \text{ that vanishes at  $y=y_1$ at the single root $x_1$ of $L_1(x_1)=0$}
\nonumber \\
\frac{dy}{dx} && = L_1'(x) >0
\nonumber \\
 \frac{\partial }{\partial y} && = \frac{1}{L_1'(x)}  \frac{\partial }{\partial x}
\label{defyfroml1}
\end{eqnarray}

For this specific choice, the general rules for changes of variables in Fokker-Planck dynamics discussed in the previous section
\ref{sec_ChangeVariables} lead to the following simplifications :

$\bullet$ The change of Eq. \ref{Lnchange} for the left eigenvectors yields that the first excited eigenvector $ {\cal L}_1(y) $
is simply linear in the variable $y$
\begin{eqnarray}
 {\cal L}_1(y) && = L_1(x) \bigg\vert_{x=x(y)} = y-y_1
 \nonumber \\
 {\cal L}_1'(y) && =1
  \nonumber \\
 {\cal L}_1''(y) && =0
\label{L1y}
\end{eqnarray}

$\bullet $ The change of Eq. \ref{DchangefrimL1} for the diffusion coefficients yields using ${\cal L}_1'(y)  =1 $
  that the new diffusion coefficient ${\cal D}(y) $ in the variable $y$ reads
\begin{eqnarray}
  {\cal D}(y)  && = D(x) \big( L_1'(x) \big)^2 \bigg\vert_{x=x(y)}
\label{calDy}
\end{eqnarray}

$\bullet$ The Ito force ${\cal F}_I(y) $ of Eq. \ref{FIEigenl1} computed from the eigenvalue Eq. \ref{EigenL1}
satisfied by $ {\cal L}_1(y) $ is thus also linear in $y$
\begin{eqnarray}
{\cal F}_I(y)  =  -   E_1 \frac{  {\cal L}_1(y)   }{ {\cal L}_1'(y)  } -  {\cal D}(y) \frac{ {\cal L}_1''(y)  }{ {\cal L}_1'(y)  }
= -   E_1 (y-y_1) 
\label{FItoy}
\end{eqnarray}
where the first-excited energy $E_1>0$ appears directly in the slope.

Note that this linearity of the Ito force
is also very natural from the point of view of the Ito SDE 
that generates the trajectories $Y(t)$ from the increments $dB(t)$  of the Brownian motion $B(t)$ 
\begin{eqnarray}
dY(t)  && = {\cal F}_I(Y(t)) dt + \sqrt{ 2 {\cal D}( Y(t)) } dB(t) = -   E_1 (Y(t)-y_1) + \sqrt{ 2 {\cal D}( Y(t)) } dB(t) 
\label{ItoSDEy}
\end{eqnarray}
So, independently of the diffusion coefficient ${\cal D}(y) $, the averaged value $ {\mathbb E} \left( Y(t) \right)$ of $Y(t)$
follows  the closed dynamics
\begin{eqnarray}
\partial_t {\mathbb E} \left( Y(t) \right)  =  -   E_1 \bigg({\mathbb E} \left( Y(t) \right)-y_1\bigg) 
\label{ItoSDEyav}
\end{eqnarray}
The solution corresponds to the exponential convergence as $e^{- E_1 t}$ towards $y_1$ representing the averaged value in the steady state
\begin{eqnarray}
 {\mathbb E} \left( Y(t) \right)  -  y_1 = e^{- E_1 t}  \bigg({\mathbb E} \left( Y(0) \right)-y_1\bigg) 
\label{ItoSDEyavsolu}
\end{eqnarray}

$\bullet$  Let us stress that while the Ito force $ {\cal F}_I(y) = -E_1 (y-y_1)$ of Eq. \ref{FItoy}
 is linear,
the corresponding Fokker-Planck force $ {\cal F}(y) $ and the Stratonovich forces $ {\cal F}_S(y)$ computed from Eq. \ref{forceStrato}
\begin{eqnarray}
  {\cal F}(y) && = {\cal F}_I(y)- {\cal D}'(y) = -E_1 (y-y_1)- {\cal D}'(y)
\nonumber \\
 {\cal F}_S(y) && =  {\cal F}_I(y) - \frac{{\cal D}' (y)}{2}  =  -E_1 (y-y_1) - \frac{{\cal D}'(y)}{2}
\label{FFSyys}
\end{eqnarray}
include additional contributions coming from the derivative ${\cal D}'(y) $ of the diffusion coefficient ${\cal D}(y) $,
and are thus a priori non-linear, unless 
${\cal D}'(y) $ is also linear, i.e. ${\cal D}(y)$ is a polynomial of degree 2, 
an important special case that will be discussed later in subsection \ref{subsec_pearson}.

$\bullet$ The steady state ${\cal P}^*(y) $ can be obtained
by plugging the left eigenvector ${\cal L}_1(y) = y-y_1 $ and its derivative ${\cal L}_1'(y) = 1 $
into Eq. \ref{steadyeqfroml1} 
\begin{eqnarray}
  {\cal P}^*(y)   =  \frac{ e^{  - E_1\int^y dY  \frac{ {\cal L}_1(Y)   }{ {\cal D}(Y) {\cal L}_1'(Y) } } }
  {{\cal D}(y) {\cal L}_1'(y)  Z}
  =  \frac{ e^{  - E_1\int^y dY  \frac{ Y-y_1   }{ {\cal D}(Y)  } } }  {{\cal D}(y)  Z}
 \label{steadyeqfroml1y}
\end{eqnarray}

$\bullet$ Plugging the Ito force $ {\cal F}_I(y)=-E_1 (y-y_1) $ of Eq. \ref{FItoy}
into the adjoint of Eq. \ref{adjoint}
\begin{eqnarray}
{\boldsymbol {\cal G}}^{\dagger} && = {\cal F}_I(y) \frac{\partial }{\partial y} +  {\cal D}(y) \frac{\partial^2 }{\partial y^2}
\nonumber \\
&& =  -E_1 (y-y_1) \frac{\partial }{\partial y} +  {\cal D}(y) \frac{\partial^2 }{\partial y^2}
\label{adjointyito}
\end{eqnarray}
yields that the higher left eigenvectors ${\cal L}_n(y)$
beyond ${\cal L}_{n=0}(y)=1$ and ${\cal L}_{n=1}(y) = y-y_1 $,
satisfy the eigenvalue equations 
\begin{eqnarray}
-E_n {\cal L}_n(y) && = {\boldsymbol {\cal G}}^{\dagger} {\cal L}_n(y) 
\nonumber \\
&& =  -E_1 (y-y_1) {\cal L}_n'(y) +  {\cal D}(y) {\cal L}_n''(y)
\label{Eigenleftny}
\end{eqnarray}


\subsection{ Simplifications for the partner ${\breve {\bold {\cal G}} } $ 
and for the new Fokker-Planck generator $ {\bold {\cal G}} ^{[1]} $}

Lets us now describe how the variable $y$ introduced in Eq. \ref{defyfroml1}
also simplifies the properties of the partner ${\breve {\bold {\cal G}} } $ 
and of the new Fokker-Planck generator $ {\bold {\cal G}} ^{[1]} $ :

$\bullet $ The left eigenvector $ {\breve {\cal L}}_1(y)$ of Eq. \ref{hatL1} computed via the derivative ${\cal L}_1'(y)=1 $
\begin{eqnarray}
{\breve {\cal L}}_1(y) =     {\cal L}_1'(y) =1
\label{hatL1y}
\end{eqnarray}
reduces to unity : as a consequence, 
the Doob transformation of Eq. \ref{ADoobRing} needed to construct the new Fokker-Planck generator $ {\bold {\cal G}}^{[1]} $
via the similarity transformation involving the left eigenvector ${\breve {\cal L}}_1(y) $
reduces to a simple shift of $E_1$ 
\begin{eqnarray}
 {\bold {\cal G}}^{[1]} && \equiv {\breve {\cal L}}_1(y) {\breve {\bold {\cal G}} } \frac{1}{{\breve {\cal L}}_1(y) } + E_1
 =  {\breve {\bold {\cal G}} } + E_1
  \label{DoobRingy}
\end{eqnarray}

$\bullet $ The right eigenvector ${\breve {\cal R}}_1(y)$ of Eq. \ref{RightEigenPartner1FP} 
coincides with the steady state ${\cal P}_*^{[1]}(y)={\breve {\cal L}}_1(y) {\breve {\cal R}}_1(y)$ of Eq. \ref{hatphi1square}
 since  ${\breve {\cal L}}_1(y)=1 $
\begin{eqnarray}
{\breve {\cal R}}_1(y)  &&  = \frac{  {\cal D}(y) {\cal L}_1'(y)  {\cal P}_*(y)  }{E_1}
=  \frac{ e^{  - E_1\int^y dY  \frac{ Y-y_1   }{ {\cal D}(Y)  } } }  {  Z E_1} = {\cal P}_*^{[1]}(y)
\label{RightEigenPartner1FPy}
\end{eqnarray}

$\bullet$ The new potential ${\cal U}^{[1]}(y)  $ of Eq. \ref{U1etU1prime} becomes using ${\cal L}_1'(y) =1 $
\begin{eqnarray}
{\cal U}^{[1]}(y) && =  - \ln \left(   {\cal L}_1'(y) \right) +  E_1 \int^y dY  \frac{ Y-y_1   }{ {\cal D}(Y)  } =  E_1 \int^y dY  \frac{ Y-y_1   }{ {\cal D}(Y)  }
\nonumber \\
\frac{ d {\cal U}^{[1]}(y) }{dy}   && =  E_1   \frac{ y-y_1   }{ {\cal D}(y)  }
\label{U1etU1primey}
\end{eqnarray}
The corresponding new Fokker-Planck force ${\cal F}^{[1]}(y) $ of Eq. \ref{F1}
\begin{eqnarray}
{\cal F}^{[1]}(y) \equiv - {\cal D}(y) \frac{ d {\cal U}^{[1]}(y) }{dy}  =- E_1(y-y_1) = {\cal F}^I(y)
\label{F1y}
\end{eqnarray}
thus coincides with the initial linear Ito force ${\cal F}^I(y)=- E_1(y-y_1)$ of Eq. \ref{FItoy}.
However the new Ito force $  {\cal F}_I^{[1]}(y)$ reads
\begin{eqnarray}
  {\cal F}_I^{[1]}(y)  = {\cal F}^{[1]}(y)+ {\cal D}'(y) = -E_1 (y-y_1)+ {\cal D}'(y)
\label{F1Itoy}
\end{eqnarray}
and is thus a priori non-linear, unless 
${\cal D}'(y) $ is also linear, i.e. ${\cal D}(y)$ is a polynomial of degree 2, 
as will be discussed in subsection \ref{subsec_pearson}


\subsection{ Consequences from the perspective of quantum Hamiltonians } 

In the variable $y$, the quantum Hamiltonian ${\bold {\cal H}}  $ of Eq. \ref{hsusy} 
\begin{eqnarray}
{\bold {\cal H}} &&=  - \frac{ \partial  }{\partial y} {\cal D}(y) \frac{ \partial  }{\partial y} +{\cal V}(y)
\label{hsusyY}
\end{eqnarray}
involves the scalar potential ${\cal V}(y) $ of Eq. \ref{vfromu}
that can be computed using the Fokker-Planck force $
  {\cal F}(y) = -E_1 (y-y_1)- {\cal D}'(y) $ of Eq. \ref{FFSyys}
\begin{eqnarray}
{\cal V}(y) && =\frac{ {\cal F}^2(y) }{4 {\cal D}(y) } + \frac{{\cal F}'(y)}{2}
= \frac{ \left[  -E_1 (y-y_1)- {\cal D}'(y)\right]^2 }{4 {\cal D}(y) } + \frac{\left[ -E_1 - {\cal D}''(y) \right]}{2}
\label{vfromuY}
\end{eqnarray}
while the partner potential of \ref{vpartner} involving the linear Ito force ${\cal F}_I(y) = -E_1 (y-y_1) $ reduces to
\begin{eqnarray}
\breve{ {\cal V} }(y) && = \frac{ \left[ {\cal F}_I(y) \right]^2 }{4 {\cal D}(y) } - \frac{ {\cal F}_I'(y)  }{2} 
= \frac{  E_1^2 (y-y_1)^2 }{4 {\cal D}(y) } + \frac{ E_1  }{2} 
\label{vpartnery}
\end{eqnarray}

As a consequence, the Riccati Eq. \ref{Riccati}
\begin{eqnarray}
 \breve{ {\cal V} }(y) && = {\cal V}^{[1]}(y) +E_1
\nonumber \\
\text{ i.e.} \ \ \ \ \ \ \ \  \frac{  E_1^2 (y-y_1)^2 }{4 {\cal D}(y) } + \frac{ E_1  }{2}  && 
=  \frac{ \left[ {\cal F}^{[1]}(y) \right]^2 }{4 {\cal D}(y) } + \frac{\frac{d {\cal F}^{[1]}(y)}{dy}}{2}   + E_1
\label{susyrecursiony}
\end{eqnarray}
has the obvious solution
\begin{eqnarray}
{\cal F}^{[1]}(y)= -E_1(y-y_1)
\label{F1toy}
\end{eqnarray}
in agreement with Eq \ref{F1y}.


\subsection{ Simplifications for Pearson models when the diffusion coefficient ${\cal D}(y)$
 is a polynomial of degree 2 } 
 
 \label{subsec_pearson}

When the diffusion coefficient ${\cal D}(y)$
 is a positive polynomial of degree 2, with its derivative $ {\cal D}'(y)$ a polynomial of degree 1
 \begin{eqnarray}
{\cal D}(y) && = d_2 y^2+d_1 y +d_0 \geq 0
\nonumber \\
{\cal D}'(y) && = 2d_2 y+d_1 
\label{dyppearson}
\end{eqnarray}
then the Fokker-Planck force $ {\cal F}(y) $ and the Stratonovich forces $ {\cal F}_S(y)$
of Eq. \ref{FFSyys} are also linear but with different coefficients than the Ito force ${\cal F}_I(y) = -E_1 (y-y_1) $ of Eq. \ref{F1Itoy}
\begin{eqnarray}
  {\cal F}(y) && = {\cal F}_I(y)- {\cal D}'(y) = - (E_1 +  2d_2) y + (E_1 y_1-d_1 )
\nonumber \\
 {\cal F}_S(y) && =  {\cal F}_I(y) - \frac{{\cal D}' (y)}{2}  =  - (E_1 +  d_2) y  + \left( E_1 y_1 -\frac{d_1 }{2}\right)
\label{FFSyyspearson}
\end{eqnarray}

The iterated Ito force of Eq. \ref{F1Itoy} is then also linear
\begin{eqnarray}
  {\cal F}_I^{[1]}(y)  =  -E_1 (y-y_1)+ {\cal D}'(y) = - (E_1 -  2d_2) y + (E_1 y_1+d_1 ) \equiv - e_2 (y-y_2)
\label{F1Itoypearson}
\end{eqnarray}
where the new gap $e_2=E_2-E_1$ and the new root $y_2$
read in terms of the initial gap $e_1=E_1=E_0=E_1$ and in terms of the initial root $y_1$
\begin{eqnarray}
e_2= e_1 -  2d_2
\nonumber \\
e_2 y_2= e_1 y_1+d_1
\label{iterationey}
\end{eqnarray}

The generalization for the Ito force after $p$ iterations
\begin{eqnarray}
  {\cal F}_I^{[p]}(y)   \equiv - e_p (y-y_p)
\label{FpItoypearson}
\end{eqnarray}
leads to the simple solutions via recurrence
\begin{eqnarray}
e_p && = e_{p-1}  -  2d_2 = ... = e_1 -  2 d_2 (p-1) 
\nonumber \\
e_p y_p &&= e_{p-1} y_{p-1}+d_1 = ... = e_1 y_1 +d_1 (p-1)
\label{gape2e1n}
\end{eqnarray}

This analysis based on the iterated Ito force ${\cal F}_I^{[p]}(y) $
is in agreement with the direct analysis based on the
 adjoint ${\boldsymbol {\cal G}}^{\dagger} $ of Eq. \ref{adjointyito} with the diffusion coefficient ${\cal D}(y)   $ of Eq. \ref{dyppearson}
\begin{eqnarray}
{\boldsymbol {\cal G}}^{\dagger}  =  -E_1 (y-y_1) \frac{\partial }{\partial y} + \bigg(d_2 y^2+d_1 y +d_0 \bigg) \frac{\partial^2 }{\partial y^2}
\label{adjointyitopearson}
\end{eqnarray}
The left eigenvector ${\cal L}_n(y)$  of Eq. \ref{Eigenleftny} is then a polynomial of order $n$
and its dominant term $x^n$ is sufficient to determine $E_n$ via the eigenvalue equation
\begin{eqnarray}
-E_n \bigg( y^n  + ... \bigg) =  -E_1  \bigg( n y^n  + ... \bigg) +  \bigg(d_2 n (n-1) y^n+... \bigg)
\label{Eigenleftnypoly}
\end{eqnarray}
that leads to the solution
\begin{eqnarray}
E_n  =  n E_1  -   d_2 n (n-1) 
\label{energyfromEigenleftnypoly}
\end{eqnarray}
so that the consecutive gaps
\begin{eqnarray}
e_{n+1} = E_{n+1}-E_n  = (n+1) E_1  -   d_2 (n+1)n    -n E_1  +   d_2 n (n-1) = E_1 - 2 d_2 n 
\nonumber \\
e_n = E_{n}-E_{n-1}   = E_1 - 2 d_2 (n-1) 
\label{gappearson}
\end{eqnarray}
satisfy
\begin{eqnarray}
e_{n+1} - e_n =  - 2 d_2 n + 2 d_2 (n-1) = - 2 d_2
\label{gappearsoniter}
\end{eqnarray}
in agreement with Eq \ref{gape2e1n}.

In summary, the change of variable towards the variable $y$ is also useful to identify the models related to 
the Pearson family of exactly-soluble models with quadratic diffusion coefficients ${\cal D}(y)$ and linear forces 
that will not be rediscussed here (see  \cite{pearson1895,pearson_wong,diaconis,autocorrelation,pearson_class,pearson2012,PearsonHeavyTailed,pearson2018,c_pearson}
and references therein for the six basic examples where it is essential to take into account the boundary conditions that are not discussed in the present paper).



\section{ Variable $z$ with the diffusion coefficient $d(z)=1$  } 

\label{sec_z}

In this section, we describe the simplifications in the variable $z$ with the diffusion coefficient $d(z)=1$.

\subsection{ Change of variables towards the variable $z$ with the diffusion coefficient $d(z)=1$  }

$ \bullet $ The change of variables either from the variable $x$, or from the variable $y$, is defined via the rule of Eq. \ref{partialderichange}
for the partial derivative that produces the diffusion coefficient $d(z)=1$
\begin{eqnarray}
 \frac{\partial }{\partial z} && = \sqrt{D(x) } \frac{\partial }{\partial x} = \sqrt{ {\cal D}(y) } \frac{\partial }{\partial y}
\label{partialderiz}
\end{eqnarray}

$\bullet$ The change of variables of Eq. \ref{Lnchange} for the first left eigenvector yields using  
$ {\cal L}_1(y) = y-y_1 $ of Eq. \ref{L1y}
\begin{eqnarray}
y-y_1 && =  {\cal L}_1(y)  = l_1(z) \bigg\vert_{z=z(y)} \ \ \text{ that vanishes at  $y=y_1$ at the single root $z_1$ of $l_1(z_1)=0$}
\label{Lnchangey}
\end{eqnarray}

$\bullet$ The change of Eq. \ref{DchangefrimL1} for the diffusion coefficients yields using ${\cal L}_1'(y)  =1 $
and $d(z)=1$  that the first derivative $l_1'(z)$ reads
\begin{eqnarray}
 l_1'(z)    = \sqrt{ {\cal D}(y) } \bigg\vert_{y=y(z)}
\label{linkl1pzDy}
\end{eqnarray}

$ \bullet $ In the variable $z$ where $d(z)=1$, the three forces coincide $f(z)=f_I(z)=f_S(z)$
and can be rewritten in terms of $l_1(z)$ and its first two derivatives via Eq. \ref{FIEigenl1}
\begin{eqnarray}
f(z)=f_I(z)=f_S(z)  = - E_1 \frac{  l_1(z)  }{ l_1'(z)} -  \frac{  l_1''(z)  }{ l_1'(z)}  
\label{fzfromy}
\end{eqnarray}

$\bullet$ The steady state $p^*(z)  $ of Eq. \ref{steadyeqfroml1} reads
\begin{eqnarray}
  p^*(z)  && =  \frac{ e^{  - E_1\int^z d{z'}  \frac{ l_1(z')   }{  l_1' (z') } } }
  { l_1'(z)  Z} = \frac{ e^{- u(z) } }{ Z} 
 \label{steadyeqfroml1z}
\end{eqnarray}
with the potential
\begin{eqnarray}
u(z) && =E_1\int^z d{z'}  \frac{ l_1(z')   }{  l_1' (z') }  + \ln \big(  l_1'(z) \big)
\nonumber \\
u'(z) && = E_1 \frac{  l_1(z)  }{ l_1'(z)} +  \frac{  l_1''(z)  }{ l_1'(z)}  = -f(z)
\label{uzfromy}
\end{eqnarray}

$\bullet$ The adjoint of Eq. \ref{adjoint} reduces to
\begin{eqnarray}
{\boldsymbol g}^{\dagger} && =f(z) \frac{\partial }{\partial z} +   \frac{\partial^2 }{\partial z^2}
\label{adjointz}
\end{eqnarray}

$\bullet$ The iterated force of Eq. \ref{F1} reads
\begin{eqnarray}
f^{[1]}(z)    =   - E_1 \frac{  l_1(z)  }{ l_1'(z)} +  \frac{  l_1''(z)  }{ l_1'(z)}  
\label{f1z}
\end{eqnarray}

$ \bullet $ The quantum Hamiltonian ${\bold h}  $ of Eq. \ref{hsusy} 
\begin{eqnarray}
{\bold h} &&=  - \frac{ \partial^2  }{\partial z^2}  +v(z)
\label{hsusyz}
\end{eqnarray}
has the standard kinetic term $  \frac{ \partial^2  }{\partial z^2}$
and involves the scalar potential $v(z)$ of Eq. \ref{vfromu} 
\begin{eqnarray}
v(z) && =\frac{ f^2(z) }{4  } + \frac{f'(z)}{2}
\label{vfromuz}
\end{eqnarray}
while the partner potential of \ref{vpartner} reads
\begin{eqnarray}
\breve{ v }(z) && = \frac{ f^2(z) }{4  } - \frac{f'(z)}{2}
\label{vpartnerz}
\end{eqnarray}
As a consequence, the Riccati Eq. \ref{Riccati} becomes
\begin{eqnarray}
 \breve{v }(z) && = v^{[1]}(z)+E_1
\nonumber \\
\text{ i.e.} \ \ \ \ \ \ \ \  \frac{ f^2(z) }{4  } - \frac{\frac{d f(z)}{dz}}{2}  && 
=  \frac{ \left[ f^{[1]}(z) \right]^2 }{4  } + \frac{\frac{d f^{[1]}(z)}{dz}}{2}   + E_1
\label{susyrecursionz}
\end{eqnarray}


\subsection{ Translation of the Pearson models in the variable $y$ towards the variable $z$ with $d(z)=1$  }

In the variable $y$, we have described the simplifications
of the Pearson family when  the diffusion coefficient ${\cal D}(y)$
 is a positive polynomial of degree 2 around Eq. \ref{dyppearson}.
 It is thus interesting to translate these models in the variable $z$ 
by plugging $y=y(z)=y_1 + l_1(z) $ of Eq. \ref{Lnchangey}
into  Eq. \ref{linkl1pzDy} 
\begin{eqnarray}
\bigg( l_1'(z)  \bigg)^2 && = {\cal D}(y)  \bigg\vert_{y=y(z)} = d_2  \bigg(y_1 + l_1(z) \bigg)^2+d_1  \bigg(y_1 + l_1(z) \bigg) +d_0  
\label{linkl1pzDypearson}
\end{eqnarray}
that can be derived with respect to $z$ to obtain after dividing by $2 l_1'(z)$
\begin{eqnarray}
 l_1 ''(z)   =   d_2  \bigg(y_1 + l_1(z) \bigg)+\frac{d_1}{2} =    d_2  l_1(z) +   \bigg(d_2 y_1 + \frac{d_1 }{2} \bigg) 
\label{linkl1pzDypearsonderi}
\end{eqnarray}

For the special case $z=z_1$ where $l_1(z_1)=0$, these equations read
\begin{eqnarray}
\bigg( l_1'(z_1)  \bigg)^2 && = d_2  y_1^2+d_1 y_1   +d_0  
\nonumber \\
l_1 ''(z_1)  &&  =     \bigg(d_2 y_1 + \frac{d_1 }{2} \bigg) 
\label{linkl1pzDypearsonz1}
\end{eqnarray}

In conclusion, the left eigenvector $l_1(z)$ satisfies the second-order linear differential equation of Eq. \ref{linkl1pzDypearsonderi}
that only involves constant coefficients
\begin{eqnarray}
 l_1''(z)    = d_2  l_1(z) +  c \ \ \ \text{ with } \ \ c \equiv l_1''(z_1) = \bigg(d_2 y_1 + \frac{d_1 }{2} \bigg)
\label{linkl1pzDyderipearsonbis}
\end{eqnarray}

The form of the solutions depends on the coefficient $d_2$  as follows :
\begin{eqnarray}
d_2=0 : \ \ \ \ \ \  l_1(z) &&  = c  \frac{ z^2 }{2} + c_1 z +  c_2
\nonumber \\
d_2>0 : \ \ \ \ \ \  l_1(z) &&  =  - \frac{c}{d_2} +  b_+  e^{ z \sqrt{d_2}} + b_-   e^{ - z \sqrt{d_2}}  
\nonumber \\
d_2<0 : \ \ \ \ \ \  l_1(z) &&  =  - \frac{c}{d_2} +  b_+  \cos( z \sqrt{- d_2} ) + b_-   \sin ( z \sqrt{- d_2})  
\label{pearsonsolul1d2plus}
\end{eqnarray}
and one recovers the standard translations of the Pearson models of the previous section involving the variable $y$
into the variable $z$ where $l_1(z)$ is either polynomial, or hyperbolic or trigonometric 
(see  \cite{c_pearson} and references therein for the six basic examples).


\section{ Conclusion}

\label{sec_conclusion}

In this paper, we have revisited the construction of Quasi-Exactly-Solvable quantum Hamiltonians with two explicit eigenstates $\Phi_0(x)$ and $\Phi_1(x)$ of energies $E_0$ and $E_1$ from the point of view of one-dimensional Markov processes satisfying detailed-balance, whose generators are related to quantum Hamiltonians via similarity transformations. Here the lowest energy vanishes $E_0=0$ and is associated with the conservation of probability and with the steady state $P_*(x)$, while $E_1>0$ is the rate that governs the exponential relaxation towards the steady-state, and is associated with the slowest observable $L_1(x)$. 
We have explained in detail how the Markov perspective leads to interesting re-interpretations and simplifies the construction of quasi-exactly-solvable models with $N=2$ explicit levels when one takes the slowest observable $L_1(x)$ as the central object from which all the other properties can be reconstructed. This general approach has been applied to Fokker-Planck dynamics in continuous space,
where generators correspond to second-order differential operators. We have also explained how to make changes of variables
towards variables that simplify the properties, namely the variable $y$ where the slowest observable ${\cal L}_1(y)$ is linear,
and the variable $z$ with the diffusion coefficient $d(z)=1$.

The case of Markov jump processes on the lattice is considered in the following Appendix in order to stress the similarities and the differences
with the diffusion processes considered in the main text.

As  final remark, let us mention that the results of the present paper concerning one-dimensional Markov
processes satisfying detailed-balance are also useful for one-dimensional Markov
processes breaking detailed-balance, when a non-vanishing steady current is produced by periodic boundary conditions 
or by external reservoirs at the boundaries: 
in both cases, as discussed in detail in the respective recent works \cite{c_lyapunov} and \cite{c_boundarydriven},
it is nevertheless still possible and very useful to continue to use the quantum Hamiltonian $H=Q^{\dagger} Q$ 
even if $Q$ does not annihilate the ground state $\Phi_0(x)$,
while the partner $H=QQ^{\dagger}  $ has a zero-energy ground state that is annihilated by $Q^{\dagger}$,
so that the partner is associated with an equilibrium Markov process (see \cite{c_lyapunov} and \cite{c_boundarydriven}
for more details).


\appendix

\section { Application to one-dimensional Markov jump dynamics }

\label{app_Jump}

In this section, we focus on Markov jump dynamics on the one-dimensional lattice $x$,
where the operator $\nabla$ introduced in Eq. \ref{Adefnabla} is 
the finite-difference operator 
 \begin{eqnarray}
\nabla \equiv  e^{ \frac{1}{2} \frac{\partial}{\partial x} } - e^{ - \frac{1}{2} \frac{\partial}{\partial x }} 
 \equiv  e^{ \frac{\partial}{2}  } - e^{ -  \frac{\partial}{2}} 
 \label{nablafinitediff}
\end{eqnarray}
where the translation operators 
$e^{ b \frac{\partial }{ \partial x} }$ will be denoted by the simplified notation $e^{ b \partial  }$.


\subsection{ Specific properties of the translation operators $e^{ b \partial }$ in discrete space  }

The translation operators $e^{ b \partial }$ by $b$
acts on any function $O( x)$ via the Taylor formula
\begin{eqnarray}
e^{ b \partial  } O(x) = \sum_{p=0}^{+\infty} \frac{ b^p }{p! } \left( \frac{\partial }{ \partial x}\right)^p O( x)
= O \left(  x +b \right)
\label{finitedefop}
\end{eqnarray}

The application to product of two functions $O(x)$ and $\Omega(x)$
\begin{eqnarray}
e^{ b \partial  } O(x) \Omega(x) = O ( x +b )  \Omega (x+b)
 \label{TranslationProduitFunctions}
\end{eqnarray}
yields the following rule at the operator level 
for the exchange of the translation operator $e^{ b \partial  } $ and the multiplication by a function $O(x)$
\begin{eqnarray}
e^{ b \partial  } O(x)   = O (x+b)  e^{ b \partial  } 
 \label{EchangeTranslationO}
\end{eqnarray}

Finally, the product of two translations operators satisfy the simple composition rule 
\begin{eqnarray}
e^{  b_1  \partial  } e^{  b_2  \partial  } =e^{  (b_1+b_2)  \partial  }
 \label{ProductTranslation}
\end{eqnarray}



\subsection{ Explicit forms of the generator ${\bold G} $ and its partner ${\breve {\bold G}} $  as tridiagonal matrices}

\subsubsection{ Generator ${\bold G} $ as a tridiagonal Markov matrix parametrized by the two potentials $U(.)$ and $U_I\left( .+ \frac{1}{2} \right) $}

Using the finite-difference operator of Eq. \ref{nablafinitediff},
the continuity Eq. \ref{Acontinuitynabla}
becomes
\begin{eqnarray}
 \partial_t P_t(x)   && =   - \nabla J_t(x) = - \left( e^{ \frac{\partial}{2}  } - e^{ -  \frac{\partial}{2}} \right)  J_t(x)
 = J_t\left( x- \frac{1}{2} \right)  -  J_t\left( x+ \frac{1}{2} \right)  
 \label{Acontinuitydiscrete}
\end{eqnarray}
while the current of Eq. \ref{AdefJop} with \ref{AJUIU} reads
\begin{eqnarray}
 J_t(x) = {\bold J} P_t(x) && =  - e^{- U_I(x)} \left( e^{ \frac{\partial}{2}  } - e^{ -  \frac{\partial}{2}} \right) e^{U(x)} P_t(x)
\nonumber \\
&&  =  - e^{- U_I(x)+U\left( x+ \frac{1}{2} \right)} P_t\left( x+ \frac{1}{2} \right)
  + e^{- U_I(x)+U\left( x- \frac{1}{2} \right)} P_t\left( x- \frac{1}{2} \right)
 \label{AdefJopdiscretesite}
\end{eqnarray}
i.e. the current on the bond $\left( x+ \frac{1}{2} \right) $ between the two sites $x$ and $(x+1)$ involve the two probabilities $P_t(x)$ and $P_t(x+1)$
\begin{eqnarray}
 J_t\left( x+ \frac{1}{2} \right)  
   =  - e^{- U_I\left( x+ \frac{1}{2} \right)+U(x+1)} P_t(x+1)
   + e^{- U_I\left( x+ \frac{1}{2} \right)+U(x)} P_t(x)
 \label{AdefJopdiscrete}
\end{eqnarray}

So the master equation for the probability $P_t(x)$
\begin{eqnarray}
 \partial_t P_t(x)   && = J_t\left( x- \frac{1}{2} \right)   -  J_t\left( x+ \frac{1}{2} \right)  
 \nonumber \\
 && =   e^{- U_I\left( x- \frac{1}{2} \right)+U(x-1)} P_t(x-1)
   + e^{- U_I\left( x+ \frac{1}{2} \right)+U(x+1)} P_t(x+1)
   - \left[ e^{- U_I\left( x- \frac{1}{2} \right)+U(x)} +e^{- U_I\left( x+ \frac{1}{2} \right)+U(x)} \right] P_t(x)
   \nonumber \\
 && \equiv {\bold G} (x,x-1)P_t(x-1) + {\bold G} (x,x+1)P_t(x+1) + {\bold G} (x,x)P_t(x)
 \label{masterdiscrete}
\end{eqnarray}
is governed by the tridiagonal matrix $ {\bold G}$ with the matrix elements
\begin{eqnarray}
{\bold G} (x,x-1) && \equiv e^{- U_I\left( x- \frac{1}{2} \right)+U(x-1)} ={\bold G}^{\dagger} (x-1,x)
\nonumber \\
{\bold G} (x,x+1) && \equiv e^{- U_I\left( x+ \frac{1}{2} \right)+U(x+1)} ={\bold G}^{\dagger} (x+1,x)
\nonumber \\
{\bold G} (x,x) && \equiv - \left[ e^{- U_I\left( x- \frac{1}{2} \right)+U(x)} +e^{- U_I\left( x+ \frac{1}{2} \right)+U(x)} \right]
= - \left[ {\bold G}(x-1,x) + {\bold G}(x+1,x) \right] ={\bold G}^{\dagger} (x,x)
 \label{Gjumprates}
\end{eqnarray}

So the generator ${\bold G} $ is a tridiagonal Markov matrix, with positive off-diagonal elements ${\bold G} (x,x\pm1) >0$
 representing jump rates into $x$ from its two neighbors $(x \pm 1)$, while the diagonal element ${\bold G} (x,x) $
is negative and represents the opposite of the total jump rate out of site $x$.


The eigenvalue equations for the left eigenvectors $L_n(x)$ read using Eq. \ref{Gjumprates}
\begin{eqnarray}
- E_n L_n(x) && = {\bold G}^{\dagger}(x,x+1) L_n(x+1) 
+ {\bold G}^{\dagger}(x,x-1) L_n(x-1) + {\bold G}^{\dagger}(x,x) L_n(x) 
 \nonumber \\
&& =    e^{U(x) - U_I\left( x+ \frac{1}{2} \right)} 
\big[ L_n(x+1) - L_n(x) \big]
- e^{U(x)- U_I\left( x- \frac{1}{2} \right)} 
\big[ L_n(x) - L_n(x-1) \big]
\nonumber \\
&& \equiv {\bold G}^{\dagger}(x,x+1) 
\big[ L_n(x+1) - L_n(x) \big]
- {\bold G}^{\dagger}(x,x-1)
\big[ L_n(x) - L_n(x-1) \big]
\label{EigenLeftJump}
\end{eqnarray}
that displays the factorization of Eq. \ref{ChoiceTwoleftviaNabla}
\begin{eqnarray}
  {\breve L}_1\left( x+ \frac{1}{2} \right) && = L_1 (x+1) - L_1 (x)
 \nonumber \\
-E_n  L_n(x) && =  \left[ e^{ U(x)  } \nabla e^{- U_I(x) }  \right] {\breve L}_n(x)   
 = e^{ U(x)  } \left( e^{ \frac{\partial}{2}  } - e^{ -  \frac{\partial}{2}} \right) e^{- U_I(x) }  {\breve L}_n(x) 
\nonumber \\
&&  = e^{U(x) - U_I\left( x+ \frac{1}{2} \right)}  {\breve L}_n \left( x+ \frac{1}{2} \right)
 - e^{U(x)- U_I\left( x- \frac{1}{2} \right)}{\breve L}_n \left( x- \frac{1}{2} \right)
 \nonumber \\
 && \equiv {\bold G}^{\dagger}(x,x+1) 
{\breve L}_n \left( x+ \frac{1}{2} \right)
- {\bold G}^{\dagger}(x,x-1)
{\breve L}_n \left( x- \frac{1}{2} \right)
\label{ChoiceTwoleftviaNablajump}
\end{eqnarray}
when one introduces the left eigenvectors $ {\breve L}_n(x) $ of the partner ${\breve {\bold G}} $
written in the next section.


\subsubsection{ Explicit form of the partner ${\breve {\bold G}} $ as a tridiagonal matrix }

The dynamics of the current $J_t\left( x+ \frac{1}{2} \right)$ of Eq. \ref{AdefJopdiscrete} 
obtained using the continuity Eq. \ref{Acontinuitydiscrete}
\begin{eqnarray}
\partial_t J_t\left( x+ \frac{1}{2} \right)  
 &&  =  - e^{- U_I\left( x+ \frac{1}{2} \right)+U(x+1)} \partial_t P_t(x+1)
   + e^{- U_I\left( x+ \frac{1}{2} \right)+U(x)} \partial_t P_t(x)
   \nonumber \\
   && =   - e^{- U_I\left( x+ \frac{1}{2} \right)+U(x+1)} \left[J_t\left( x+ \frac{1}{2} \right)  -  J_t\left( x+ \frac{3}{2} \right) \right]
   + e^{- U_I\left( x+ \frac{1}{2} \right)+U(x)} \left[J_t\left( x- \frac{1}{2} \right)  -  J_t\left( x+ \frac{1}{2} \right) \right]
     \nonumber \\
 && \equiv \breve{ {\bold G} }\left( x+ \frac{1}{2},x+ \frac{3}{2} \right) J_t\left( x+ \frac{3}{2} \right)
 + \breve{ {\bold G} }\left( x+ \frac{1}{2},x- \frac{1}{2} \right) J_t\left( x- \frac{1}{2} \right)
 + \breve{ {\bold G} }\left( x+ \frac{1}{2},x+ \frac{1}{2} \right) J_t\left( x+ \frac{1}{2} \right)
  \nonumber \\
 \label{dynJdiscrete}
\end{eqnarray}
is governed by the tridiagonal matrix $ \breve{ {\bold G} }$ with the matrix elements
\begin{eqnarray}
\breve{ {\bold G} }\left( x+ \frac{1}{2},x+ \frac{3}{2} \right) 
 &&  = e^{- U_I\left( x+ \frac{1}{2} \right)+U(x+1)} ={\breve {\bold G}}^{\dagger}\left( x+ \frac{3}{2},x+ \frac{1}{2} \right)
   \nonumber \\
\breve{ {\bold G} }\left( x+ \frac{1}{2},x- \frac{1}{2} \right) 
  && =   e^{- U_I\left( x+ \frac{1}{2} \right)+U(x)} ={\breve {\bold G}}^{\dagger}\left( x- \frac{1}{2},x+ \frac{1}{2} \right)
     \nonumber \\
\breve{ {\bold G} }\left( x+ \frac{1}{2},x+ \frac{1}{2} \right) 
&& =- e^{- U_I\left( x+ \frac{1}{2} \right)+U(x+1)}-e^{- U_I\left( x+ \frac{1}{2} \right)+U(x)}
= {\breve {\bold G}}^{\dagger}\left( x+ \frac{1}{2},x+ \frac{1}{2} \right)
 \label{Partnerdiscrete}
\end{eqnarray}

The eigenvalue equation for ${\breve L}_n(.) $
\begin{eqnarray}
-E_n {\breve L}_n\left( x+ \frac{1}{2} \right) 
&&= 
{\breve {\bold G}}^{\dagger}\left( x+ \frac{1}{2},x+ \frac{3}{2} \right){\breve L}_n\left( x+ \frac{3}{2} \right)
+{\breve {\bold G}}^{\dagger}\left( x+ \frac{1}{2},x- \frac{1}{2} \right){\breve L}_n\left( x- \frac{1}{2} \right)
+ {\breve {\bold G}}^{\dagger}\left( x+ \frac{1}{2},x+ \frac{1}{2} \right){\breve L}_n\left( x+ \frac{1}{2} \right)
\nonumber \\
&&=  e^{- U_I\left( x+ \frac{3}{2} \right)+U(x+1)}{\breve L}_n\left( x+ \frac{3}{2} \right)
+e^{- U_I\left( x- \frac{1}{2} \right)+U(x)}{\breve L}_n\left( x- \frac{1}{2} \right)
\nonumber \\
&&- \left[ e^{- U_I\left( x+ \frac{1}{2} \right)+U(x+1)}+e^{- U_I\left( x+ \frac{1}{2} \right)+U(x)} \right] {\breve L}_n\left( x+ \frac{1}{2} \right)
\label{EigenLeftbreveDiscrete}
\end{eqnarray}
in agreement with the factorization of Eq. \ref{ChoiceTwoleftviaNablajump}.



\subsection{ Explicit forms of the quantum Hamiltonians
${\bold H} = {\bold Q}^{\dagger} {\bold Q}$ and ${\breve{\bold H}} =  {\bold Q} {\bold Q}^{\dagger} $ 
and ${\bold H}^{[1]} = {\breve{\bold H}} - E_1=\left( {\bold Q}^{[1]} \right)^{\dagger} {\bold Q}^{[1]} $}

\subsubsection{ Explicit form of the quantum Hamiltonian ${\bold H} = {\bold Q}^{\dagger} {\bold Q}$ as a tridiagonal matrix}

The similarity transformation of Eq. \ref{AgeneratorSimilarity}
yields that the Hamiltonian ${\bold H} $ is a symmetric tridiagonal matrix,
whose matrix elements can be directly obtained from the matrix elements of ${\bold G} $ of Eq. \ref{Gjumprates}
\begin{eqnarray}
{\bold H}(x+1,x) &&= - e^{ \frac{U(x+1) }{2} } {\bold G} (x+1,x)e^{ - \frac{U(x) }{2} }   
= - e^{ \frac{U(x+1) +U(x)}{2}  - U_I\left( x+ \frac{1}{2} \right)} = {\bold H}(x,x+1)
 \nonumber \\
 {\bold H}(x-1,x)&& = - e^{ \frac{U(x-1) }{2} } {\bold G} (x-1,x)e^{ - \frac{U(x) }{2} }   
 =- e^{ \frac{U(x)+U(x-1)  }{2} - U_I\left( x- \frac{1}{2} \right)} = {\bold H}(x,x-1)
 \nonumber \\
 {\bold H} (x,x) && = -  {\bold G} (x,x)=e^{U(x) - U_I\left( x+ \frac{1}{2} \right)} + e^{U(x)- U_I\left( x- \frac{1}{2} \right)} 
\label{Htridiago}
\end{eqnarray}

To make the link with the quantum Hamiltonian of Eq. \ref{hsusy} in continuous space where the diffusion coefficient $D(x)$ 
is the amplitude of the kinetic term, it is useful to define the diffusion coefficient $D\left( x+ \frac{1}{2} \right) $ on each bond of the discrete model 
from the off-diagonal elements of Eq. \ref{Htridiago}
\begin{eqnarray}
D\left( x+ \frac{1}{2} \right) \equiv -  {\bold H}(x+1,x) =  e^{ \frac{U(x+1) +U(x)}{2}  - U_I\left( x+ \frac{1}{2} \right)} 
= \sqrt{  {\bold G}(x+1,x)  {\bold G}(x,x+1)} 
\label{Ddiscrete}
\end{eqnarray}
where the last expression gives the re-interpretation in terms of the two jump rates ${\bold G}(x+1,x) $ and ${\bold G}(x,x+1) $
of Eq. \ref{Gjumprates}
between the two sites $x$ and $(x+1)$ in the two directions.

This equation can be used to write the replacement
\begin{eqnarray}
U_I\left( x+ \frac{1}{2} \right) =\frac{U(x+1) +U(x)}{2}  - \ln D\left( x+ \frac{1}{2} \right)
\label{DdiscreteReplacementUI}
\end{eqnarray}
that is the discrete counterpart of the continuous relation of Eq. \ref{UIfromU}.

So the parametrisation in terms of the two potentials $[U(.),U_I(.)]$ can be changed into 
the parametrization in terms of $[U(.),D(.)]$, where the matrix elements of Eq. \ref{Htridiago} become
\begin{eqnarray}
{\bold H}(x+1,x) &&= - D\left( x+ \frac{1}{2} \right)= {\bold H}(x,x+1)
 \nonumber \\
 {\bold H} (x,x) && 
 =D\left( x+ \frac{1}{2} \right)  e^{  \frac{U(x)-U(x+1)}{2}   } 
 + D\left( x- \frac{1}{2} \right)  e^{  \frac{U(x)-U(x-1)}{2}   }
\label{HtridiagoD}
\end{eqnarray}


\subsubsection{ Explicit form of the partner ${\breve{\bold H}} =  {\bold Q} {\bold Q}^{\dagger} $ as a tridiagonal matrix}

The similarity transformation of Eq. \ref{AFPpartnersimilarity}
yields that the partner ${\breve{\bold H}} $ is a symmetric tridiagonal matrix,
whose matrix elements can be directly obtained from the matrix elements of ${\breve {\bold G}} $ of Eq. \ref{Partnerdiscrete}
\begin{eqnarray}
 {\breve{\bold H}} \left( x+ \frac{1}{2},x+ \frac{3}{2} \right)&& 
 = -  e^{ \frac{U_I\left( x+ \frac{1}{2} \right) }{2} }
 {\breve {\bold G}} \left( x+ \frac{1}{2},x+ \frac{3}{2} \right)
 e^{ -\frac{U_I\left( x+ \frac{3}{2} \right) }{2} }  
 = -  e^{ U(x+1) -\frac{U_I\left( x+ \frac{1}{2} \right)+U_I\left( x+ \frac{3}{2} \right) }{2} }
 =  {\breve{\bold H}} \left( x+ \frac{3}{2},x+ \frac{1}{2} \right)
 \nonumber \\
 {\breve{\bold H}} \left( x+ \frac{1}{2},x- \frac{1}{2} \right)&& 
 = -  e^{ \frac{U_I\left( x+ \frac{1}{2} \right) }{2} }
 {\breve {\bold G}} \left( x+ \frac{1}{2},x- \frac{1}{2} \right)
 e^{ -\frac{U_I\left( x- \frac{1}{2} \right) }{2} }  
 =  -  e^{ U(x) -\frac{U_I\left( x+ \frac{1}{2} \right)+U_I\left( x- \frac{1}{2} \right) }{2} } 
 =  {\breve{\bold H}} \left( x- \frac{1}{2},x+ \frac{1}{2} \right)
 \nonumber \\
\breve{ {\bold H} }\left( x+ \frac{1}{2},x+ \frac{1}{2} \right)  && = - \breve{ {\bold G} }\left( x+ \frac{1}{2},x+ \frac{1}{2} \right) 
 = e^{- U_I\left( x+ \frac{1}{2} \right)+U(x+1)}+e^{- U_I\left( x+ \frac{1}{2} \right)+U(x)}
=\breve{ {\bold H} }\left( x+ \frac{1}{2},x+ \frac{1}{2} \right)
\label{Hpartnertridiago}
\end{eqnarray}

So the corresponding diffusion coefficient analog to Eq. \ref{Ddiscrete} given by the opposite of the off-diagonal matrix elements read
\begin{eqnarray}
{\breve D}(x) \equiv -   {\breve{\bold H}}  \left( x+ \frac{1}{2},x- \frac{1}{2} \right) 
=  e^{ U(x) -\frac{U_I\left( x+ \frac{1}{2} \right)+U_I\left( x- \frac{1}{2} \right) }{2} }
= \sqrt{ \breve{ {\bold G} }\left( x+ \frac{1}{2},x- \frac{1}{2} \right) \breve{ {\bold G} }\left( x- \frac{1}{2},x+ \frac{1}{2} \right)} 
\label{Ddiscretebreve}
\end{eqnarray}

This equation can be used to write the replacement
\begin{eqnarray}
  U(x) = \frac{U_I\left( x+ \frac{1}{2} \right)+U_I\left( x- \frac{1}{2} \right) }{2} + \ln {\breve D}(x)
\label{DdiscretebreveReplacementUI}
\end{eqnarray}
that will change the parametrisation in terms of the two potentials $[U(.),U_I(.)]$ by
the parametrization in $[U_I(.),{\breve D}(.)]$, where the matrix elements of Eq. \ref{Hpartnertridiago} become
\begin{eqnarray}
 {\breve{\bold H}} \left( x+ \frac{1}{2},x- \frac{1}{2} \right)&& 
 = - {\breve D}(x)=  {\breve{\bold H}} \left( x- \frac{1}{2},x+ \frac{1}{2} \right)
 \nonumber \\
\breve{ {\bold H} }\left( x+ \frac{1}{2},x+ \frac{1}{2} \right)  && 
 ={\breve D}(x+1) e^{\frac{U_I\left( x+ \frac{3}{2} \right)-U_I\left( x+ \frac{1}{2} \right) }{2} }
 +{\breve D}(x)e^{\frac{U_I\left( x- \frac{1}{2} \right)-U_I\left( x+ \frac{1}{2} \right) }{2} }
\label{HpartnertridiagoDbreve}
\end{eqnarray}


\subsubsection{ Construction of the new Hamiltonian  ${\bold H}^{[1]} \equiv \left( {\bold Q}^{[1]} \right)^{\dagger} {\bold Q}^{[1]}= {\breve{\bold H}} - E_1$   }

On one hand, the new Hamiltonian ${\bold H}^{[1]}={\breve{\bold H}} - E_1 $ 
can be directly obtained from the tridiagonal partner ${\breve{\bold H}} $ 
written in Eqs \ref{Hpartnertridiago} and \ref{HpartnertridiagoDbreve}
\begin{small}
\begin{eqnarray}
{\bold H}^{[1]}\left( x+ \frac{1}{2},x- \frac{1}{2} \right)  
&& = {\breve{\bold H}} \left( x+ \frac{1}{2},x- \frac{1}{2} \right) 
= - e^{ U(x) -\frac{U_I\left( x+ \frac{1}{2} \right)+U_I\left( x- \frac{1}{2} \right) }{2} }
 = - {\breve D}(x) 
 \nonumber \\
{\bold H}^{[1]}\left( x+ \frac{1}{2},x+ \frac{1}{2} \right) +E_1
= \breve{ {\bold H} }\left( x+ \frac{1}{2},x+ \frac{1}{2} \right)   
&& = e^{- U_I\left( x+ \frac{1}{2} \right)+U(x+1)}+e^{- U_I\left( x+ \frac{1}{2} \right)+U(x)}
\nonumber \\
&& ={\breve D}(x+1) e^{\frac{U_I\left( x+ \frac{3}{2} \right)-U_I\left( x+ \frac{1}{2} \right) }{2} }
 +{\breve D}(x)e^{\frac{U_I\left( x- \frac{1}{2} \right)-U_I\left( x+ \frac{1}{2} \right) }{2} } 
\label{H1frompartner}
\end{eqnarray}
\end{small}

On the other hand, one wishes to write the parametrization 
of the new Hamiltonian ${\bold H}^{[1]} $ that is the analog of the initial Hamiltonian
${\bold H} $ of Eqs \ref{Htridiago} and \ref{HtridiagoD},
except for the exchange of the role of sites and bonds 
\begin{small}
\begin{eqnarray}
{\bold H}^{[1]}\left( x+ \frac{1}{2},x- \frac{1}{2} \right) &&=  - e^{ \frac{U^{[1]}\left( x+ \frac{1}{2} \right) +U^{[1]}\left( x- \frac{1}{2} \right)}{2}  - U_I^{[1]}(x)}= - D^{[1]}(x)
 \nonumber \\
 {\bold H}^{[1]} \left( x+ \frac{1}{2},x+ \frac{1}{2} \right)&& 
 =e^{U^{[1]}\left( x+ \frac{1}{2} \right) - U_I^{[1]}(x+1 )} 
 + e^{U^{[1]}\left( x+ \frac{1}{2} \right)- U_I^{[1]}(x)}
 =D^{[1]}(x+1)  e^{  \frac{U^{[1]}\left( x+ \frac{1}{2} \right)-U^{[1]}\left( x+ \frac{3}{2} \right)}{2}   } 
 + D^{[1]}(x)  e^{  \frac{U^{[1]}\left( x+ \frac{1}{2} \right)-U^{[1]}\left( x- \frac{1}{2} \right)}{2}   }
  \nonumber \\
\label{H1parametrization}
\end{eqnarray}
\end{small}

The identification between the two expressions of Eqs \ref{H1frompartner} and \ref{H1parametrization} leads to the following discussion :

 (i) The identification of the off-diagonal matrix elements ${\bold H}^{[1]}\left( x+ \frac{1}{2},x- \frac{1}{2} \right) $
 yields that the new diffusion coefficient $ D^{[1]}(x)$ coincides with the diffusion coefficient ${\breve D}(x)  $ of the partner 
 \begin{eqnarray}
D^{[1]}(x)  && = {\breve D}(x) 
 \nonumber \\
\text{ or equivalently} \ \ \ \ e^{ \frac{U^{[1]}\left( x+ \frac{1}{2} \right) +U^{[1]}\left( x- \frac{1}{2} \right)}{2}  - U_I^{[1]}(x)}
&& = e^{ U(x) -\frac{U_I\left( x+ \frac{1}{2} \right)+U_I\left( x- \frac{1}{2} \right) }{2} }
\label{H1offdiagidentif}
\end{eqnarray}

From the point of view of the potentials, this equation can be used to compute the potential $U_I^{[1]}(.) $
in terms of the three other potentials $[U^{[1]}(.);U(.);U_I(.)]$
\begin{eqnarray}
U_I^{[1]}(x) && =  \frac{U^{[1]}\left( x+ \frac{1}{2} \right) +U^{[1]}\left( x- \frac{1}{2} \right)}{2}  - 
  U(x) +\frac{U_I\left( x+ \frac{1}{2} \right)+U_I\left( x- \frac{1}{2} \right) }{2} 
  \nonumber \\
  && =  \frac{U^{[1]}\left( x+ \frac{1}{2} \right) +U^{[1]}\left( x- \frac{1}{2} \right)}{2}  -  \ln \left( {\breve D}(x)  \right) 
\label{UI1Elimination}
\end{eqnarray}
where the last expression is the discrete counterpart of the continuous expression of Eq. \ref{UI1fromU1}.

(ii) The identification of the diagonal matrix elements ${\bold H}^{[1]} \left( x+ \frac{1}{2},x+ \frac{1}{2} \right) $ reads
\begin{small}
\begin{eqnarray}
D^{[1]}(x+1)  e^{  \frac{U^{[1]}\left( x+ \frac{1}{2} \right)-U^{[1]}\left( x+ \frac{3}{2} \right)}{2}   } 
 + D^{[1]}(x)  e^{  \frac{U^{[1]}\left( x+ \frac{1}{2} \right)-U^{[1]}\left( x- \frac{1}{2} \right)}{2}   }
 && ={\breve D}(x+1) e^{\frac{U_I\left( x+ \frac{3}{2} \right)-U_I\left( x+ \frac{1}{2} \right) }{2} }
 +{\breve D}(x)e^{\frac{U_I\left( x- \frac{1}{2} \right)-U_I\left( x+ \frac{1}{2} \right) }{2} } - E_1
 \nonumber \\
\text{ or equivalently} 
e^{U^{[1]}\left( x+ \frac{1}{2} \right) - U_I^{[1]}(x+1 )} 
 + e^{U^{[1]}\left( x+ \frac{1}{2} \right)- U_I^{[1]}(x)}
&& =  e^{- U_I\left( x+ \frac{1}{2} \right)+U(x+1)}+e^{- U_I\left( x+ \frac{1}{2} \right)+U(x)}
 - E_1
\label{H1diagidentif}
\end{eqnarray}
\end{small}

One can then take into account the output of (i) for the off-diagonal elements in two ways :

(ii-a) If one wishes to work with diffusion coefficients, one can plug the first line $D^{[1]}(x)   = {\breve D}(x)  $ of Eq. \ref{H1offdiagidentif}
into the first line of Eq. \ref{H1diagidentif} to obtain
\begin{small}
\begin{eqnarray}
{\breve D}(x+1)  e^{  \frac{U^{[1]}\left( x+ \frac{1}{2} \right)-U^{[1]}\left( x+ \frac{3}{2} \right)}{2}   } 
 + {\breve D}(x)  e^{  \frac{U^{[1]}\left( x+ \frac{1}{2} \right)-U^{[1]}\left( x- \frac{1}{2} \right)}{2}   }
 && ={\breve D}(x+1) e^{\frac{U_I\left( x+ \frac{3}{2} \right)-U_I\left( x+ \frac{1}{2} \right) }{2} }
 +{\breve D}(x)e^{\frac{U_I\left( x- \frac{1}{2} \right)-U_I\left( x+ \frac{1}{2} \right) }{2} } - E_1
\nonumber \\
\label{H1diagidentifD}
\end{eqnarray}
\end{small}
that can be considered as the discrete counterpart of the Riccati Eq. \ref{Riccati} concerning the continuous model,
and that can be rewritten as
\begin{eqnarray}
{\breve D}(x+1)  e^{ - \frac{W^{[1]}(x+1)}{2}   } 
 + {\breve D}(x)  e^{  \frac{W^{[1]}(x)}{2}   }
 && ={\breve D}(x+1) e^{\frac{W_I(x+1) }{2} }
 +{\breve D}(x)e^{-\frac{W_I(x) }{2} } - E_1
\nonumber \\
\label{H1diagidentifDW}
\end{eqnarray}
in terms of the local potential differences
\begin{eqnarray}
W^{[1]}(x) && \equiv U^{[1]}\left( x+ \frac{1}{2} \right)-U^{[1]}\left( x- \frac{1}{2} \right)
\nonumber \\
W_I(x) && \equiv U_I\left( x+ \frac{1}{2} \right)-U_I\left( x- \frac{1}{2} \right)
\label{Wpotdiff}
\end{eqnarray}

(ii-b) If one prefers to work with the potentials, one can plug the potential $U_I^{[1]}(.) $ of Eq. \ref{UI1Elimination}
into the second line of Eq. \ref{H1diagidentif} to obtain the equation that determines the new potential $U^{[1]} $ in terms of
the two initial potentials $[U(.);U_I(.)]$ 
\begin{small}
\begin{eqnarray}
&& e^{  \frac{ U^{[1]}\left( x+ \frac{1}{2} \right)- U^{[1]}\left( x+ \frac{3}{2} \right) }{2} 
+U(x+1) -\frac{U_I\left( x+ \frac{3}{2} \right)+U_I\left( x+ \frac{1}{2} \right) }{2} }
 + e^{  \frac{U^{[1]}\left( x+ \frac{1}{2} \right) -U^{[1]}\left( x- \frac{1}{2} \right)}{2} 
+U(x) -\frac{U_I\left( x+ \frac{1}{2} \right)+U_I\left( x- \frac{1}{2} \right) }{2} }
\nonumber \\
&& =  e^{- U_I\left( x+ \frac{1}{2} \right)+U(x+1)}+e^{- U_I\left( x+ \frac{1}{2} \right)+U(x)}
 - E_1
\label{H1diagidentifpot}
\end{eqnarray}
\end{small}
that is also an equation for the local potential differences $W^{[1]}(.) $ introduced in Eq. \ref{Wpotdiff}
\begin{eqnarray}
&& e^{ - \frac{ W^{[1]}(x+1) }{2} 
+U(x+1) -\frac{U_I\left( x+ \frac{3}{2} \right)+U_I\left( x+ \frac{1}{2} \right) }{2} }
 + e^{  \frac{W^{[1]}(x)}{2} 
+U(x) -\frac{U_I\left( x+ \frac{1}{2} \right)+U_I\left( x- \frac{1}{2} \right) }{2} }
\nonumber \\
&& =  e^{- U_I\left( x+ \frac{1}{2} \right)+U(x+1)}+e^{- U_I\left( x+ \frac{1}{2} \right)+U(x)}
 - E_1
\label{H1diagidentifpotW}
\end{eqnarray}



\subsection{ Construction of the new Markov matrix ${\bold G}^{[1]}$
via the Doob-transformation of the partner ${\breve {\bold G} } $  }

The Doob-transformation of Eq. \ref{ADoobRingDagger} 
yields that the matrix elements of $\left( {\bold G}^{[1]}  \right)^{\dagger} $ can be constructed from the 
matrix elements of $ {\breve {\bold G} }^{\dagger}$ written in Eq. \ref{Partnerdiscrete}
and from the left eigenvector ${\breve L}_1(.) $ via 
\begin{eqnarray}
 \left( {\bold G}^{[1]}  \right)^{\dagger}\left( x- \frac{1}{2},x+ \frac{1}{2} \right)  && 
 = \frac{1}{ {\breve L}_1\left( x- \frac{1}{2} \right) } {\breve {\bold G} }^{\dagger}\left( x- \frac{1}{2},x+ \frac{1}{2} \right)  {\breve L}_1\left( x+ \frac{1}{2} \right) 
 = \frac{{\breve L}_1\left( x+ \frac{1}{2} \right)}{ {\breve L}_1\left( x- \frac{1}{2} \right) }
 e^{- U_I\left( x+ \frac{1}{2} \right)+U(x)}
 \nonumber \\
 \left( {\bold G}^{[1]}  \right)^{\dagger}\left( x+ \frac{1}{2},x- \frac{1}{2} \right) && 
 = \frac{1}{ {\breve L}_1\left( x+ \frac{1}{2} \right) } {\breve {\bold G} }^{\dagger}\left( x+ \frac{1}{2},x- \frac{1}{2} \right) {\breve L}_1\left( x- \frac{1}{2} \right)  
 =  \frac{{\breve L}_1\left( x- \frac{1}{2} \right)}{ {\breve L}_1\left( x+ \frac{1}{2} \right) }
 e^{- U_I\left( x- \frac{1}{2} \right)+U(x)}
  \nonumber \\
 \left( {\bold G}^{[1]}  \right)^{\dagger} \left( x+ \frac{1}{2},x+ \frac{1}{2} \right)&& =  {\breve {\bold G} }^{\dagger} \left( x+ \frac{1}{2},x+ \frac{1}{2} \right) + E_1
 = - e^{- U_I\left( x+ \frac{1}{2} \right)+U(x+1)}-e^{- U_I\left( x+ \frac{1}{2} \right)+U(x)} +E_1
  \label{ADoobRingDaggerjump}
\end{eqnarray}

The identification with the parametrization of the genuine Markov matrix ${\bold G}^{[1]} $ analog to Eq. \ref{Gjumprates}
in terms of the two potentials $U^{[1]}(.)$ and $U^{[1]}_I(.)$
\begin{eqnarray}
 \left( {\bold G}^{[1]}  \right)^{\dagger}\left( x- \frac{1}{2},x+ \frac{1}{2} \right) && 
 = e^{- U_I^{[1]}(x)+U^{[1]}\left( x- \frac{1}{2} \right)} 
 ={\bold G}^{[1]}\left( x+ \frac{1}{2},x- \frac{1}{2} \right)
\nonumber \\
 \left( {\bold G}^{[1]}  \right)^{\dagger}\left( x+ \frac{1}{2},x- \frac{1}{2} \right) && 
 = e^{- U_I^{[1]}(x)+U^{[1]}\left( x+ \frac{1}{2} \right)} 
 = {\bold G}^{[1]}\left( x- \frac{1}{2},x+ \frac{1}{2} \right)
\nonumber \\
\left( {\bold G}^{[1]}  \right)^{\dagger} \left( x+ \frac{1}{2},x+ \frac{1}{2} \right) && 
= - \left[ e^{- U_I^{[1]}(x)+U^{[1]}\left( x+ \frac{1}{2} \right)} +e^{- U_I^{[1]}(x+1)+U^{[1]}\left( x+ \frac{1}{2} \right)} \right]
= {\bold G}^{[1]}  \left( x+ \frac{1}{2},x+ \frac{1}{2} \right)  
 \label{G1jumprates}
\end{eqnarray}
leads to the following discussion :

(1) The identification of the two types of off-diagonal elements of $\left( {\bold G}^{[1]}  \right)^{\dagger} \left( x\pm \frac{1}{2},x\mp \frac{1}{2} \right)$ 
between Eqs \ref{ADoobRingDaggerjump}
and \ref{G1jumprates}
 can be rewritten as equations for the same ratio $\frac{{\breve L}_1\left( x+ \frac{1}{2} \right)}{ {\breve L}_1\left( x- \frac{1}{2} \right) } $ of two consecutive components of the left eigenvector ${\breve L}_1(.) $
\begin{eqnarray}
\frac{{\breve L}_1\left( x+ \frac{1}{2} \right)}{ {\breve L}_1\left( x- \frac{1}{2} \right) }
&& =\frac{ \left( {\bold G}^{[1]}  \right)^{\dagger}\left( x- \frac{1}{2},x+ \frac{1}{2} \right) }
{ {\breve {\bold G} }^{\dagger}\left( x- \frac{1}{2},x+ \frac{1}{2} \right)}  
= \frac{ e^{- U_I^{[1]}(x)+U^{[1]}\left( x- \frac{1}{2} \right)}  }
{e^{- U_I\left( x+ \frac{1}{2} \right)+U(x)}  }
 \nonumber \\
 \frac{{\breve L}_1\left( x+ \frac{1}{2} \right)}{{\breve L}_1\left( x- \frac{1}{2} \right)}  
 && 
 = \frac{ {\breve {\bold G} }^{\dagger}\left( x+ \frac{1}{2},x- \frac{1}{2} \right) }
 {\left( {\bold G}^{[1]}  \right)^{\dagger}\left( x+ \frac{1}{2},x- \frac{1}{2} \right) }
 =  \frac{ e^{- U_I\left( x- \frac{1}{2} \right)+U(x)} }
 { e^{- U_I^{[1]}(x)+U^{[1]}\left( x+ \frac{1}{2} \right)} }
  \label{ADoobRingDaggerjumpoffdiag}
\end{eqnarray}

The compatibility between these two equations yields
\begin{eqnarray}
 \left( {\bold G}^{[1]}  \right)^{\dagger}\left( x- \frac{1}{2},x+ \frac{1}{2} \right)
 \left( {\bold G}^{[1]}  \right)^{\dagger}\left( x+ \frac{1}{2},x- \frac{1}{2} \right)
 && = {\breve {\bold G} }^{\dagger}\left( x- \frac{1}{2},x+ \frac{1}{2} \right)
 {\breve {\bold G} }^{\dagger}\left( x+ \frac{1}{2},x- \frac{1}{2} \right)
 \nonumber \\
\text{ i.e. } \ \   e^{- 2 U_I^{[1]}(x)+U^{[1]}\left( x- \frac{1}{2} \right)+U^{[1]}\left( x+ \frac{1}{2} \right)} 
&&  =e^{- U_I\left( x+ \frac{1}{2} \right)- U_I\left( x- \frac{1}{2} \right)+2U(x)} 
 \nonumber \\
\text{ i.e. } \ \      D^{[1]}(x)   = {\breve D}(x)
  \label{compatibilityleftDoob}
\end{eqnarray}
which is equivalent to Eq. \ref{H1offdiagidentif} concerning the 
off-diagonal matrix elements of the quantum Hamiltonians parametrized by the diffusion coefficients $ D^{[1]}(x)   = {\breve D}(x) $

The remaining independent equation can be chosen to be the product of the two equations of Eq. \ref{ADoobRingDaggerjumpoffdiag}
\begin{eqnarray}
\left( \frac{{\breve L}_1\left( x+ \frac{1}{2} \right)}{ {\breve L}_1\left( x- \frac{1}{2} \right) } \right)^2
&&  
= \frac{ e^{
U_I\left( x+ \frac{1}{2} \right)-U^{[1]}\left( x+ \frac{1}{2} \right)}  }
{e^{U_I\left( x- \frac{1}{2} \right)- U^{[1]}\left( x- \frac{1}{2} \right)}  }
  \label{ADoobRingDaggerjumpoffdiagprod}
\end{eqnarray}
that corresponds to Eq. \ref{AU1xbreve} of the general analysis 
 \begin{eqnarray}
   {\breve L}_1\left( x+ \frac{1}{2} \right) = e^{ \frac{U_I\left( x+ \frac{1}{2} \right)-U^{[1]}\left( x+ \frac{1}{2} \right)}{2} }
\label{AU1xbrevediscrete}
\end{eqnarray}

(2) The identification of the diagonal elements of $\left( {\bold G}^{[1]}  \right)^{\dagger} $ 
in Eqs \ref{ADoobRingDaggerjump}
and \ref{G1jumprates}
\begin{eqnarray}
 \left( {\bold G}^{[1]}  \right)^{\dagger} \left( x+ \frac{1}{2},x+ \frac{1}{2} \right)&& =  {\breve {\bold G} }^{\dagger} \left( x+ \frac{1}{2},x+ \frac{1}{2} \right) + E_1
 \nonumber \\
\text{ i.e. } \ \ 
   e^{- U_I^{[1]}(x)+U^{[1]}\left( x+ \frac{1}{2} \right)} +e^{- U_I^{[1]}(x+1)+U^{[1]}\left( x+ \frac{1}{2} \right)}
&& =  e^{- U_I\left( x+ \frac{1}{2} \right)+U(x+1)} +e^{- U_I\left( x+ \frac{1}{2} \right)+U(x)} -E_1
  \label{IdDiagDiscrete}
\end{eqnarray}
is equivalent to 
Eq. \ref{H1diagidentif} concerning the 
diagonal matrix elements of the quantum Hamiltonians.

To better understand the meaning of this equation \ref{IdDiagDiscrete} ,
it is useful to compare with the eigenvalue Eq. \ref{EigenLeftbreveDiscrete} for ${\breve L}_1(.) $
that can be rewritten using Eqs \ref{ADoobRingDaggerjumpoffdiag}
\begin{eqnarray}
E_1  + {\breve {\bold G}}^{\dagger}\left( x+ \frac{1}{2},x+ \frac{1}{2} \right)
&&= 
- {\breve {\bold G}}^{\dagger}\left( x+ \frac{1}{2},x+ \frac{3}{2} \right) 
\frac{{\breve L}_1\left( x+ \frac{3}{2} \right)}{{\breve L}_1\left( x+ \frac{1}{2} \right) }
-{\breve {\bold G}}^{\dagger}\left( x+ \frac{1}{2},x- \frac{1}{2} \right) 
\frac{{\breve L}_1\left( x- \frac{1}{2} \right)}{{\breve L}_1\left( x+ \frac{1}{2} \right)}
\nonumber \\
&&= - \left( {\bold G}^{[1]}  \right)^{\dagger}\left( x+ \frac{1}{2},x+ \frac{3}{2} \right)
-\left( {\bold G}^{[1]}  \right)^{\dagger}\left( x+ \frac{1}{2},x- \frac{1}{2} \right)
\nonumber \\
&&=  \left( {\bold G}^{[1]}  \right)^{\dagger}\left( x+ \frac{1}{2},x+ \frac{1}{2} \right)
\label{EigenLeftbreveDiscrete1}
\end{eqnarray}
to conclude that this eigenvalue Eq. for ${\breve L}_1(.) $ coincides with Eq. \ref{IdDiagDiscrete}.


\subsection{ Discussion  }

In conclusion, the Doob-transformation of the Markov perspective is equivalent to the supersymmetric recursion
of the quantum perspective described in the previous subsection, but reveals the role of the left eigenvector ${\breve L}_1(.) $
as already explained in the general analysis of subsection \ref{sub_generalSusyDoobEquiv},
but with the following more precise discussion adapted to the discrete space, where the various operators are
tridiagonal matrices :

(1) the analysis of the off-diagonal elements of ${\bold G}^{[1]} $
gives both  the relation of Eq. \ref{AU1xbrevediscrete}
corresponding to Eq. \ref{AU1xbreve} of the general analysis that can be used to rewrite $U^{[1]}(.) $
 \begin{eqnarray}
 U^{[1]}\left( x+ \frac{1}{2} \right) 
  =  U_I\left( x+ \frac{1}{2} \right)   -  2 \ln {\breve L}_1\left( x+ \frac{1}{2} \right)
\label{AU1xbrevediscreteElim}
\end{eqnarray}
and the relation of Eq. \ref{H1offdiagidentif} concerning the 
off-diagonal matrix elements of the quantum Hamiltonians parametrized 
by the diffusion coefficients $ D^{[1]}(x)   = {\breve D}(x) $
that can be used to compute $U_I^{[1]}(x) $ in terms of the three other potentials $U^{[1]} $ , $ U_I(.)$ and $U(.) $
via Eq. \ref{UI1Elimination}. As a consequence, one can plug the potential $U^{[1]} $ of Eq. \ref{AU1xbrevediscreteElim}
into Eq. \ref{UI1Elimination} to obtain the expression of 
 $U_I^{[1]}(x) $ in terms of the two initial potentials $ U_I(.)$ and $U(.) $ and of the left eigenvector ${\breve L}_1(.) $
\begin{eqnarray}
  U_I^{[1]}(x) && =\frac{U^{[1]}\left( x- \frac{1}{2} \right)+ U_I\left( x- \frac{1}{2} \right)
  + U^{[1]}\left( x+ \frac{1}{2} \right) + U_I\left( x+ \frac{1}{2} \right)
  }{2} -U(x) 
\nonumber \\
   && = U_I\left( x- \frac{1}{2} \right)   
  +U_I\left( x+ \frac{1}{2} \right)    -U(x) 
  - \ln {\breve L}_1\left( x- \frac{1}{2} \right)-\ln {\breve L}_1\left( x+ \frac{1}{2} \right)
  \label{EmimUI1fromDL1}
\end{eqnarray}

In summary, the left eigenvector ${\breve L}_1(.) $ is useful to write the two new potentials $ U^{[1]}(.)$ and $U^{[1]}_I(.) $
in terms of the two initial potentials $ U_I(.)$ and $U(.) $ via Eqs \ref{AU1xbrevediscreteElim}
and \ref{EmimUI1fromDL1}.

(2) the eigenvalue Eq \ref{EigenLeftbreveDiscrete}
for the left eigenvector ${\breve L}_1(.) $ is equivalent to the Doob transformation for the diagonal elements
$\left( {\bold G}^{[1]}  \right)^{\dagger} \left( x+ \frac{1}{2},x+ \frac{1}{2} \right)
={\bold G}^{[1]}  \left( x+ \frac{1}{2},x+ \frac{1}{2} \right)$
and to Eq. \ref{H1diagidentif} concerning the 
diagonal matrix elements of the quantum Hamiltonians.

This eigenvalue equation can be rewritten via the natural factorization of Eq. \ref{ChoiceTwoleftviaNablajump} 
involving $L_1(x)$ 
\begin{eqnarray}
-E_1  L_1(x) &&   = e^{U(x) - U_I\left( x+ \frac{1}{2} \right)}  {\breve L}_1 \left( x+ \frac{1}{2} \right)
 - e^{U(x)- U_I\left( x- \frac{1}{2} \right)}{\breve L}_1 \left( x- \frac{1}{2} \right)
 \nonumber \\
  {\breve L}_1\left( x+ \frac{1}{2} \right) &&
 = L_1 (x+1) - L_1 (x)
\label{FactotizedEigenLeftDiscrete}
\end{eqnarray}
The first equation can be used to write either $U(x)$ or $U_I(x)$
as a function of the other potential and the two left eigenvectors $L_1(.)$ and ${\breve L}_1\left( x+ \frac{1}{2} \right)  = L_1 (x+1) - L_1 (x)$
as follows

(2a) The potential $U(x)$ can be obtained from
$U_I(.)$ and the two left eigenvectors via
\begin{eqnarray}
e^{U(x)} =
E_1  \frac{ L_1(x) } { e^{- U_I\left( x- \frac{1}{2} \right)}{\breve L}_1 \left( x- \frac{1}{2} \right) 
- e^{ - U_I\left( x+ \frac{1}{2} \right)}  {\breve L}_1 \left( x+ \frac{1}{2} \right)  } 
\label{UDiscreteUILeft}
\end{eqnarray}

(2b) The rewriting of the first equation of Eq. \ref{FactotizedEigenLeftDiscrete} as the difference equation
\begin{eqnarray}
-E_1  L_1(x) e^{- U(x) }&&   =  e^{ - U_I\left( x+ \frac{1}{2} \right)}  {\breve L}_1 \left( x+ \frac{1}{2} \right)
 - e^{- U_I\left( x- \frac{1}{2} \right)}{\breve L}_1 \left( x- \frac{1}{2} \right) 
\label{UDiscreteUILeftderitotR1}
\end{eqnarray}
leads to the inversion
\begin{eqnarray}
e^{ - U_I\left( x+ \frac{1}{2} \right)}  {\breve L}_1 \left( x+ \frac{1}{2} \right) &&= -E_1 \sum_{X=..}^x L_1(X) e^{- U(X) }  
\label{UDiscreteUILeftderitotR1inv}
\end{eqnarray}
and to the expression of $U_I(.)$ as a function of $U(.)$ and the two left eigenvectors via
\begin{eqnarray}
e^{ - U_I\left( x+ \frac{1}{2} \right)}   &&= - \frac{ E_1 }{ {\breve L}_1 \left( x+ \frac{1}{2} \right)} \sum_{X=..}^x L_1(X) e^{- U(X) }  
\label{UDiscreteUILeftderitotR1UIfromU}
\end{eqnarray}



\end{document}